\documentclass{aa}

\usepackage{graphicx}
\usepackage{txfonts}
\usepackage{dsfont}
\usepackage{hyperref}
\usepackage{xcolor}
\usepackage[normalem]{ulem}
\usepackage{natbib}
\bibpunct{(}{)}{;}{a}{}{,}

\newcommand{\ha}{H$\alpha$}
\newcommand{\hii}{H\,{\sc ii}}
\newcommand{\hi}{H\,{\sc i}}

\usepackage{amstext}

\begin{document}

   \title{Tracking down the origin of superbubbles and supergiant shells in the Magellanic Clouds with Minkowski tensor analysis}

   \titlerunning{Minkowski tensor analysis of MCs}
   
\author{Caroline Collischon \inst{1,2} \and Manami Sasaki \inst{2} \and Klaus Mecke\inst{1} \and Sean~D.~Points\inst{3} \and Michael A. Klatt\inst{4}
}

\authorrunning{Collischon et al.}

\institute{
 Institut f\"ur Theoretische Physik, Universit\"at Erlangen-N\"urnberg, Staudtstr. 7, 91058 Erlangen, Germany 
\and
Remeis Observatory and ECAP, Universit\"{a}t Erlangen-N\"{u}rnberg, Sternwartstr. 7, 96049 Bamberg, Germany
\and Cerro Tololo Inter-American Observatory/NSF's NOIRLab, Casilla 603, La Serena, Chile
\and Department of Physics, Princeton University, Princeton, NJ 08544, USA
}

\abstract{}{
We develop an automatic bubble-recognition routine based on Minkowski functionals (MF) and tensors (MT) to detect bubble-like interstellar structures in optical emission line images.
}{
Minkowski functionals and MT are powerful mathematical tools for parameterizing the shapes of bodies. Using the papaya2-library, we created maps of the desired MF or MT of structures at a given window size. We used maps of the irreducible MT $\psi_2$, which is sensitive to elongation, to find filamentary regions in H$\alpha$, [\ion{S}{II}], and [\ion{O}{III}] images of the Magellanic Cloud Emission Line Survey (MCELS). Using the phase of $\psi_2$, we were able to draw lines perpendicular to each filament and thus obtain line-density maps. This allowed us to find the center of a bubble-like structure and to detect structures at different window sizes.
}{
The detected bubbles in all bands are spatially correlated to the distribution of massive stars, showing that we indeed detect interstellar bubbles without large spatial bias. 
Eighteen out of 59 supernova remnants in the Large Magellanic Cloud (LMC) and 13 out of 20  superbubbles are detected in at least one wavelength. The lack of detection is mostly due to surrounding emission that disturbs the detection, a too small size, or the lack of a (circular) counterpart in our emission line images. In line-density maps at larger scales, maxima can be found in regions with high star formation in the past, often inside supergiant shells (SGS). In SGS LMC 2, there is a maximum west of the shell where a collision of large gas clouds is thought to have occurred. In the Small Magellanic Cloud (SMC), bubble detection is impaired by the more complex projected structure of the galaxy. Line maps at large scales show large filaments in the SMC in a north-south direction, especially in the [\ion{S}{II}] image. The origin of these filaments is unknown. 
}{}


   \maketitle

\section{Introduction}

Stars are formed out of dense cores of cold interstellar molecular clouds through gravitational collapse.
Young massive stars inject energy into the interstellar medium (ISM) through radiation and stellar winds, partly ionizing their environment and thus forming \hii\ regions. 
The shock waves of stellar winds can further heat and ionize the ambient gas, and create interstellar bubbles filled with hot thin plasma. These stars complete their evolution in supernova (SN) explosions, which again heat and ionize the interstellar medium and create structures called supernova remnants (SNRs). 
Because stars are formed in associations or clusters, the interaction of many stars with their ambient interstellar medium results in large superbubbles with sizes of typically 100 -- 1000 pc.
Superbubbles are usually embedded in warm ionized gas, which can be observed as large \hii\ regions.
The ISM thus has multiple phases, from cold molecular clouds and neutral gas with temperatures of 10 -- 100 K to hot low-density plasma with 10$^7$ K, and comprises sites of stellar birth, evolution, and death.

The Large Magellanic Cloud (LMC) and the Small Magellanic Cloud (SMC) are the largest satellite galaxies of the 
Milky Way. They are located at a distance of $\sim$50~kpc \citep{pietr2019} and 60~kpc \citep{degrijs2015}, respectively.
They are found off the Galactic plane of the Milky Way, so that the extinction
toward the LMC and SMC is relatively low.
Their proximity, position in the sky, and the modest absorption in the line 
of sight (Galactic foreground $N_{\rm H}$ = 0.6 $\times$ 10$^{21}$~cm$^{-2}$) make 
them an ideal laboratory for a detailed study of the ISM in a galaxy.
Observations of the 21cm line of atomic hydrogen toward the LMC and the SMC has
been carried out since the 1960s \citep{1964IAUS...20..289M,1967AuJPh..20..147H} and revealed an extended distribution of atomic interstellar gas.
Emission line images in the optical have shown that the Magellanic Clouds host a large population of \hii\ regions, bubbles, and superbubbles of various sizes \citep{1956ApJS....2..315H,1976MmRAS..81...89D}, many of which also show X-ray emission from the hot low-density plasma in their interiors. 
The combination of newer \textit{XMM-Newton} observations with optical images of the Magellanic Clouds Emission Line Survey \citep[MCELS,][]{2004AAS...20510108S} and additional radio data have allowed studies of the properties of superbubbles in the Magellanic Clouds and their origin \citep{2011A&A...528A.136S,2012A&A...547A..19K,2015A&A...573A..73K,2019A&A...621A.138K}.
Moreover, optical H$\alpha+$[N\,II] images have revealed that there are
large filamentary structures in the distribution of warm matter that
form large shells in the ISM 
\citep{1978A&A....68..189G,1980MNRAS.192..365M}. 
These shells are called supergiant shells (SGSs) and are believed
to be created out of matter that was swept up by expanding gas. 

The immediate effect of the stars on the ambient ISM is best observed in the optical narrow-band images of emission lines such as H$\alpha$, [S\,II], and [O\,III] from recombinations and collisional excitations in the denser, but still low-density warm interstellar gas. The study of the distribution and morphological properties of these structures is crucial for the understanding of the physics of massive stars and the evolution of the ISM, and hence of the entire galaxy.

With this aim, we analyze the distributions of spatial structures in the MCELS data
using Minkowski tensors (MTs), which are powerful shape descriptors from integral geometry
that characterize additive shape information~\citep{schroederturk2010,schroder-turk_minkowski_2011}.
We use them to develop an automatic detection of superbubbles and apply it to both the LMC and SMC.
We then perform a statistical data analysis to confirm spatial correlations between SNRs and superbubbles.

In astronomy, morphometric analyses based on Minkowski functionals (MFs)
and MTs have been successfully applied for a broad range of length scales,
including the detection of extended sources in gamma-ray sky maps of the H.E.S.S. telescope \citep{goering2013,klatt2019} and Fermi Gamma-ray Space Telescope~\citep{ebner_goodness--fit_2018},
analyzing the large-scale distribution of galaxies~\citep{1994a&a...288..697m,kerscher2001a,kerscher2001b,wiegand2017,sullivan2019},
exploring non-Gaussianity of the cosmic microwave background (e.g., \citealt{hikage2006,hikage2008,matsubara2012,Gay2012,Ducout:2013,santos2016,novaes2016,buchert2017}),
and possible anomalous alignments within it \citep{joby2019},
classifying the shape of galaxies \citep{rahman2003,rahman2004},
analyzing the morphological evolution of the intergalactic medium at the epoch of reionization \citep{yoshiura2017,kapahtia2019}, or
the formation of nuclear matter in SN explosions~\citep{physrevc.77.035806,schuetrumpf_time-dependent_2013,schuetrumpf_appearance_2015}.
An early conceptual work demonstrating the use of MTs has been done by \cite{beisbart2002}.

A manual search for bubbles lacks objectivity and is impractical because of the sheer number of objects.
In the past, other automatic methods have been developed, in particular, for infrared (IR) data (e.g., \citealt{williams1994,stutzki1990,motte1998}; \citealt{menshchikov2010, wachter2010, menshchikov2012} and the review by \citealt{andre2014}). These studies aimed to find and characterize filaments, clumps at their intersection, and clumps that have a simple shape and are located in a region with complex background, as can be found, for example, in the dust emission around protostars. 

More complex interstellar shells have been characterized in the citizen-science Milky Way Project \citep{kendrew2012, jayasinghe2019}
and machine-learning (ML) projects, which in part use the citizen-science project data as training data
\citep[e.g.,][]{beaumont2014, vanoort2019, xu2020}.
Both methods are resource-intensive: citizen-science projects require a large number of volunteers and the corresponding infrastructure, while ML methods require extensive training sets for more complex objects. Some models \citep{xu2020} also incorporate velocity information. 

The MT-based bubble detection we present here is able to find any round structure, whether it is a shell or a filled bubble, in a single emission-line image in a comprehensible and quantitative way and runs on any standard personal computer. Further analysis then allows automatic calculation of the structure orientation and, more generally, the spatial origin of filaments. 

In Section 2 we present the optical data that are analyzed here.
In Section 3 we briefly discuss the mathematical background and algorithms of the MTs and their irreducible representation.
They allow the automatic detection of superbubbles that we develop in Section 4.1,
which is applied to the LMC and SMC in Sections 4.2 and 4.3, respectively.
In Section 5.1 we perform a hypothesis test for spatial correlations between detected bubbles and massive stars. 
In Section 5.2 we compare the detected objects to known SNRs and superbubbles. 
We discuss the results of our analysis and the astrophysical implications in Section 5.3, and we summarize our findings in Section 6.

\section{Data}

The optical data for this investigation used mosaicked images from the
MCELS.
These images were taken at the University of Michigan/Cerro Tololo
Inter-American Observatory (UM/CTIO) Curtis Schmidt telescope.  The detector, a Tek 
$2048 \times 2048$ CCD with 24~$\mu$m pixels, gave a scale of 2.3
arcsec per pixel and a resulting angular resolution of approximately 4.6 arcsec.
The narrowband images were taken with filters centered on the [O\,III] ($\lambda 5012$\AA, FWHM=30\AA), H$\alpha$ 
($\lambda 6568$\AA, FWHM=30\AA) and [S II] ($\lambda 6729$\AA, FWHM=50\AA) 
emission lines along with green ($\lambda 5130$\AA, FWHM=155\AA) and red ($\lambda$ 6850\AA, 
FWHM=95\AA) continuum filters ($\lambda$ 6850\AA).  The optical data were reduced using the IRAF\footnote{IRAF is 
distributed by the National Optical Astronomy Observatories, which is operated 
by the Association of Universities for Research in Astronomy, Inc. (AURA) 
under cooperative agreement with the National Science Foundation.} software 
package for bias subtraction and flat-field correction.  The astrometry was 
derived from stars in the Two Micron All Sky Survey (2MASS) J-band catalog 
\citep{skrutskie2006}.  After an astrometric solution was obtained for the
individual pointings in each filter, the data were reprojected to have a scale
of 2 arcsec pixel$^{-1}$.  The data were flux-calibrated using images obtained 
of spectrophotometric standard stars 
\citep{1994PASP..106..566H,1992PASP..104..533H} and were then continuum subtracted.

In order to perform our analysis of the data using MTs, we 
binned the mosaics by a factor of 5 (LMC) or 3 (SMC). 
In the binned LMC data, 10\,px corresponds to $\sim$1.67', which
again corresponds to $\sim$24\,pc at a distance of (49.59$\pm$0.09 [stat]
$\pm$0.54 [sys])\,kpc \citep{pietr2019}.
For the SMC, 10\,px in the binned images corresponds to $\sim$1' and
$\sim$18\,pc at distance of (61.9$\pm $0.6)\,kpc \citep{degrijs2015}.

\section{Minkowski tensors}

\subsection{Definition and properties}
Minkowski functionals and MTs are versatile
shape descriptors from integral geometry~\citep{schneider_stochastic_2008-1}.
They sensitively quantify the shape, orientation, and position of complex spatial structures.
The MFs and MTs were originally defined for convex bodies and finite unions of convex bodies (which importantly includes pixelated images), but they have also been generalized to (sufficiently) smooth domains.
For a smooth two-dimensional body $A$, the MFs can be defined by area and contour integrals,
\begin{equation}
W_0(A)   := \int_A\,\text d^2r
\text{ and }
W_\nu(A) :=\frac{1}{2}\int_{\partial A} G_\nu\text dl
\text{ with }
\nu\in\{1,\,2\},
\end{equation}
where $G_1=1$ and $G_2=\kappa$ (i.e., the local curvature of the body boundary).
The three MFs are therefore up to prefactors given by the area, perimeter, and Euler characteristic.
For a compact domain, the latter is the number of clusters minus the number of holes.
More generally, in $d$ dimensions, there are $d+1$ MFs (including the volume, surface area, and Euler characteristic).

The MTs are a straightforward generalization of MFs using position vectors $\vec{r}$ and normal vectors $\vec{n}$ of the contour $\partial K$.
Let
\begin{equation}
\vec{r}^a\otimes\vec{n}^b := \underbrace{\vec{r}\otimes \ldots\otimes\vec{r}}_{a\text{ times}}\otimes \underbrace{\vec{n} \ldots\otimes \vec{n}}_{b\text{ times}}
\end{equation}
using the symmetric tensor product $(\vec{x}\otimes\vec{x})_{ij} = x_ix_j$ and 
\begin{equation}
(\vec{r}^a\otimes\vec{n}^b)_{i_1\ldots i_{a+b}} =\frac{1}{(a+b)!}\sum_{\sigma\in S_{a+b}} r_{i_{\sigma(1)}}\ldots r_{i_{\sigma(a)}}\cdot n_{i_{\sigma(a+1)}} \ldots n_{i_{\sigma(a+b)}},
\end{equation}
where $S_n$ is the permutation group of $n$ elements. 
Then the MTs of rank $a+b$ are given in 2D by 
\begin{equation}
W_0^{a,0}(K) := \int_K\vec{r}^a\,\text d^2 r~\text{ and }~ W_\nu^{a,b}(K) := \frac{1}{2} \int_{\partial K} \vec{r}^a\otimes\vec{n}^b G_\nu\, \text d r~.
\end{equation}
Here, $W_\nu^{0,0}=W_\nu$.
In contrast to the MFs, linear relations exist between MTs.

For convex bodies with sharp edges or corners, the parallel (dilated) body $A_\varepsilon$ can be constructed,
\begin{equation}
A_\varepsilon :=\{ \vec{x}\in\mathds{R}^2|\,\exists \vec{y}\in A:\,\|\vec{x}-\vec{y}\|\leq\varepsilon \}\,. 
\end{equation}
Because $A_\varepsilon$ has a finite local curvature everywhere,
the MTs of $A$ are given by $W_\nu^{a,b}(A) = \lim_{\varepsilon\rightarrow 0} W_\nu^{a,b}(A_\varepsilon)$. 
In integral geometry, the area of the parallel body as a function of $\varepsilon$ is used to define the MF for convex bodies~(\citealt{schneider_stochastic_2008-1}).
The definition for convex bodies immediately generalizes to a definition for finite unions of convex bodies using the property described below.

Minkowski tensors and MFs are additive, that is,
$W_\nu^{a,b}(A\cup A') = W_\nu^{a,b}(A) + W_\nu^{a,b}(A') - W_\nu^{a,b}(A\cap A')$
 for two convex bodies $A$ and $A'$.
According to Hadwiger's characterization theorem~\citep{hadwiger1957},
any additive, motion-invariant, and continuous functional on convex bodies can be expressed as a linear combination of MFs.
An analogous theorem for MTs was proven by \citet{alesker_description_1999}.
This shows that MFs and MTs are comprehensive shape descriptors for complex spatial structures.
Both MFs and MTs have been successfully applied to describe a broad range of physical systems (see, e.g.,  \citealt{schroederturk2013}).

\begin{figure}
    \centering
    \includegraphics[width=0.49\hsize]{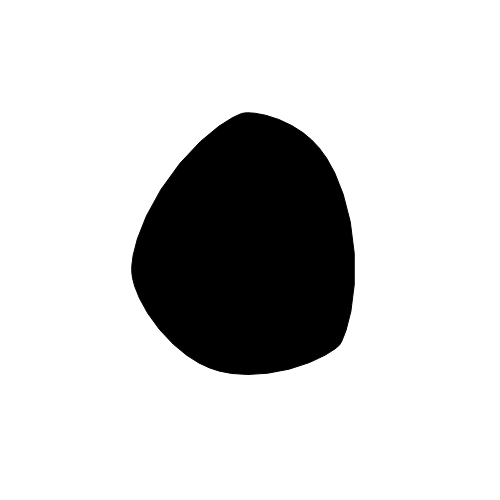}
    \includegraphics[width=0.49\hsize]{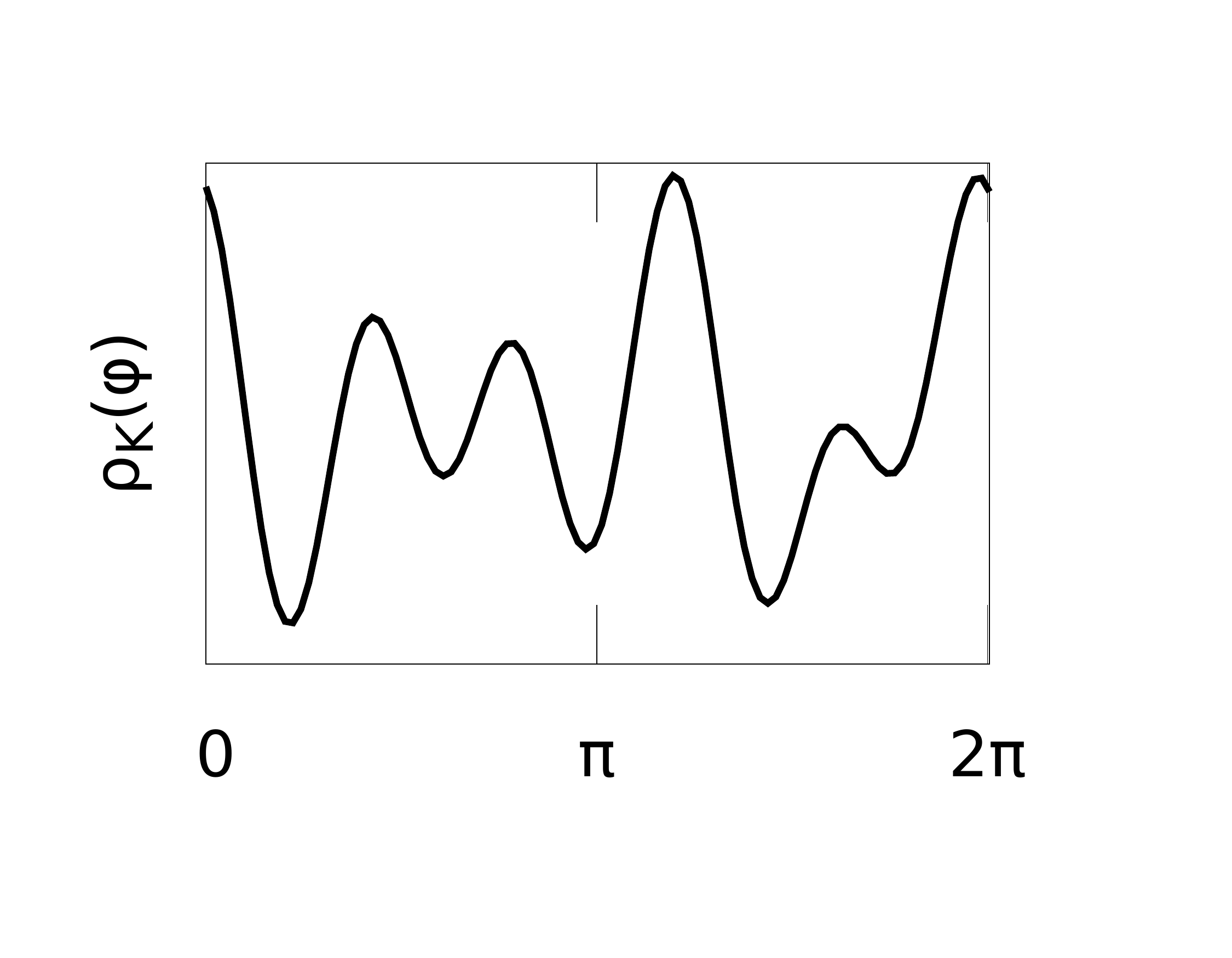}
    \caption[An example shape $S$ with corresponding normal density $\rho_A(\varphi)$.]{Example shape $A$ with corresponding normal density $\rho_A(\varphi)$. Low and high $\rho_A(\varphi)$ indicate corners and flat surfaces, respectively. $\varphi = 0$ is defined to be pointing to the right. 
    }
    \label{fig:shape_example}
\end{figure}

Different MTs characterize different geometrical aspects.
For practical reasons, we focused on the translation-invariant MT $W^{0,s}_1$.
Instead of the Cartesian representation given above,
we used the irreducible representation in 2D that provides a convenient
access to the shape information of high-rank MTs.
The irreducible representation of MTs in 2D has already 
been proposed for a sensitive and robust characterization of 
anisotropy in medical physics, demonstrating the benefit of higher-rank anisotropy information~\citep{klatt_mean-intercept_2017}.
The irreducible Minkowski Tensors (IMT) in 2D can be defined in the following way.
Let $A$ be a 2D convex body.
Then, its normal density $\rho_A(\varphi)$ is proportional to the fraction of normal vectors pointing in direction $-\varphi$, or, in other words,  if, for instance, the direction to the right has been defined as the reference, $\rho_A(\varphi)$ is the fraction pointing to the right when the body is rotated by $\varphi$. 
An example shape and its normal density are shown in Fig.~\ref{fig:shape_example}.
For a polygon with edges labeled by $k$ of length $L_k$ , we have $\rho_A(\varphi) = \sum_k L_k\delta(\varphi - \varphi_k)$.
The integral of the normal density over all angles is equal to the perimeter of $A$.

\begin{figure}
\begin{center}
\includegraphics[width=0.24\hsize]{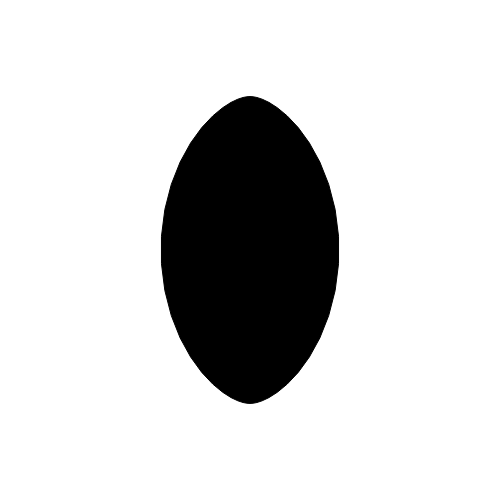}
\includegraphics[width=0.24\hsize]{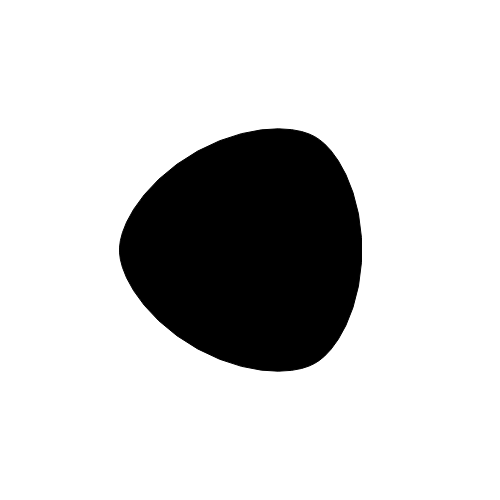}
\includegraphics[width=0.24\hsize]{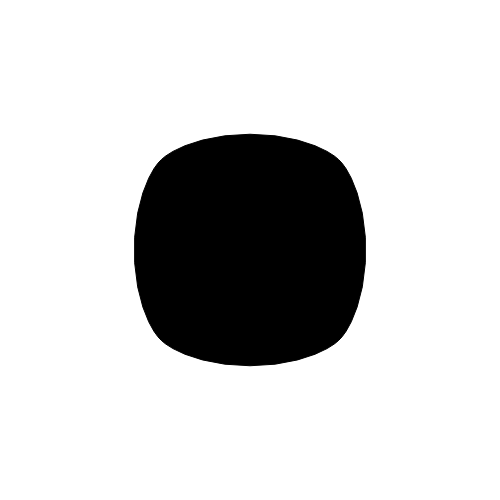}
\includegraphics[width=0.24\hsize]{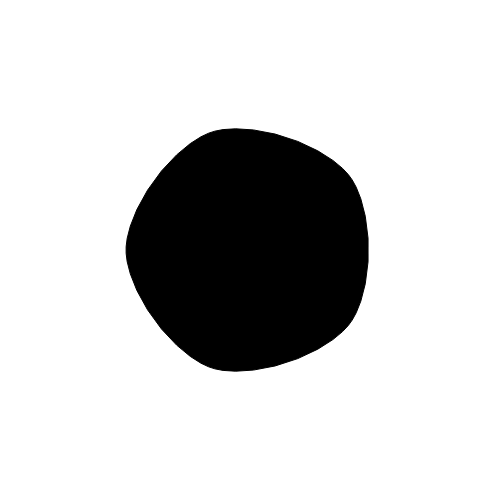}
\includegraphics[trim = 1.2cm 0 1cm 0,clip,width=0.24\hsize]{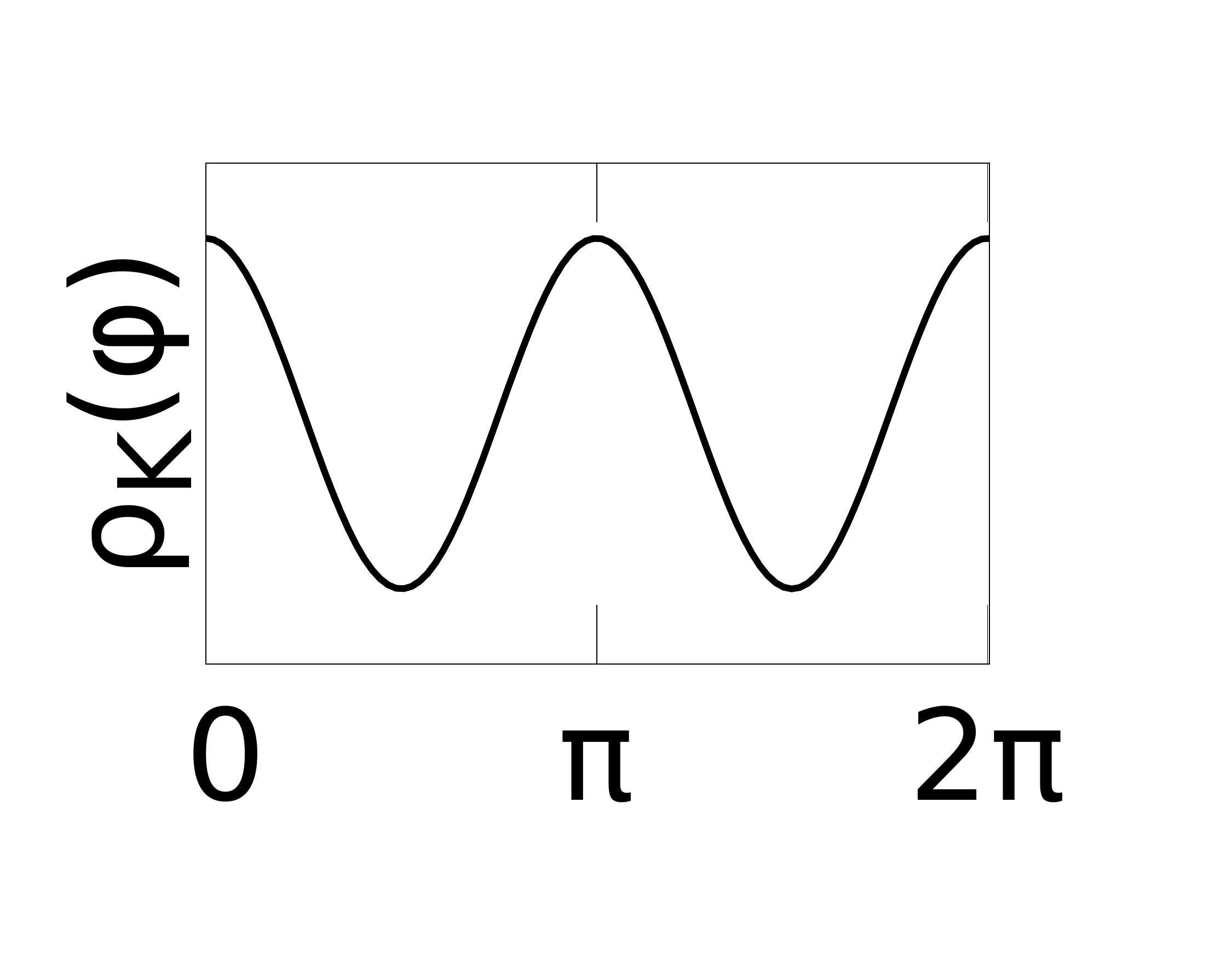}
\includegraphics[trim = 1.2cm 0 1cm 0,clip,width=0.24\hsize]{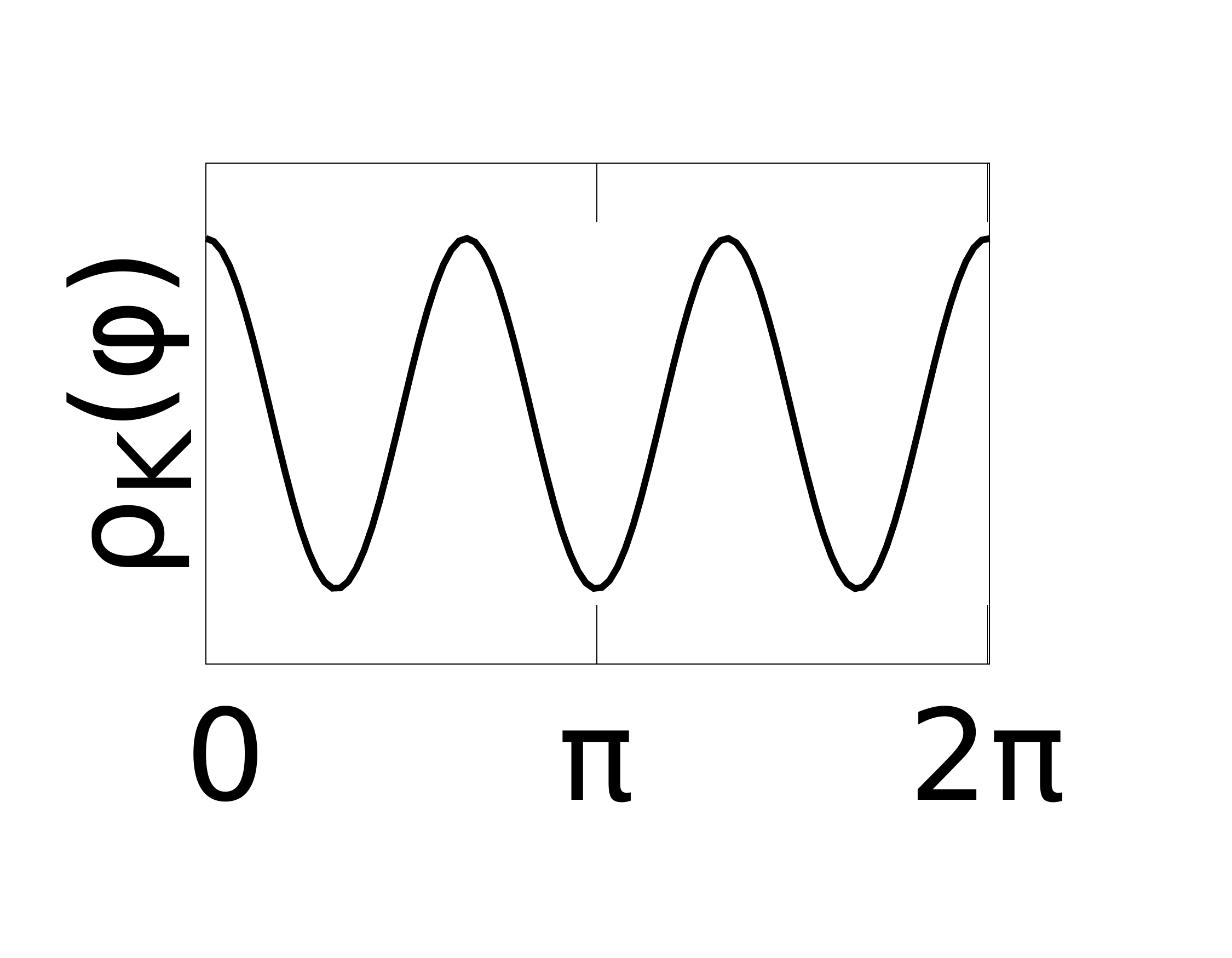}
\includegraphics[trim = 1.2cm 0 1cm 0,clip,width=0.24\hsize]{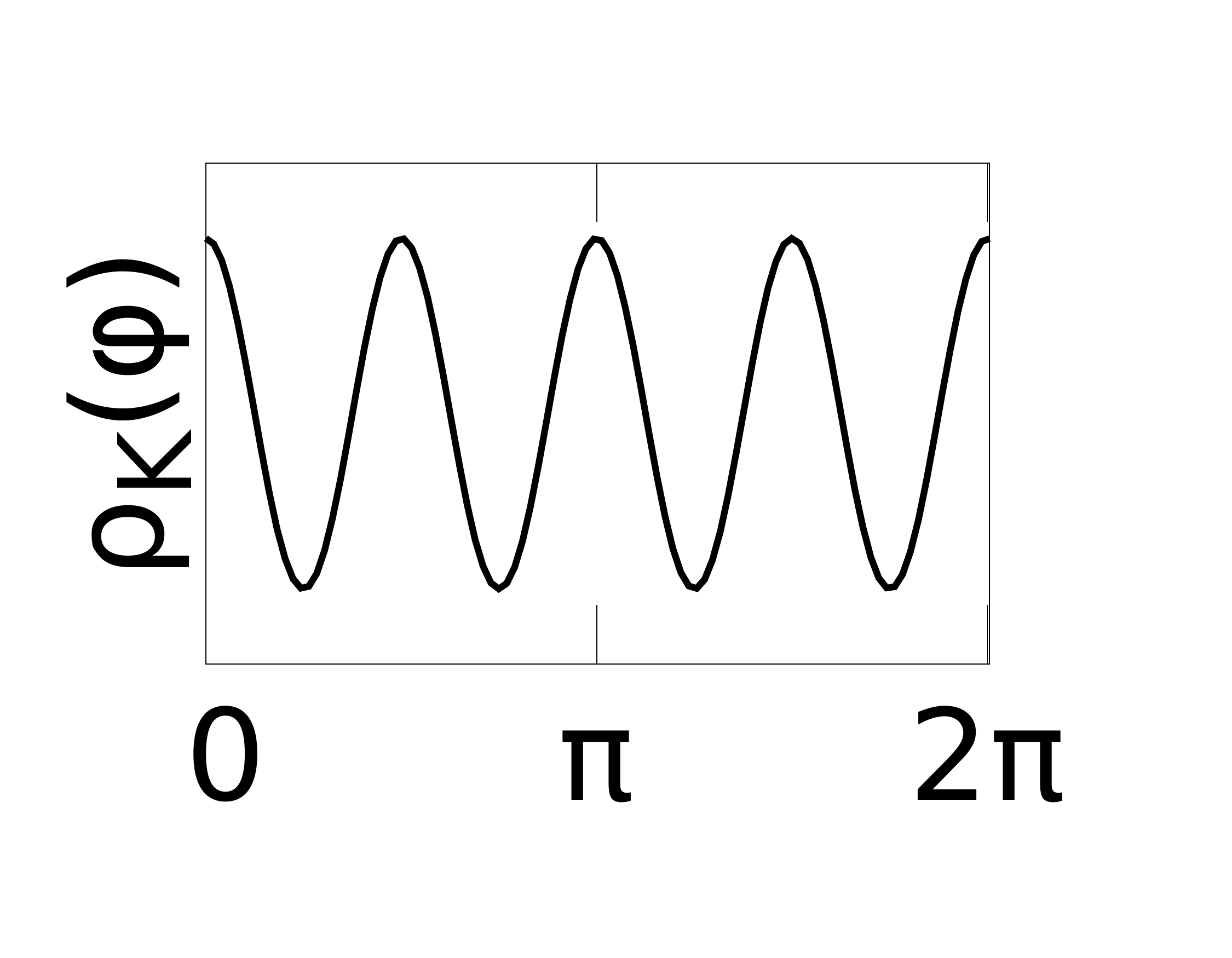}
\includegraphics[trim = 1.2cm 0 1cm 0,clip,width=0.24\hsize]{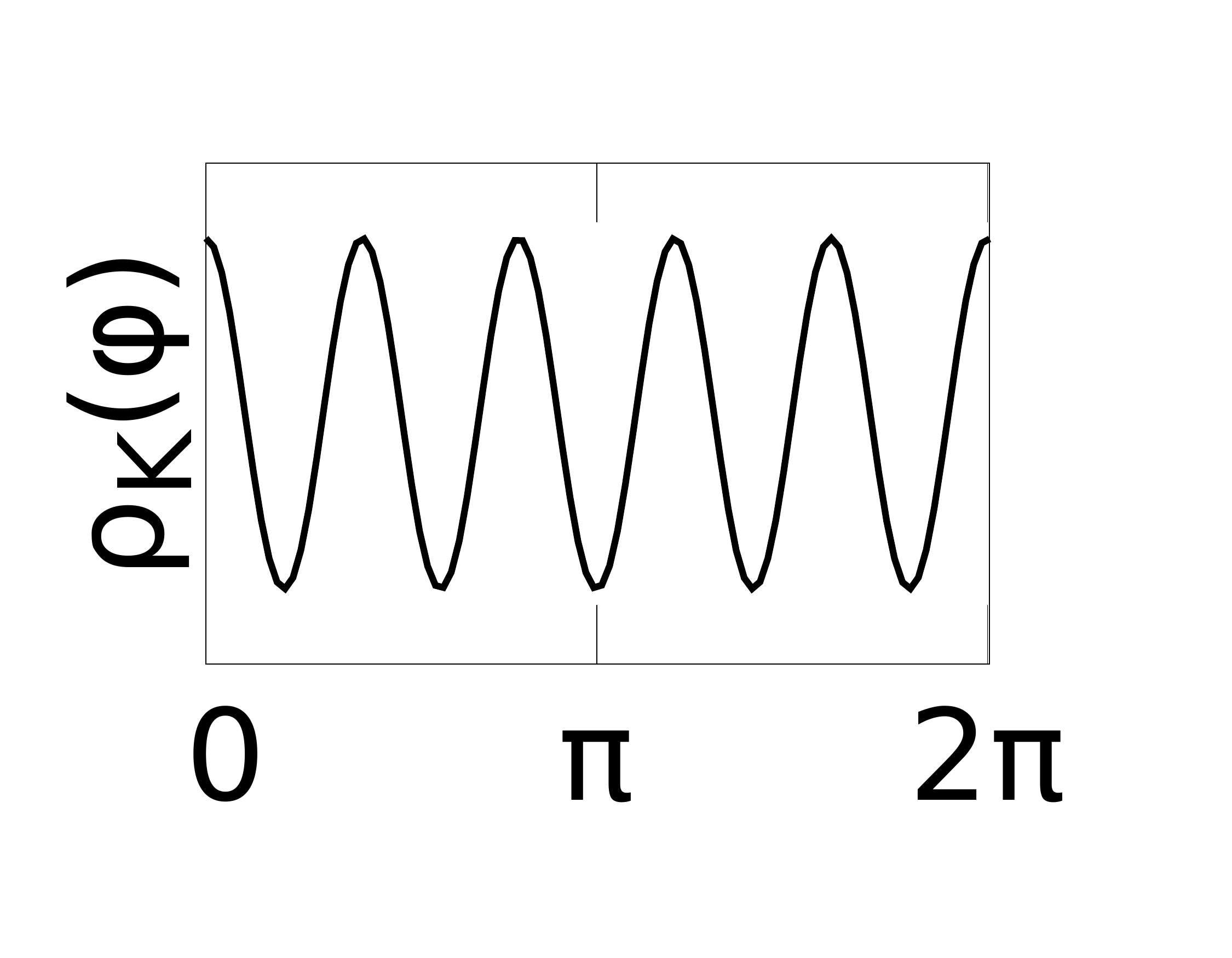}
\end{center}
    \caption[Shapes where $|\psi_s|>0$ for only one $s>1$ and phase zero along with their normal densities in arbitrary units.]{Shapes where $|\psi_s|>0$ for only one $s>1$ and phase zero along with their normal densities in arbitrary units. The $\rho_K$ are harmonic waves because the $\psi_s$ are their Fourier coefficients with increasing frequencies. From left to right: $s = 2,~3,~4,\text{ and}~5$. }
    \label{fig:psi_s}
\end{figure}

The IMTs of $A$ are then given by a Fourier transform of $\rho_A$,
\begin{equation}
    \label{eq:psi_s}
    \psi_s (A) = \int_0^{2\pi}\,  \exp(is\varphi)\,\rho_A(\varphi) \text d\varphi \stackrel{\text{polygon}}{=} \sum_{k} L_k\,\exp(is\varphi_k),
\end{equation}
where $\psi_0$ is the perimeter of the body, $\psi_1 = 0$ for closed bodies, and the higher-order IMTs correspond to a symmetry decomposition of $A$.
Figure~\ref{fig:psi_s} shows convex
bodies where $|\psi_s|>0$ for only one $s>1$ together with their normal densities.

Under rotation $R(\alpha)$ at an angle $\alpha$, the phase of the IMT changes by $-s\alpha$, 
\begin{align}
\psi_s(R(\alpha)A) &= \int_0^{2\pi} \text d \varphi \,  \exp(is\varphi)\,\rho_A(\varphi + \alpha)\\
&= \int_0^{2\pi} \text d \varphi \,  \exp(is(\varphi- \alpha))\,\rho_A(\varphi)\\
&= \psi_s(A) \text e^{-is\alpha.}
\end{align}
This becomes clear in Fig~\ref{fig:psi_s}. 
Because of the $s$-fold symmetry of the bodies, a rotation of ${2\pi}/{s}$ maps them onto their initial position, so  $\psi_s(R({2\pi}/{s})A) = \psi_s(A)$. Additionally, it is clear that there are $s$ preferred directions of the bodies at 
\begin{equation}
\varphi_n = \frac{2\pi n + \text{arg}(\psi_s)}{s},
\end{equation}
where $n \in \{0,\ldots, s-1\}$.

Because the IMTs scale linearly with the size of the body ($\psi_s(\lambda A) = \lambda \psi_s(A)$, $\lambda A$ meaning point-wise multiplication), a scale-independent \textit{\textup{anisotropy index}} is often used, where the $\psi_s$ are divided by the perimeter,
\begin{equation}
    q_s \coloneqq \frac{|\psi_s|}{\psi_0}.
\end{equation}

\subsection{Minkowski maps}
\begin{figure}
\centering
\includegraphics[width=0.49\hsize]{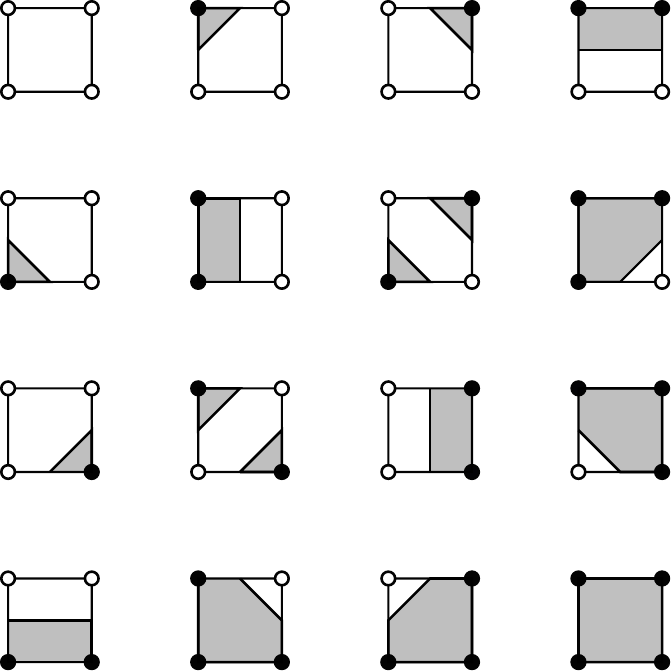}
\caption{Possible configurations in the marching 2x2\,px square. Black dots represent pixels above a threshold, and open dots pixels below this threshold. The gray area is the body that is assumed to be present in the window. 
The positions of the corners located between pixel centers are interpolated based on the actual pixel values \citep{mantz2008}. When diagonally opposite pixels are above the threshold, we make the arbitrary choice to view this as two disconnected triangles. }
\label{fig:marchingsquare}
\end{figure}

The astrophysical images to be analyzed are grayscale images with a range of values, whereas the MFs and MTs are defined for bodies in the plane.
Therefore we binarized the images by choosing a threshold for the pixel value, 
separating the image into regions that belong to the body (true) and those outside (false). We chose several thresholds to analyze structures at different flux levels (see below and Sect.~\ref{bubbles_results}).
We constructed body contours using a variation of the marching-squares algorithm \citep{mantz2008}. The desired MFs and MTs are calculated for a 2x2\,px moving window, conceptually using the polygon version of Eq.~\eqref{eq:psi_s} for the IMT. This window is visualized in Fig.~\ref{fig:marchingsquare}: The 4 pixels can be above or below the threshold, leading to 16 possible shapes in a window. 
Grayscale information is included to adjust the position of the vertices, such that contours are not restricted to the pixel grid. 
For the analysis, we used the papaya2 library \citep{Schaller2020}.

This algorithm leads to a complex-valued map that shows the MFs and MTs for structures at a scale of two pixels (Fig.~\ref{fig:bubbleflowchart}, step 1). This is called a Minkowski map \citep{schroederturk2010}. The MFs and MTs of a whole section of an image can be obtained by summing this map in the relevant region. Alternatively, Minkowski maps can be generated at larger scales by smoothing the 2x2 map (Fig.~\ref{fig:bubbleflowchart}, step 3).
We generated smoothed Minkowski maps by convolving with a kernel of circular shape, with weights linearly decaying from the center.
This emphasizes the central region while still taking structures farther out into account.

Typically, $|\psi_s|$ decreases with increasing smoothing window size (due to the average over a larger area).
Single objects are smeared out over larger areas and generate lower maximum intensities in the Minkowski maps.
To make the absolute values comparable at different window sizes, each point was normalized by multiplying with the square root of the window area. This yields similar orders of magnitude of $|\psi_s|$  for local comparison.
To reduce computation time, the smoothing window was not calculated at every point of the 2x2-map, but was rather moved in steps of a size of one-sixth of the window diameter. This provides sufficient accuracy, and the computation time is independent of the window size.

By binarizing the image at one threshold, we effectively analyze structure at this flux level and erase all other information in the image. We therefore averaged over the 2x2 Minkowski maps that result from several thresholds  (Fig.~\ref{fig:bubbleflowchart}, steps 1 and 2). The lowest threshold was chosen such that relevant faint structures were included. 
The highest threshold was chosen to capture the brightest regions of interest, in our case, the brightest \hii\ regions.  The intermediate thresholds were spaced logarithmically. About ten thresholds suffice to capture all relevant structures. More thresholds do not change the conclusions qualitatively for the images we used.

\begin{figure*}
    \centering
    \includegraphics[trim=0 1.7cm 0 0,clip,width=\hsize]{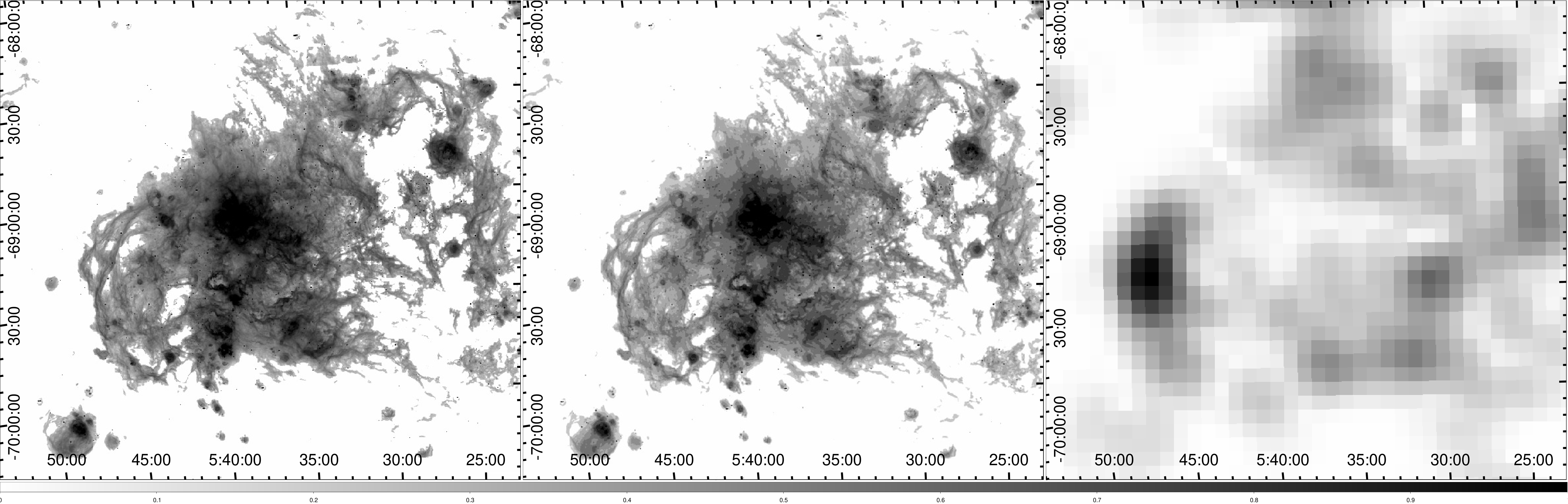}
    \caption{\ha{} image of a region in the LMC (left, logarithmic scale), thresholded image used to generate the Minkowski map (center), and the corresponding Minkowski map of $|\psi_2|$ (right). The smoothing window size  was six times the tile size in the right image. The first two images are limited to the minimum and maximum flux listed in Table~\ref{tab:thresh_lmc}. }
    \label{fig:lmc_psi2_example}
\end{figure*}

An example Minkowski map of $|\psi_2|$ of the LMC is shown in Fig.~\ref{fig:lmc_psi2_example}.
$\psi_2$ exposes filaments due to their elongated structure, as can be seen in the east. 
$\psi_2$ is thus a powerful measure for detecting both bubbles (which consist of shells) and large filamentary structures.

\section{Bubble detection}
\subsection{Method}
\begin{figure*}
\centering
\includegraphics[trim= 0 1.7cm 0 0, clip,width=0.8\hsize]{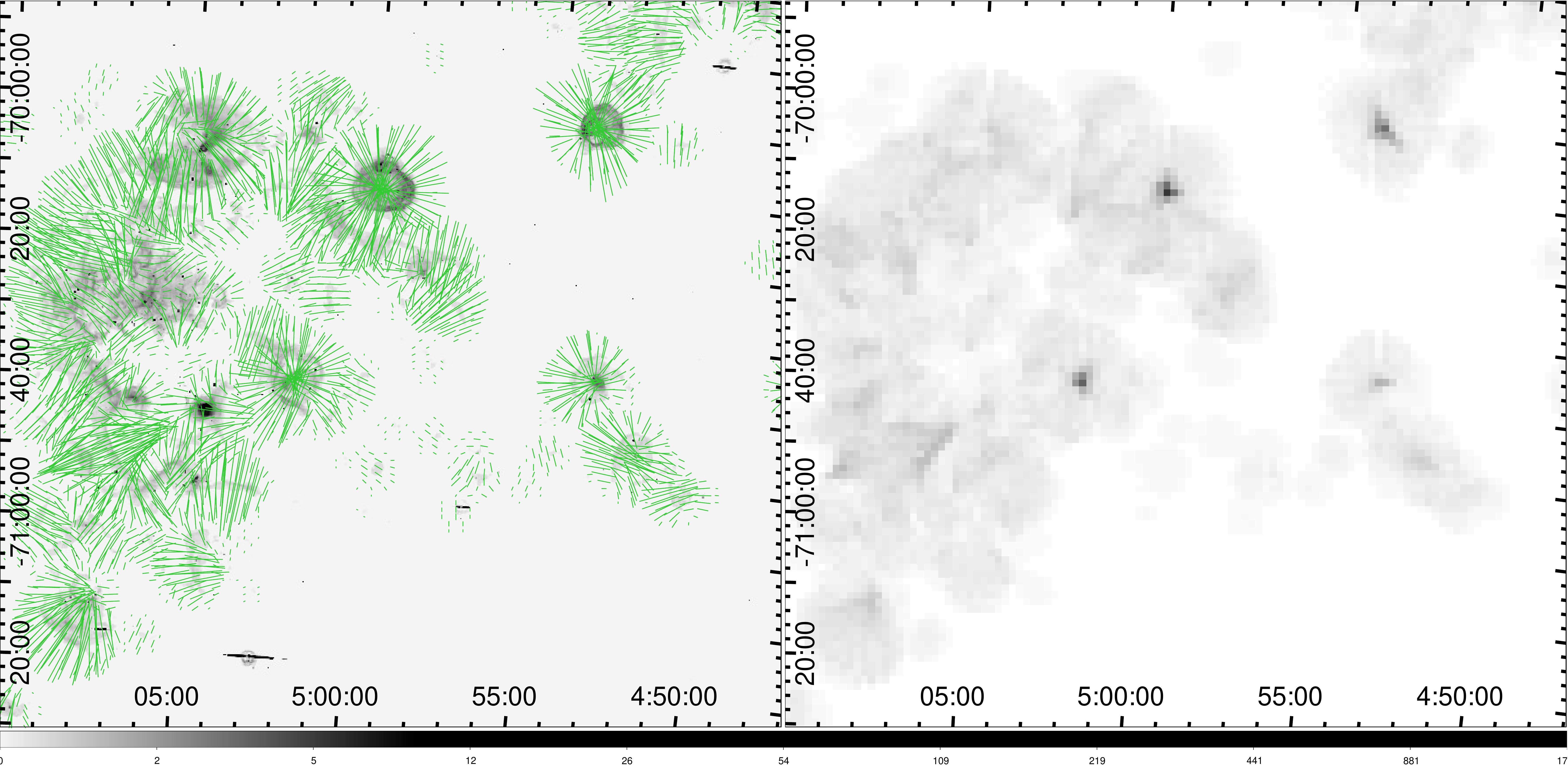}
\caption{Left: Lines perpendicular to filaments for a window size of 40\,px ($\sim 97\,$pc), determined via the phase of $\psi_2$. Right: line density of the left image}
\label{fig:linedens_example}
\end{figure*}

\begin{figure}
\centering
\includegraphics[scale=0.8]{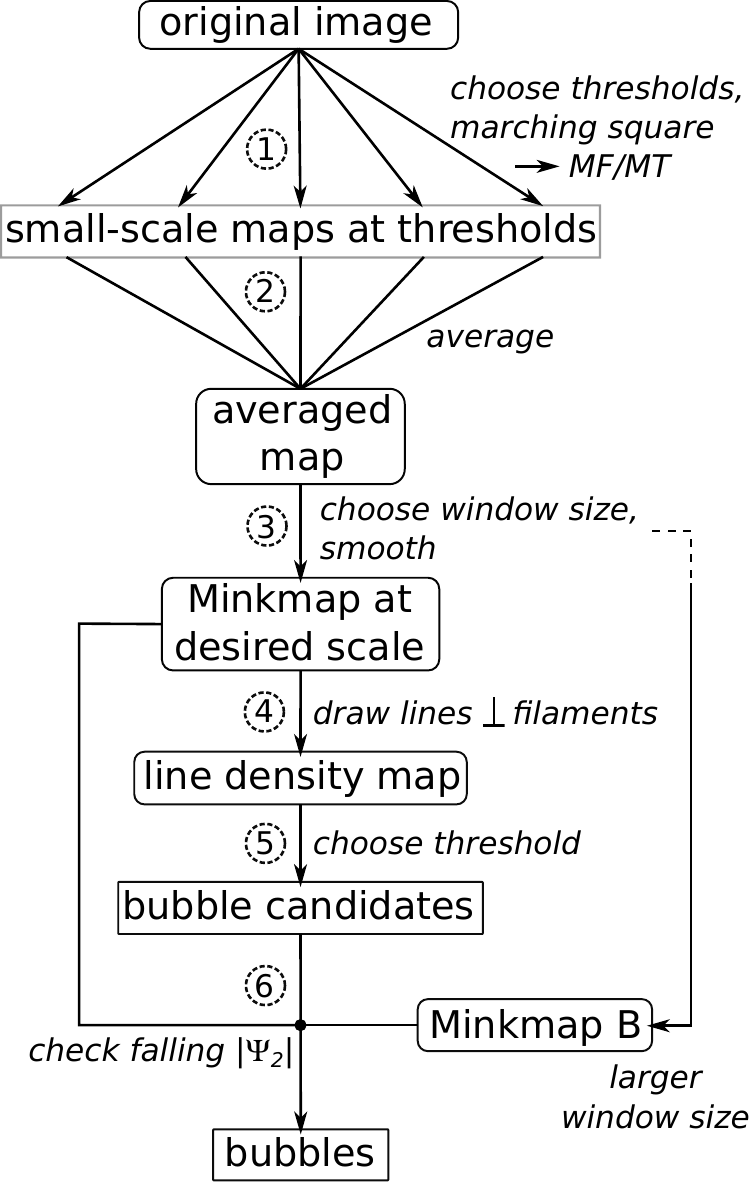}
\caption{Our bubble detection process: Create a Minkowski map by calculating marching-square maps at different thresholds (step 1), averaging them (step 2), and smoothing with a desired window size (step 3). For a $\psi_2$ Minkowski map at one window size, draw lines perpendicular to filaments using arg$(\psi_2),$ whose length is proportional to $|\psi_2|$ and a chosen length factor (step 4). Counting the lines in each position gives a line density map. By choosing a suitable threshold here, we obtain the position of the bubble candidates (step 5).  Then, $|\psi_2|$ at the bubble positions should decrease for bubbles standing alone and rise at least not significantly if there are neighboring structures. This is checked by comparing to a Minkowski map at a suitable larger window size (step 6). }
\label{fig:bubbleflowchart}
\end{figure}

We used the Minkowski maps of $\psi_2$ created for different smoothing window sizes to detect stellar bubbles and superbubbles in the MCELS mosaic images. 
The phase of $\psi_2$ contains the information about the orientation of a filament. 
To trace the origin of the filament (as a part of a bubble or a large open structure), we used the phase to draw a normal line to the filament at an angle of $0.5\cdot(\pm\pi - \text{arg}(\psi_2))$ with respect to the $x$ -axis. 
This is demonstrated in the left panel of Fig.~\ref{fig:linedens_example}.
Because bubbles are approximately round, the lines from their borders meet in the center. The lines therefore reveal the positions of the bubbles in the images. In order to detect bubbles automatically, we created line density maps by counting the number of lines per tile (Fig.~\ref{fig:bubbleflowchart}, step 4). An increase in line density above some threshold was treated as a bubble candidate (Fig.~\ref{fig:bubbleflowchart}, step 5). The line density map in Fig.~\ref{fig:linedens_example} (right) shows high line density values in the bubble centers.

This method has two free parameters at a given window size $w$: The length of a line, and the line density threshold. The length $l$ was set proportional to $w$ and $|\psi_2|$ and is given by 
\begin{equation}
    l = w\cdot \frac{|\psi_2|}{m}~,
\end{equation}
where $m$ is a free parameter corresponding to a typical value of~$|\psi_2|$. It should be chosen such that lines from a bubble border reach the center, but do not extend too far beyond it. Useful values were typically in the range of 0.25 to 0.4 in the images used here. 
The line density threshold should be chosen such that a balance between selecting many bubbles and a few non-bubbles is found. 
We performed several iterations to manually check the images. 

In addition, we took into account that $|\psi_2|$ decreases in bubbles when the window size becomes too large and there will be contaminations from the surrounding filaments.
Smaller windows will only include the edges of a bubble and yield nonzero $|\psi_2|$.
On the other hand, because the phase rotates around the bubble, $|\psi_2|$ will be averaged over a larger region and canceled out when the window size was increased: $|\psi_{2,\, w}| < |\psi_{2,\, w'}|$ for window sizes $w>w'$ (Fig.~\ref{fig:bubbleflowchart}, step 6). 

More quantitatively, a significantly larger window size is approximately an integration over $\psi_{2,\,w'}$. For a spherically symmetric bubble, $|\psi_2|$ only depends on the radial distance $r$ (at any window size). When computing $\psi_{2,\,w}$ at the center of the bubble ($r=0$), the angular part of the integral only depends on the phase that is given by twice the angle in real space and cancels out,
\begin{equation}
\psi_{2,\,w}(r=0) \sim \text{(radial integral)}\cdot \int_0^{2\pi} \exp(i2\varphi)\,\text d \varphi = 0~.
\end{equation}

\begin{figure*}
\centering
\includegraphics[trim=0 0 0 0, clip, width=\hsize]{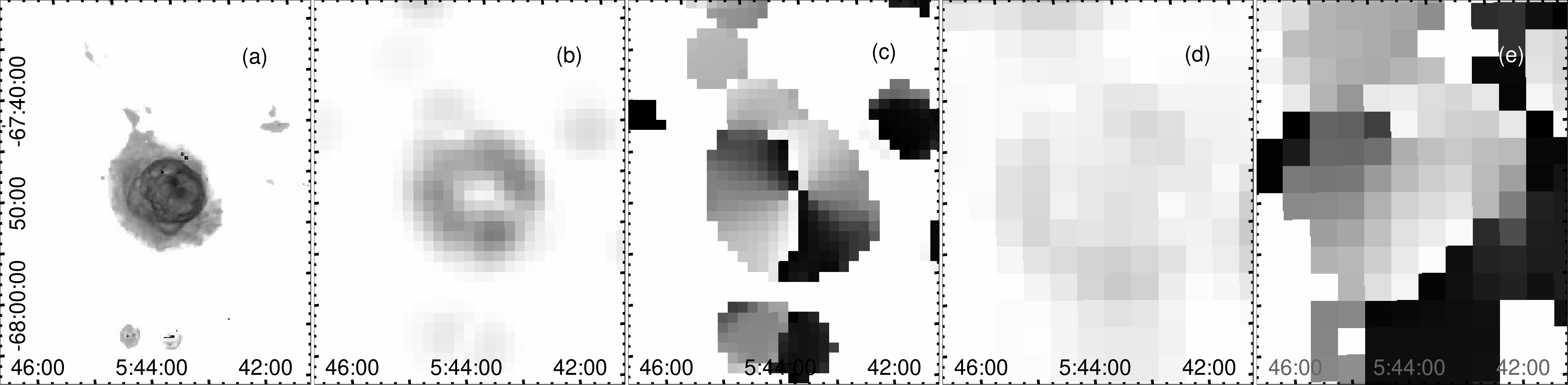}

\caption{Region LHA\,120-N\,70 in \ha{} (a, logarithmic scale as in Table~\ref{tab:thresh_lmc}) and the corresponding $\psi_2$ Minkowski maps for window sizes of 40\,px ($\sim$97\,pc, b: absolute value, c: phase) and 100\,px (242\,pc, d: absolute value, e: phase). For the larger window size (right panels), the contributions from all sides of the bubble cancel each other out due to the phase rotation along the bubble. The phase at 100\,px is thus dominated by the overall orientation of the bubble and the surrounding filaments.}
\label{fig:n70_raw_20_150}
\end{figure*}

This is demonstrated in Fig.~\ref{fig:n70_raw_20_150}. Here, $|\psi_2|$ at the smaller window size is significantly larger than at the larger size. 

\begin{figure}
    \centering
    \includegraphics[trim=0 2cm 5px 0,clip,width=0.7\hsize]{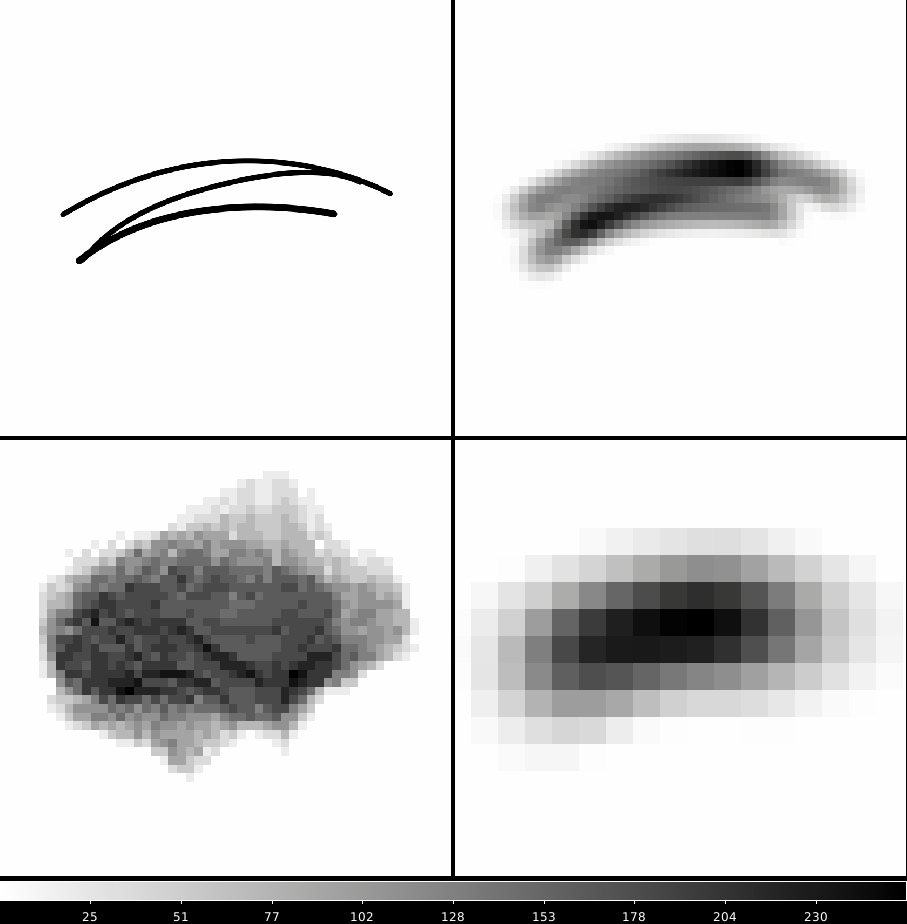}
    \caption{Several intersecting lines (top left), the $|\psi_2|$ Minkowski map at a window size large enough for an increased line density (top right), the resulting line density (bottom left), and the the $|\psi_2|$ Minkowski map at triple window size (bottom right, same scale as the other Minkowski map). In spite of setting a low line density threshold, no part of the image is detected as a bubble because the decreasing $|\psi_2|$ condition is not fulfilled.}
    \label{fig:line_demo}
\end{figure}

In practice, we found that window sizes $w' \simeq 3\cdot w$ yielded good bubble detection results. 
The values of $|\psi_{2}|$ were averaged over a central square of size $w/2$. $|\psi_{2,\, w'}|$ was allowed to be up to 0.05 higher than $|\psi_{2,\, w}|$ because otherwise, neighboring unrelated filaments disturbed the detection of many bubbles.
The importance of this decreasing $|\psi_2|$ condition for detecting bubbles (i.e., closed filaments) is shown in Fig.~\ref{fig:line_demo}.
In this case, where filaments do not form a bubble, the line density is high between the lines, but the region is rejected as a bubble because the lines cause high $|\psi_2|$ at any window size. 

Because this bubble detection was performed in a range of different window sizes, single physical objects were detected multiple times. Therefore the results needed to be combined. A simple combination of bubbles was performed the following way: For each detected bubble, a square box with an edge length of $2w$ was defined. If the box of a bubble from the analysis with a window size contained the center of the box of a bubble of the same or the next higher window size, the boxes were assumed to belong to the same physical object and were combined. The process was repeated for all window sizes. The resulting detections of bubbles are shown in the images discussed in Sects.\,\ref{sec:LMC} and \ref{sec:SMC}.
The code used for the detections, bubble combination, and various other functions is available on GitHub\footnote{\url{https://github.com/ccollischon/banana}}.


\begin{table*}
\caption{Thresholds for the Minkowski map generation for the LMC and SMC}
\begin{center}
LMC\\
\vspace{3pt}
\begin{tabular}{cccccc}
\hline
Image & Min. & Min. flux & Max. & Max. flux & No. of\\
& pixel value & [erg\,cm$^{-2}$\,s$^{-1}$] & pixel value & [erg\,cm$^{-2}$\,s$^{-1}$] & thr.\\
\hline
\ha & 0.1 & $3.1\cdot10^{-15}$ & 40 & $1\cdot10^{-12}$ & 9 \\
{}[\ion{O}{III}] & 0.1 & $3.1\cdot10^{-15}$ & 40 & $1\cdot10^{-12}$ & 9 \\
{}[\ion{S}{ii}] & 0.045 & $1.4\cdot10^{-15}$ & 40 & $1\cdot10^{-12}$ & 10\\
\hline
\end{tabular}\\
\vspace{15pt}
SMC\\
\vspace{3pt}
\begin{tabular}{cccccc}
\hline
Image & Min. & Min. flux & Max. & Max. flux & No. of\\
& pixel value & [erg\,cm$^{-2}$\,s$^{-1}$] & pixel value & [erg\,cm$^{-2}$\,s$^{-1}$] & thr.\\
\hline
\ha & 3 & $1.95\cdot10^{-13}$ & 1300 & $8.45\cdot10^{-11}$ & 12 \\
{}[\ion{O}{III}] & 6 & $3.90\cdot10^{-13}$ & 2500 & $1.62\cdot10^{-10}$ & 12 \\
{}[\ion{S}{ii}] & 3 & $1.95\cdot10^{-13}$ & 1300 & $8.45\cdot10^{-11}$  & 12\\
\hline
\end{tabular}
\end{center}
\tablefoot{Thresholds were spaced logarithmically and included the borders.}
\label{tab:thresh_lmc}
\end{table*}

\begin{figure*}
    \centering
    \includegraphics[trim=0 1.7cm 0 0,clip,width=.9\textwidth]{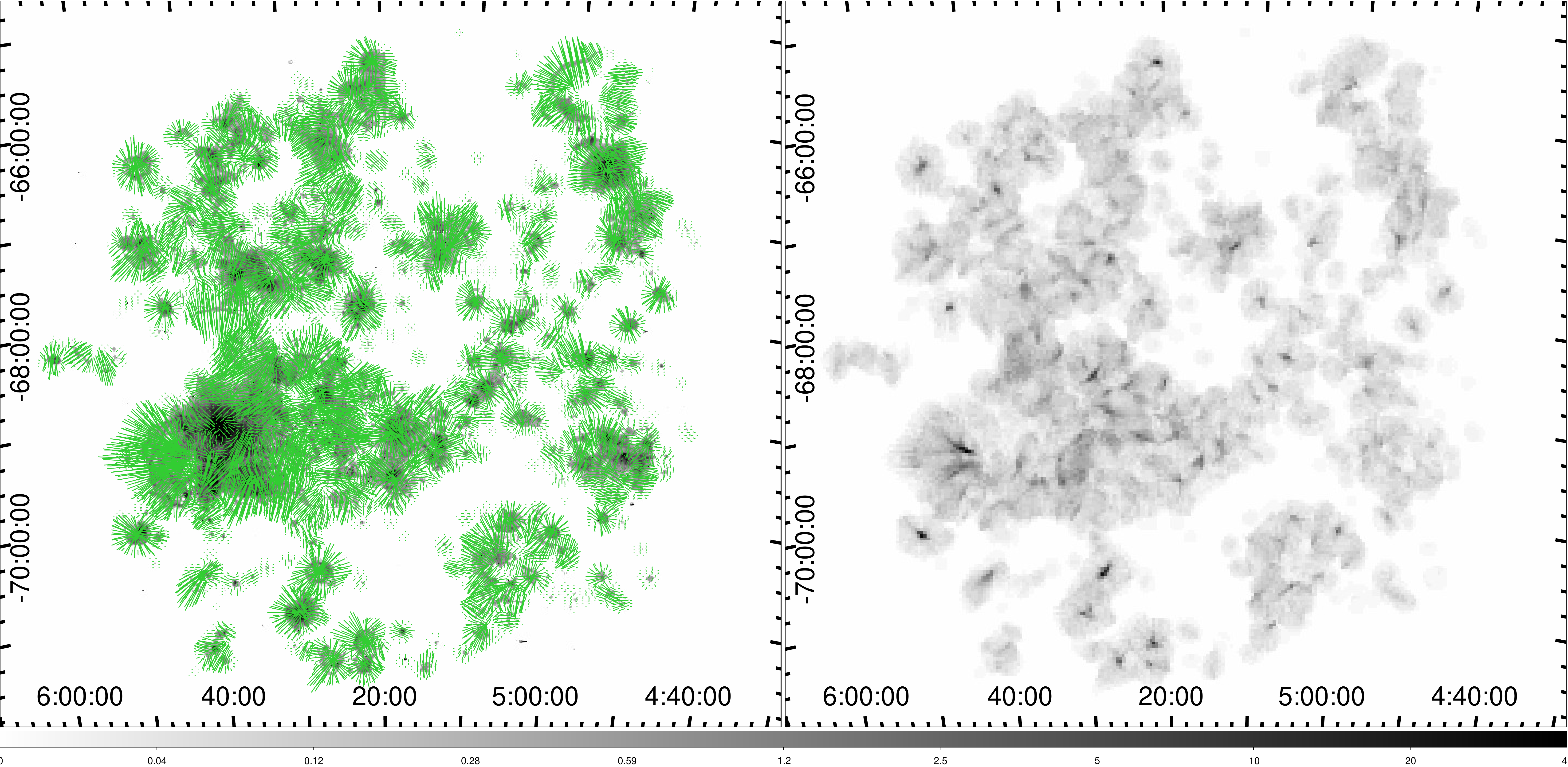}\\
    \includegraphics[trim=0 1.7cm 0 0,clip,width=.9\textwidth]{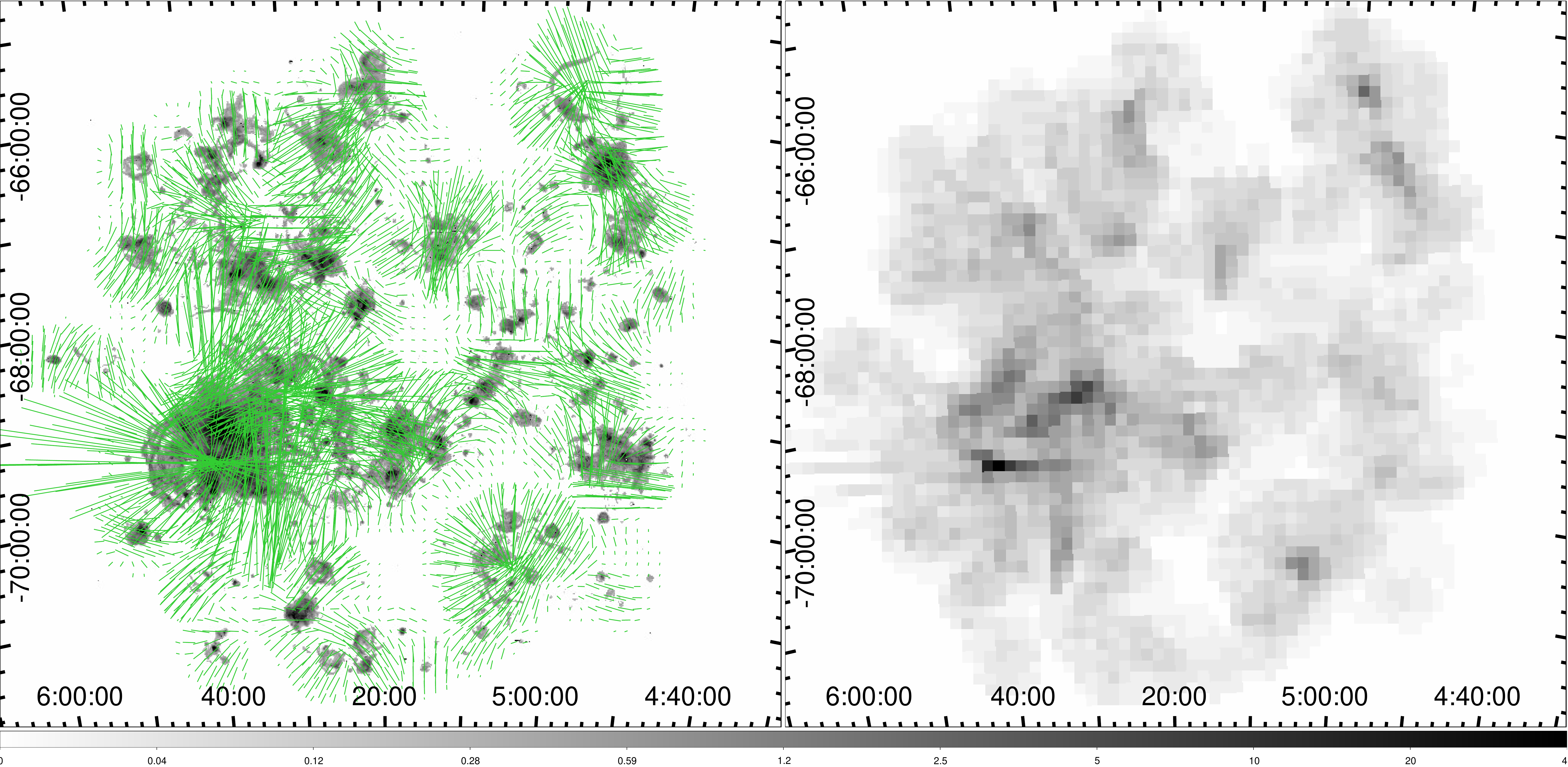}    
    \caption{Lines (left) and line densities (right) in the \ha\ image of the LMC at a window size 70\,px/$\sim$170\,pc (top) and 250\,px/$\sim$610\,pc.}
    \label{fig:lmc_lines}
\end{figure*}

\begin{figure}
    \centering
    \includegraphics[trim=0 1.7cm 0 0,clip,width=.47\textwidth]{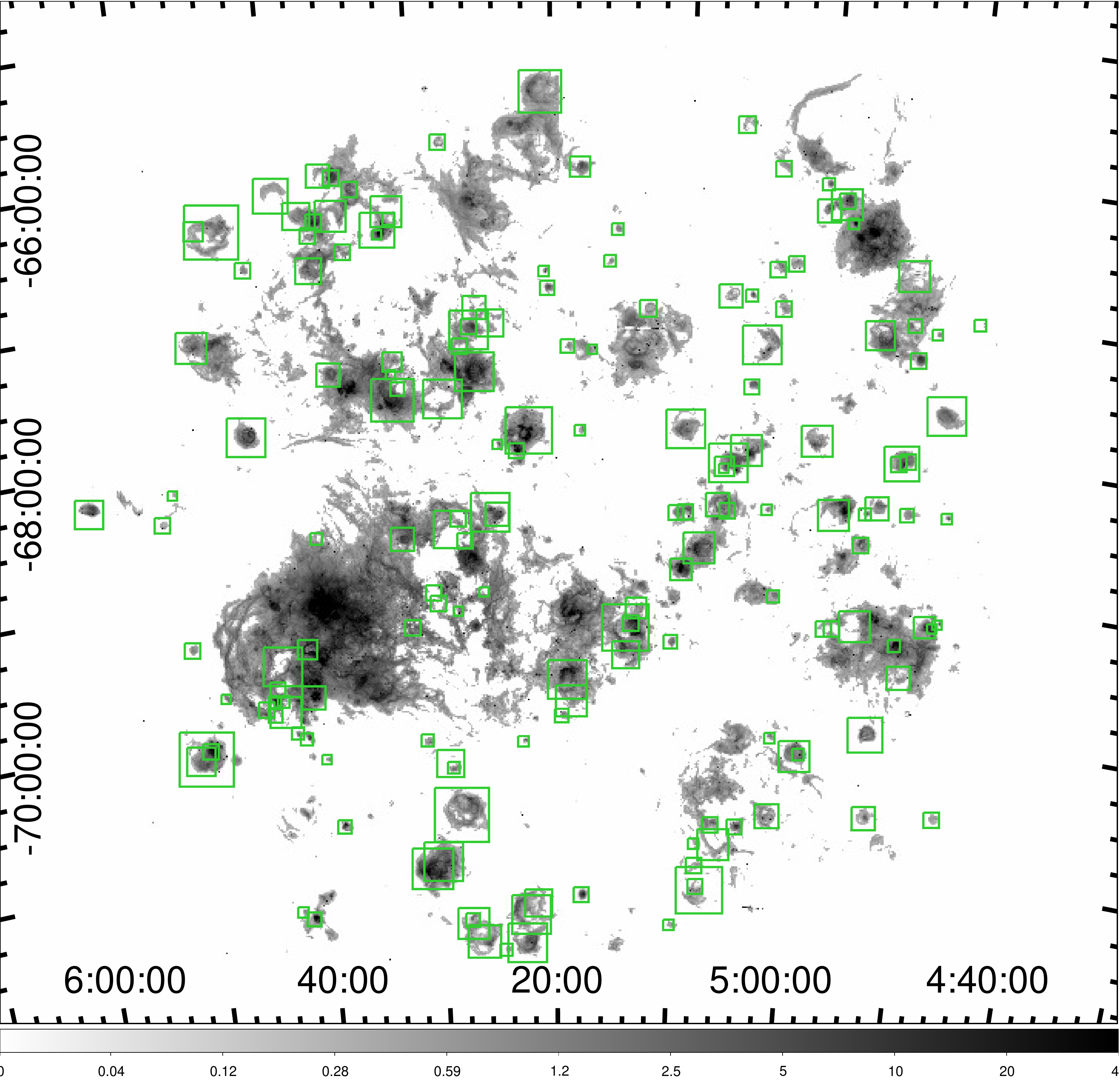}
    \includegraphics[trim=0 1.7cm 0 0,clip,width=.47\textwidth]{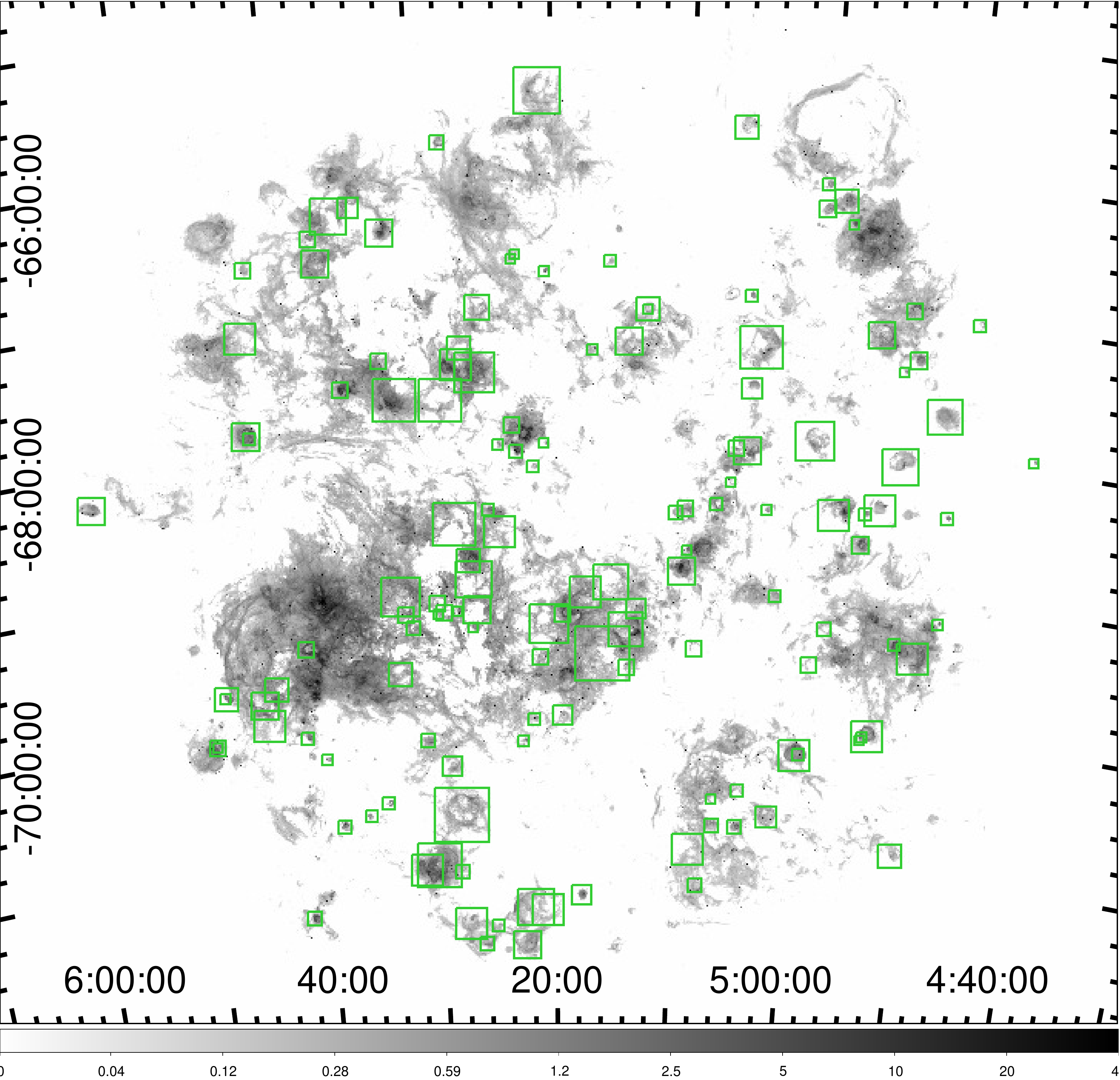}
    \includegraphics[trim=0 1.7cm 0 0,clip,width=.47\textwidth]{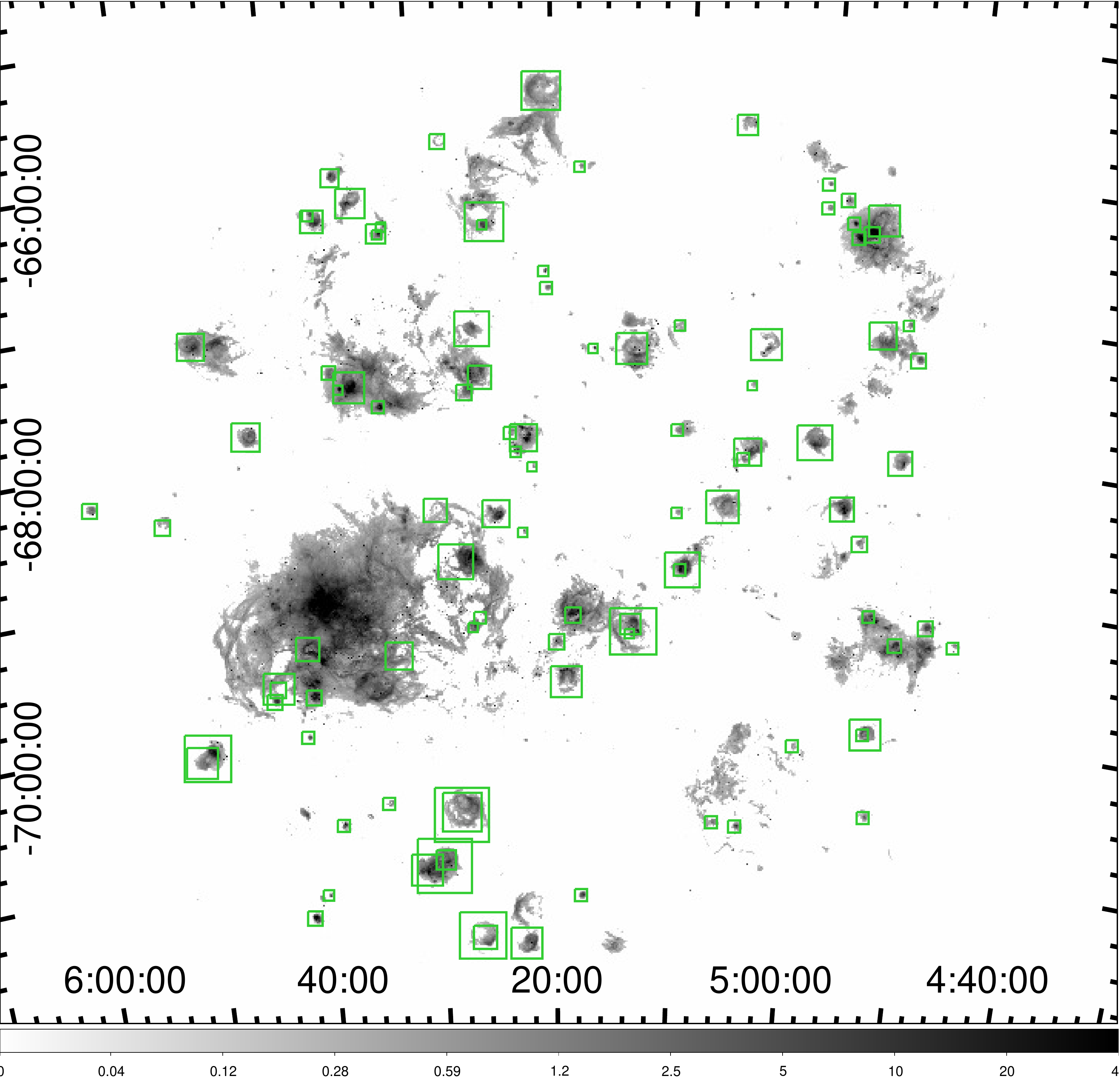}
    \caption{Bubbles detected in the LMC in the \ha\ (top), [\ion{S}{ii}] (middle), [\ion{O}{iii}] (bottom) images at window sizes ranging from 12 to 70\,px ($\sim 29 - 170\,$pc). The brightness of the line emission is shown logarithmically within the respective minimum and maximum flux thresholds used for the Minkowski maps (Table~\ref{tab:thresh_lmc}).  }
    \label{fig:lmc_bubbles}
\end{figure}

\begin{table}
\caption{Manually determined bubble detection parameters in the LMC}
\begin{center}
\ha{}/[\ion{O}{III}]\\
\begin{tabular}{cccccc}
\hline
\multicolumn{2}{c}{$w$} & $m$& line density  & \multicolumn{2}{c}{$w'$} \\
\,[pc] & [px] & & threshold & [pc] & [px]\\
\hline
29 & 12 & 0.25 & 23 & 97 & 40\\
26 & 15 & 0.25 & 24 & 121 & 50\\
48 & 20 & 0.25 & 20 & 145 & 60\\
73 & 30 & 0.3 & 16 & 242 & 100\\
97 & 40 & 0.3 & 17 & 242 & 100\\
121 & 50 & 0.3 & 16 & 364 & 150\\
145 & 60 & 0.3 & 17 & 485 & 200\\
170 & 70 & 0.3 & 24 & 606 & 250\\
\hline
\end{tabular}\\
\vspace{15pt}
[\ion{S}{ii}]\\
\begin{tabular}{cccccc}
\hline
\multicolumn{2}{c}{$w$} & $m$& line density  & \multicolumn{2}{c}{$w'$} \\
\,[pc] & [px] & & threshold & [pc] & [px]\\
\hline
29 & 12 & 0.25& 21 & 97  & 40\\
26 & 15 & 0.25& 22 & 121 & 50\\
48 & 20 & 0.25& 18 & 145 & 60\\
73 & 30 & 0.3 & 14 & 242 & 100\\
97 & 40 & 0.3 & 15 & 242 & 100\\
121& 50 & 0.3 & 14 & 364 & 150\\
145& 60 & 0.3 & 15 & 485 & 200\\
179& 70 & 0.3 & 20 & 606 & 250\\
\hline
\end{tabular}
\end{center}
\tablefoot{Parameters for H$\alpha$ and [\ion{O}{III}] images are given in the upper table and those for the [\ion{S}{II}]  image in the lower table. $w$ and $w’$ refer to the smoothing window diameter of the original Minkowski map and the map at a larger scale.}
\label{tab:bubbleparams}
\end{table}

\subsection{LMC}
\label{sec:LMC}
These methods were used to detect bubbles in the LMC at different window sizes ranging from 12\,px to 70\,px ($\sim 29 - 170\,$pc). 
Below this, the detection produced a large number of false positives because the pixel size becomes relevant. Noise was not significant in these images. The parameters applied for the generation of Minkowski maps are given in Table~\ref{tab:thresh_lmc} (top) and the bubble detection parameters are listed in Table~\ref{tab:bubbleparams}. The line density threshold in [\ion{S}{II}] had to be lower than for \ha\ and [\ion{O}{III}] because many smaller structures appear to be present in the [\ion{S}{II}] image around the larger filaments, which blurred the line images.

The overall line images and line densities of the LMC \ha{} image at window sizes 70\,px/170\,pc and 250\,px/610\,pc are shown in Fig.~\ref{fig:lmc_lines}. 
Here, larger bubbles and the large-scale structure are visible. 
Several objects detected in a range of window sizes correspond to the same physical bubble and had to be combined. The resulting detections in \ha, [\ion{S}{ii}], and [\ion{O}{iii}] are shown in Fig.~\ref{fig:lmc_bubbles}.
In total, we find 170 objects in \ha, 138 in [\ion{S}{ii}], and 100 in [\ion{O}{iii}].

\subsection{SMC}
\label{sec:SMC}

The noise behavior of these methods was tested for noisier SMC data. The unbinned images were processed using the opening technique, in which an image is smoothed by first eroding (setting the pixel value to the lowest value of its neighbors) and then dilating (setting the pixel value to the highest value in a neighborhood). 
This removes the brightest dots. The opened images were then binned to reduce the size of the files (for computer memory reasons).
For the opening method, a circular window with a diameter of three or five pixels and a binning factor of three were used. This resulted in two different noise levels, the smaller opening diameter preserving the resolution after binning. 
Additionally, the Minkowski map generation thresholds were set to either higher pixel values than most of the noise and some of the structures, or to pixel values that included some noise and fainter filaments. 

\begin{figure*}
    \centering
    \includegraphics[trim= 0 1.7cm 0 0, clip,width=\hsize]{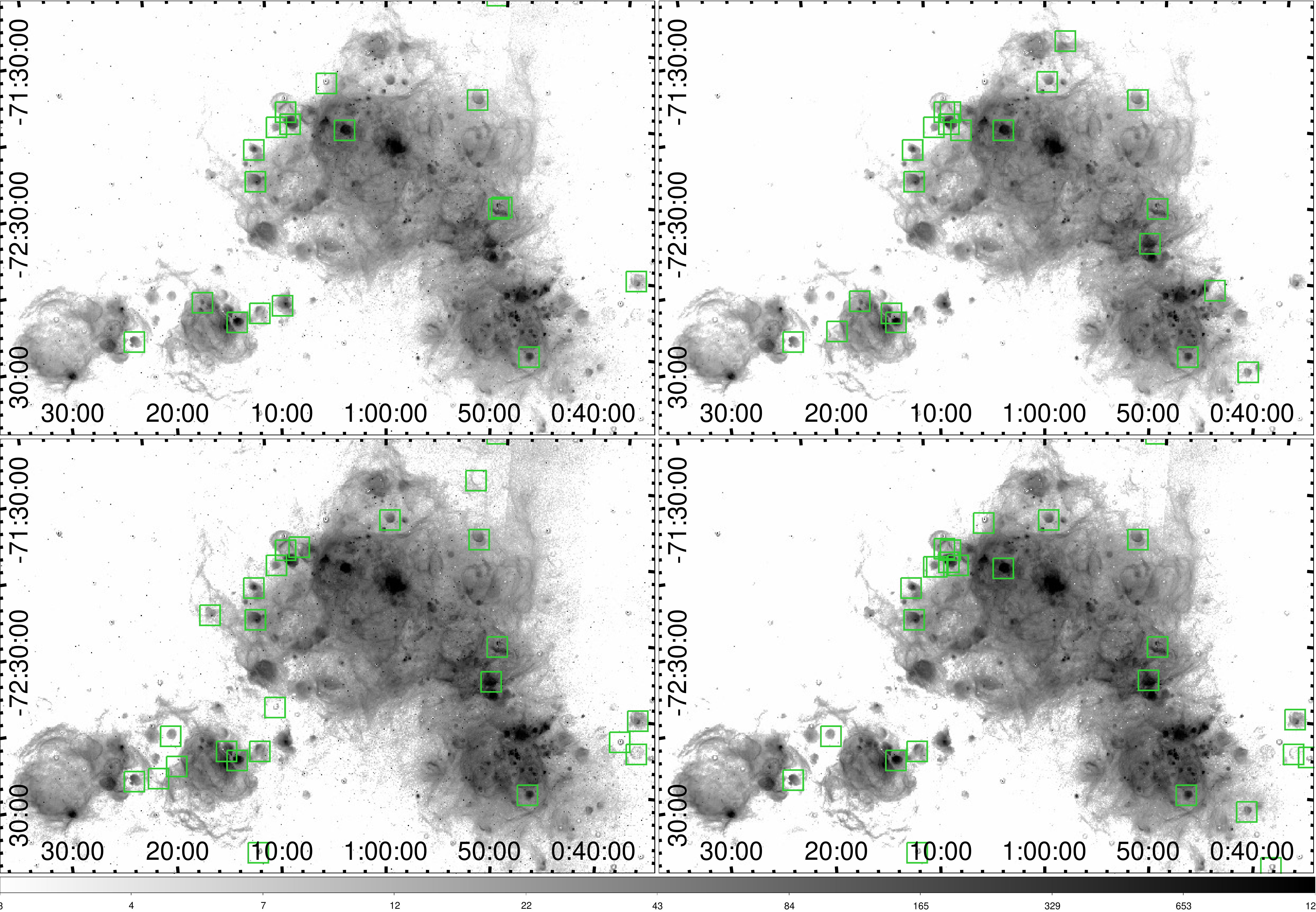}
    \caption{Bubbles detected at a window size 40\,px ($\sim 72\,$pc), $m=0.3$, line density threshold 10, in SMC \ha  \ data. Top: Twelve thresholds between 6 and 2500. Bottom: Twelve thresholds between 3 and 1300 used for the Minkowski map generation. Left: Original image opened with a diameter of 3\,px (more noise). Right: Opening diameter of 5\,px (less noise). The brightness of the line emission is shown logarithmically within the respective minimum and maximum flux thresholds used for the Minkowski maps (Table~\ref{tab:thresh_lmc}).}
    \label{fig:bubbles_smc_ha_compare}
\end{figure*}

For the SMC bubble detection, the noise analysis was carried out at a window size of 40\,px ($\sim 72\,$pc).
The differences here are not striking, but relevant for bubble detection. 
Bubbles found in \ha{} at this scale are shown in Fig.~\ref{fig:bubbles_smc_ha_compare}.
The different settings led to a comparable amount of bubbles, but partially different objects were detected. In the higher-threshold Minkowski-map images (upper panel), regions at the noisier borders are avoided.
The situation is similar in [\ion{S}{ii}]. 
Overall, fewer bubbles are detected due to noise, and the number of detected non-bubbles is high. 
In [\ion{O}{iii}], noise is a smaller problem in general, and for all settings tested here, a large number of bubbles is found. The reason most likely is that there are in general fewer faint and largely extended emissions than in the \ha\ and [\ion{S}{ii}] images. 
In all these figures the same line density threshold and line scale factor $m$ were used, and no extensive optimization of these parameters was performed. 
The analysis has shown that higher thresholds above the noise level and more denoising by opening are not universally beneficial. Noise is removed sufficiently when the Minkowski maps are created.

The results of the noise analysis were used to detect bubbles in the SMC at window sizes ranging from 12 to 70\,px (22 to 126\,pc). 
The line scale factor $m$ was taken from the LMC settings and the line density threshold was determined in few steps. This was kept simple to check whether acceptable results can be achieved without time-consuming optimization. The bubble detection was run twice with different noise settings, once with more and once with less noise.

In all cases, more objects were detected when more noise was present: In \ha{}, the number of combined bubbles rose from 153 to 287, in [\ion{S}{ii}] from 61 to 100, and in [\ion{O}{iii}] from 131 to 146. Not all of these combined objects correspond to actual bubbles, and some bubbles still have several boxes, but this gives an idea of the effects of noise. Many bubbles were found especially in the more chaotic region in the southwest.

\begin{figure*}
    \centering
    \includegraphics[trim=0 1.7cm 0 0,clip,width=\textwidth]{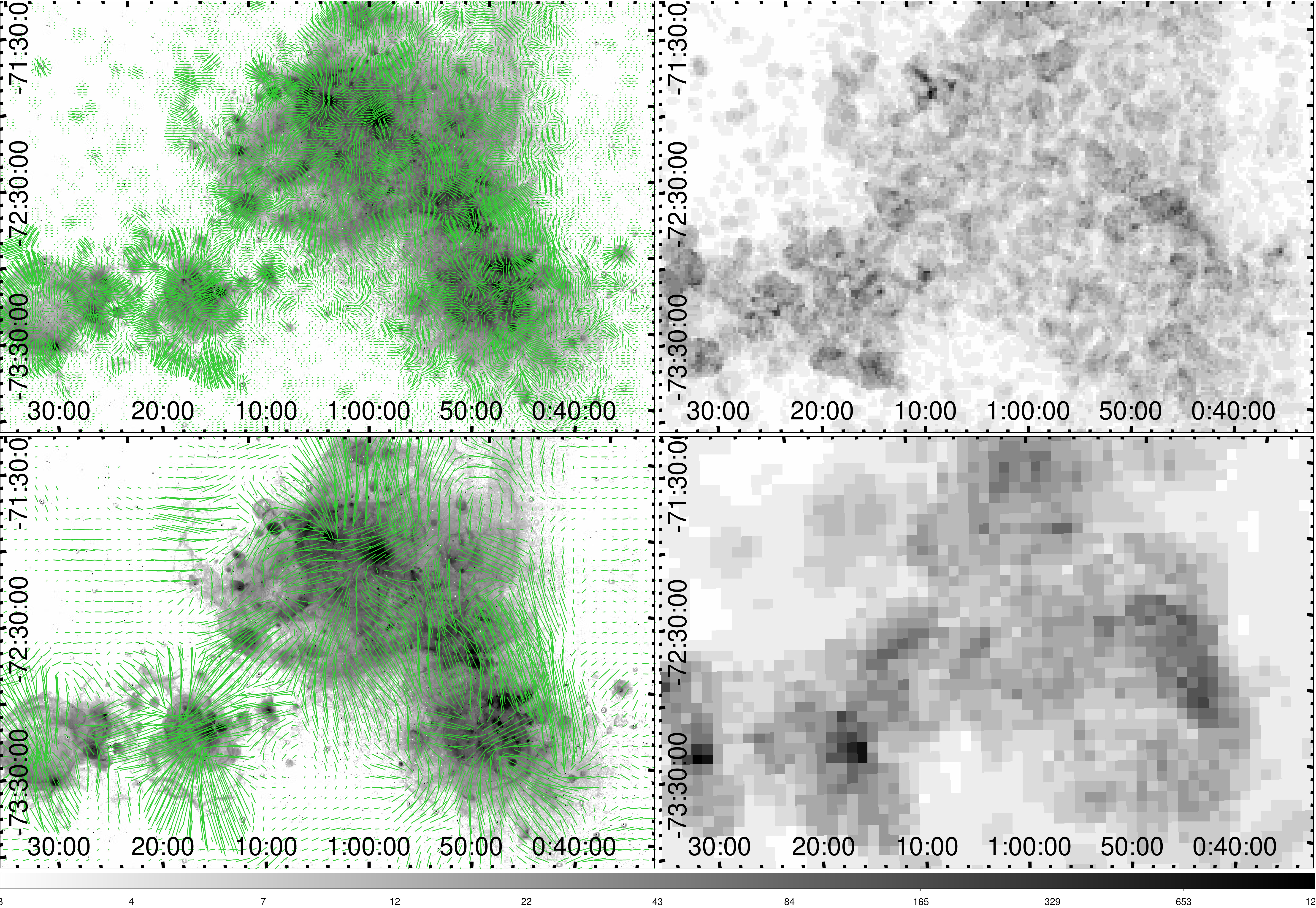}
    \caption{Lines (left) and line densities (right) in the \ha\ image of the SMC at a window size 70\,px/$\sim$130\,pc (top) and 250\,px/$\sim$450\,pc.}
    \label{fig:smc_lines}
\end{figure*}

\begin{figure*}
    \centering
    \includegraphics[trim=0 1.7cm 0 0,clip,width=\textwidth]{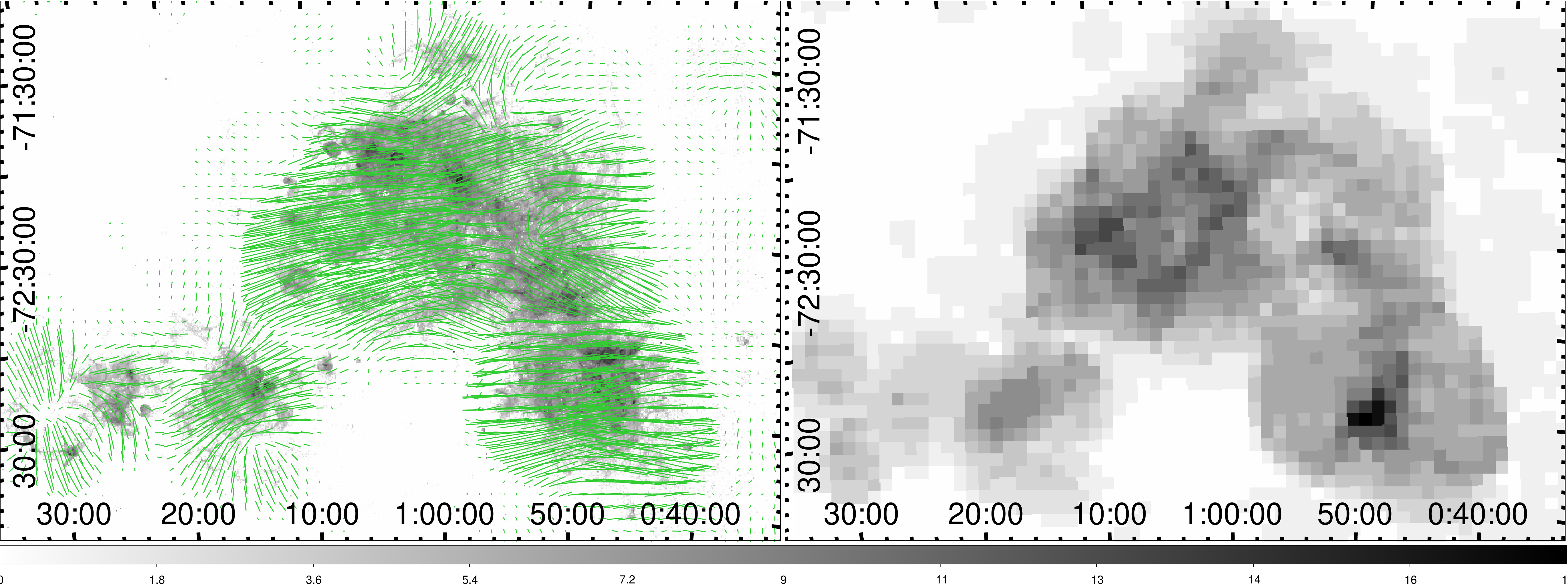}
    \caption{Lines (left) and line density (right) in the [\ion{S}{ii}] image of the SMC at a window size 250\,px/$\sim$450\,pc.}
    \label{fig:smc_lines_sii}
\end{figure*}

\begin{figure}
    \centering
    \includegraphics[trim=0 1.7cm 0 0,clip,width=.47\textwidth]{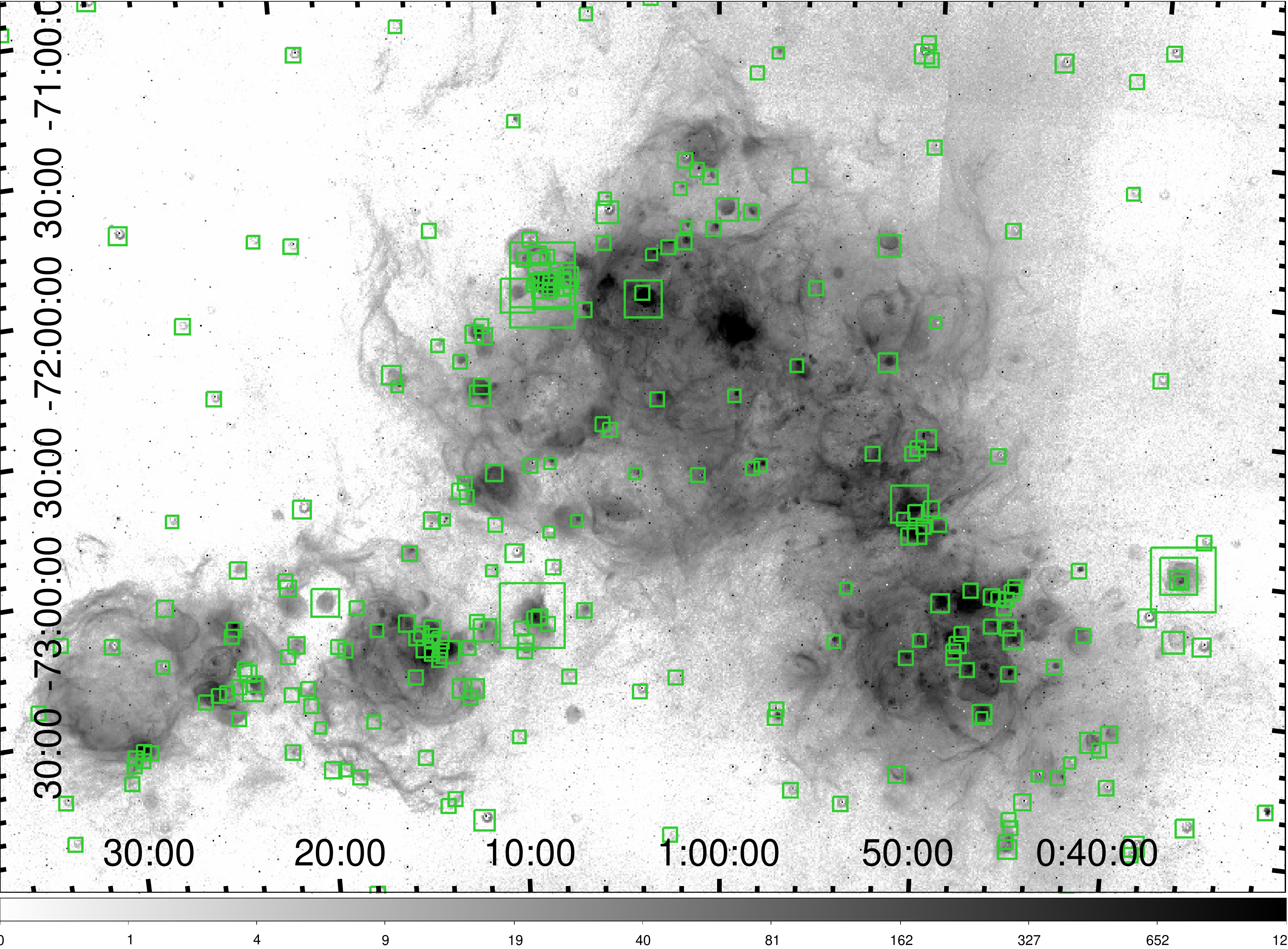}
    \includegraphics[trim=0 1.7cm 0 0,clip,width=.47\textwidth]{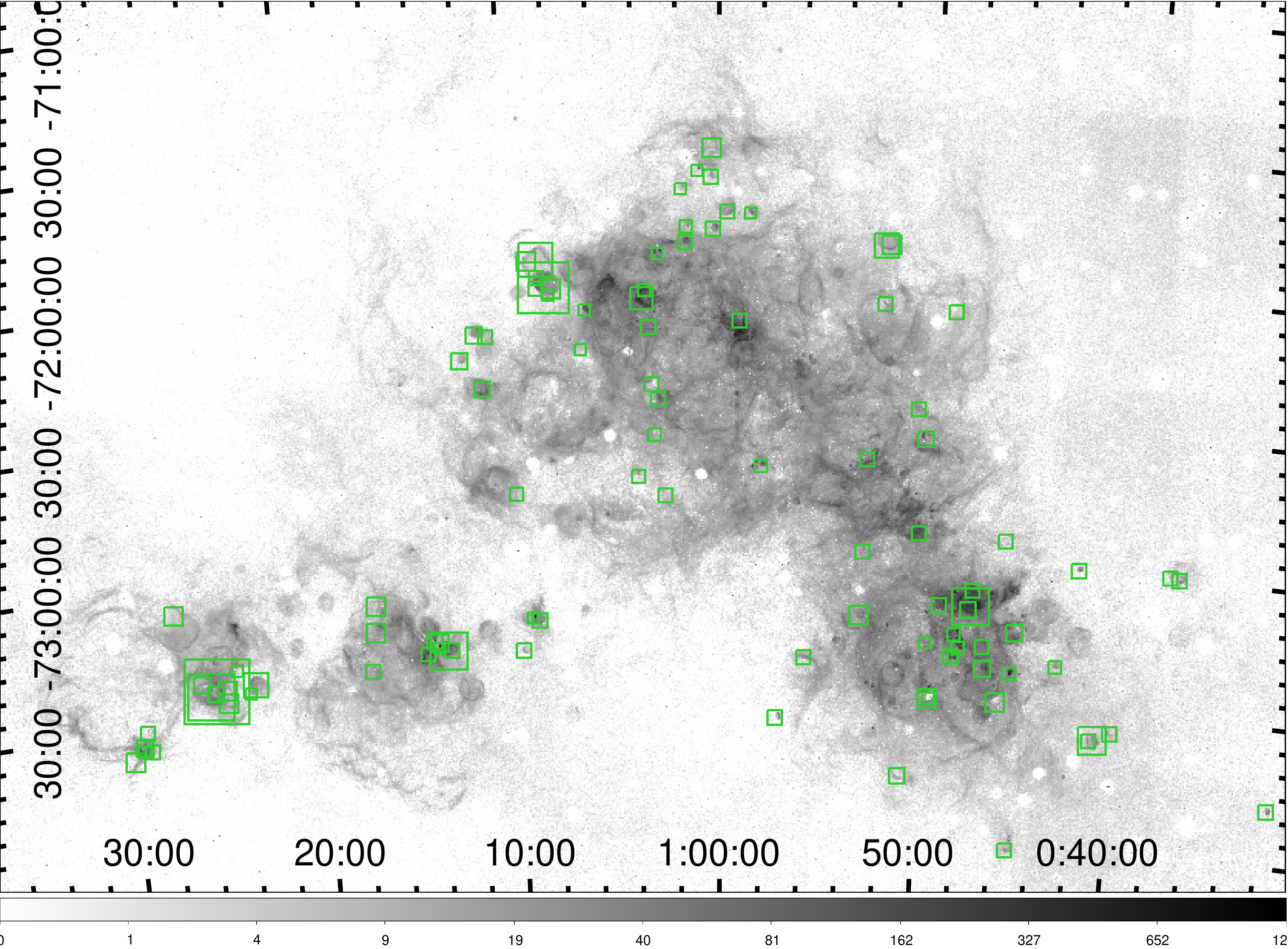}
    \includegraphics[trim=0 1.7cm 0 0,clip,width=.47\textwidth]{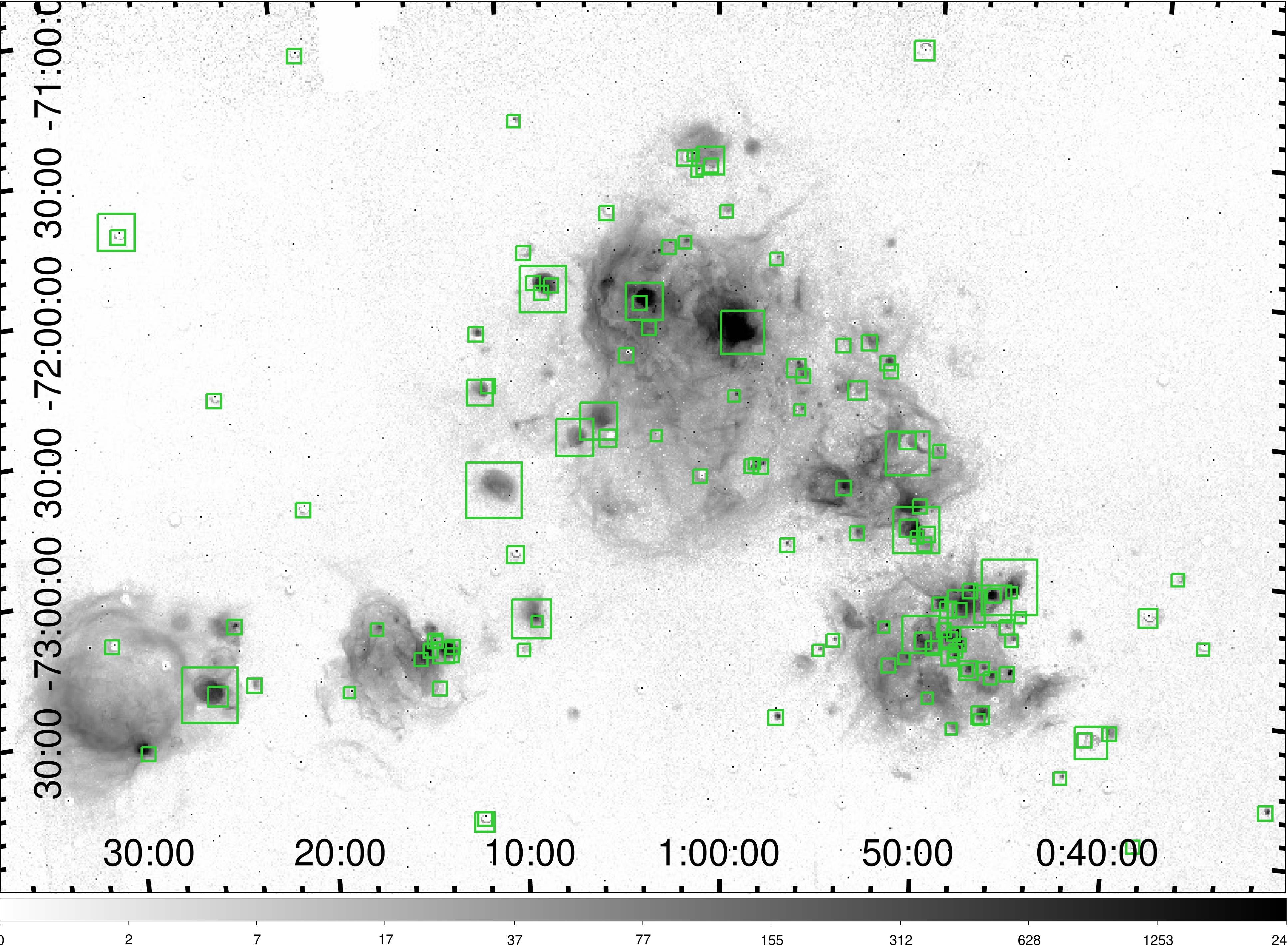}
   \caption{Combined bubbles detected in the \ha\ (top), [\ion{S}{ii}] (middle), and [\ion{O}{iii}] (bottom) images of the SMC with window sizes from 12-70\,px (22 to 126\,pc). The brightness of the line emission is shown logarithmically within the respective minimum and maximum flux thresholds used for the Minkowski maps (Table~\ref{tab:thresh_lmc}). }
    \label{fig:bubbles_smc}
\end{figure}

Table~\ref{tab:thresh_lmc} (bottom) lists the thresholds used in Minkowski map generation for the more successful detections. Table~\ref{tab:smc_bubbleparams} lists the bubble-detection settings. 
In Fig.~\ref{fig:smc_lines}, lines and line densities in the \ha{} image at window sizes of 70\,px/$\sim$130\,pc and 250\,px/$\sim$450\,pc are shown. The lack of larger bubbles (70\,px) is visible. The same data at 250\,px for the [\ion{S}{ii}] image are shown in Fig.~\ref{fig:smc_lines_sii}. Here, the lines appear largely horizontal. Upon further inspection, we were able to verify that this is caused by the larger filaments that were found around the main body of the SMC and not by image noise. The combined bubbles for the settings with more noise are shown in Fig.~\ref{fig:bubbles_smc}. 
The number of detections at window sizes of 50 and 60\,px (90 and 108\,pc) is low (zero for \ha). Bubbles at this size are located at the noisier edges of the brighter regions. We find 287 objects in \ha, 100 in [\ion{S}{ii}], and 146 in [\ion{O}{iii}] in the SMC.

\begin{table}

\caption{Manually determined bubble-detection parameters in the SMC}
    \centering
\ha{}/[\ion{O}{III}]\\
\begin{tabular}{cccccc}
\hline
\multicolumn{2}{c}{$w$} & $m$& line density  & \multicolumn{2}{c}{$w'$} \\
\,[pc] & [px] & & threshold & [pc] & [px]\\
\hline
22  & 12 & 0.25& 10 & 72  & 40\\
27  & 15 & 0.25& 10 & 90  & 50\\
36  & 20 & 0.25& 12 & 108 & 60\\
54  & 30 & 0.3 & 10 & 180 & 100\\
72  & 40 & 0.3 & 10 & 180 & 100\\
90  & 50 & 0.3 & 10 & 270 & 150\\
108 & 60 & 0.3 & 10 & 360 & 200\\
126 & 70 & 0.3 & 12 & 450 & 250\\
\hline
\end{tabular}\\
\vspace{15pt}
[\ion{S}{II}]\\
\begin{tabular}{cccccc}
\hline
\multicolumn{2}{c}{$w$} & $m$& line density  & \multicolumn{2}{c}{$w'$} \\
\,[pc] & [px] & & threshold & [pc] & [px]\\
\hline
22  & 12 & 0.25& 10 & 72  & 40\\
27  & 15 & 0.25& 10 & 90  & 50\\
36  & 20 & 0.25& 10 & 108 & 60\\
54  & 30 & 0.3 & 10 & 180 & 100\\
72  & 40 & 0.3 & 12 & 180 & 100\\
90  & 50 & 0.3 & 12 & 270 & 150\\
108 & 60 & 0.3 & 12 & 360 & 200\\
126 & 70 & 0.3 & 12 & 450 & 250\\
\hline
\end{tabular}
    \tablefoot{Parameters for H$\alpha$ and [\ion{O}{III}] images are given in the upper table and those for the [\ion{S}{II}]  image in the lower table. $w$ and $w’$ refer to the smoothing window diameter of the original Minkowski map and the map at a larger scale.}
    \label{tab:smc_bubbleparams}
\end{table}

\label{bubbles_results}

\section{Correlation analysis}
\subsection{Statistical analysis}
\label{sec:stat}
It is expected that bubbles and massive stars are spatially correlated. To verify this, an independence hypothesis was tested using the theory of point processes. Bubbles and stars are described as points in a point pattern. We chose to use Ripley's $K$ \citep{ripley1976,ripley1981} as a summary characteristic to describe the patterns. 

It is defined such that $\lambda K(r)$ is the expected number of points within a distance $r$ to a random point of the pattern, not counting the point itself. $\lambda$ is the intensity of the process.
For a point process for independent points with constant intensity (a homogeneous Poisson process), we can write $K(r) = \pi r^2$. 
If a pattern is clustered, $K(r)$ will be larger than the Poisson value because a random point will be part of a cluster and have many close neighbors. If the pattern is regular and the points are repulsive, the opposite occurs. 
Estimators of $K$ often have a high variance at large $r$ because of its cumulative nature, therefore the transform proposed by \cite{besag1977} is often used, 
\begin{equation}
    L(r) = \sqrt{\frac{K(r)}{\pi}}~.
\end{equation}
In the Poisson case, $L(r) = r$, which facilitates a visual comparison. 

It is often useful to assign further information to the points in a pattern, such as whether a point is a star or a bubble. 
This can be described by a qualitatively marked pattern. In marked patterns, each point carries an additional property (a mark) given by any mathematical object. Here, the qualitative descriptions ``star'' and ``bubble'' were used. 
The summary characteristics can be extended to qualitatively marked patterns, such that only points of one mark within a certain distance to points of the other mark are considered. Then $\lambda_j K_{\mathrm{cross},\,i, j}$ is the expected number of points of type $j$ within a distance $r$ to points of type $i$. These marked correlation functions are a standard tool described (e.g.,) in \citet[~]{statbook}.

\subsubsection{Correlation hypothesis test}
To determine whether the given samples of bubbles and stars are correlated, a random superposition test was performed. The test described here was inspired by \citet[~]{statbook} and was first described in similar form by \cite{lotwicksilverman1982}. 

Several instances of an independence null hypothesis were generated by shifting the patterns of stars and bubbles with respect to each other (with periodic boundary conditions). To minimize border effects, a rectangular region of interest encompassing most of the LMC was chosen, and only the points inside were taken into account. For each shifted pattern, a summary characteristic function was calculated. If the initial unshifted position was special, its characteristic function should deviate significantly from the shifted functions. 

The significance of the deviation here was measured using acceptance envelopes around the mean value of the shifted functions. Envelope methods were introduced by \cite{ripley1977} (for a further discussion of uses and caveats, see (e.g.,) \cite{statbook,baddeley2014}). The envelopes used here are the so-called maximum absolute deviation (MAD) envelopes: 
The patterns are shifted $n_\text{sim}$ times and the summary characteristic calculated. The mean value is taken from $n_\text{sim}/2$ of the simulations. For the other functions, the largest absolute deviation from this mean is calculated. The n-th largest of these values is chosen as a critical deviation $d_\text{crit}$. The envelope then spans the values around the mean$\pm d_\text{crit}$. The null hypothesis is rejected if the summary characteristic of the original pattern exceeds the envelope at any $r$ at significance level $\alpha = n/(1 + n_{\text{sim}}/2)$ \citep{ripley1981}. 
We used $n_{\text{sim}} = 300$ and $n = \lfloor n_{\text{sim}}/40\rfloor = 7$, giving a significance level of $\alpha \simeq 0.047$. This means that if the patterns are independent, $L(r)$ has a probability of 0.047 of exceeding the envelope by chance. The test was implemented in R using the library spatstat \citep{spatstatBook}. 

\subsubsection{Application of the correlation test}

Various combinations of bubbles, point patterns generated from star formation rates, and massive stars from the catalog of \cite{bonanos2009} were tested for correlation. 
The method was tested with bubbles detected in the three images (\ha, [\ion{O}{III}], and [\ion{S}{II}]).
As expected, a correlation is found between the images because many bubbles are located at the same positions at all three wavelengths. $L(r)$ quickly jumps to higher values than in the shifted patterns and largely exceeds the MAD envelope. The independence hypothesis is thus rejected. The plots are shown in the appendix in Fig.~\ref{appfig:bubbles_bubbles}. 
The opposite occurs when the bubbles are compared to a pattern generated from the star formation rate 1\,Gyr ago \citep{harriszaritsky2009}. Star formation at that time took place mostly in the LMC bar, and the massive stars formed then and have exploded long ago. We therefore do not expect a correlation to modern-day bubbles. This is confirmed by the correlation test, where $L$ clearly stays within the envelopes (Fig.~\ref{appfig:bubbles_sfr}). The test described here is thus able to show independence for patterns where a correlation is unlikely.

\begin{figure}
    \centering
    \includegraphics[trim=0 0 0cm 0,clip,width=\hsize]{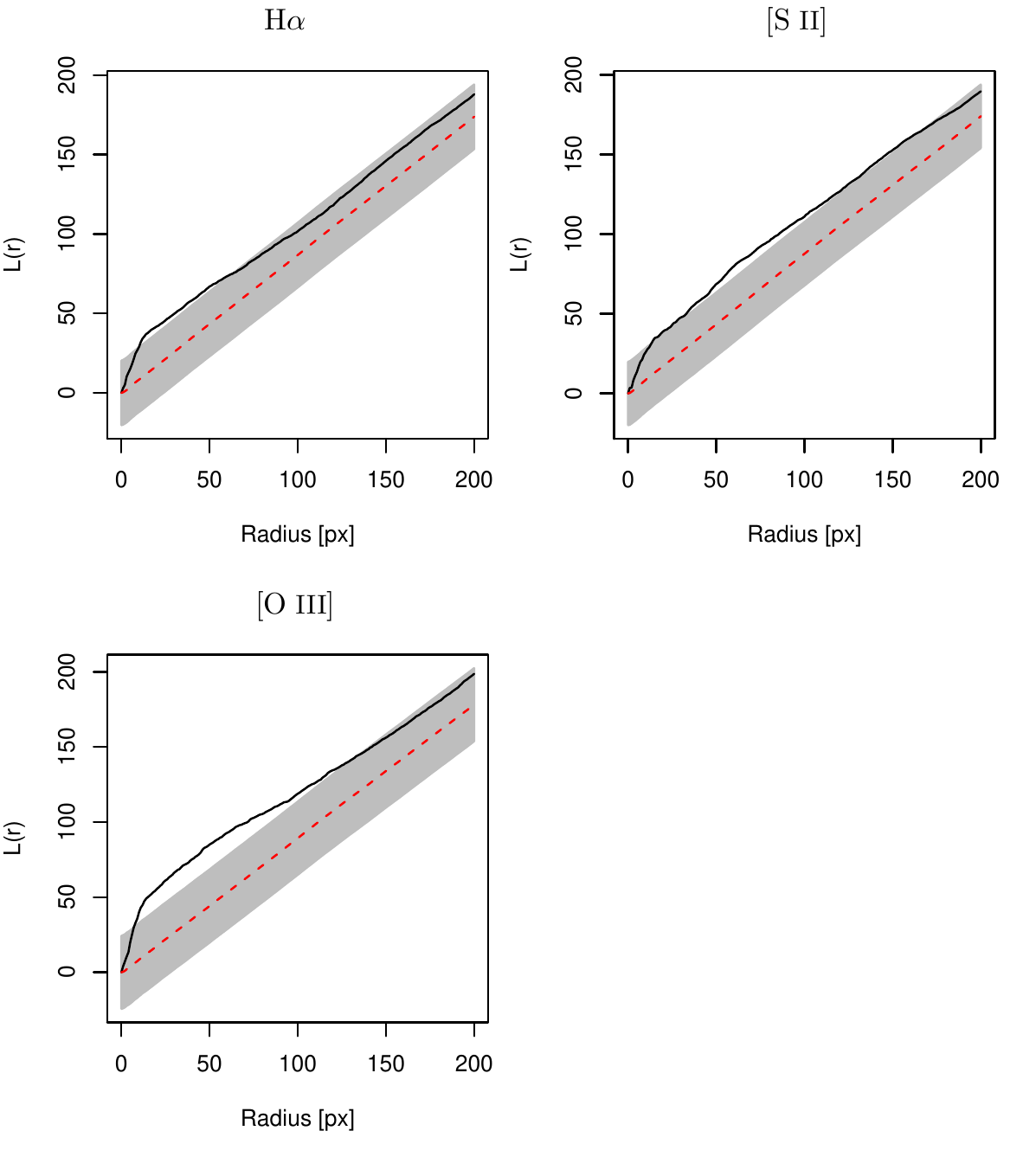}
    \caption{Bivariate Ripley’s L comparing stars in \cite{bonanos2009} and bubbles detected in \ha{} (top left), \ion{S}{ii} (top right), and \ion{O}{iii} (bottom) at window sizes of 12-70\,px. Solid black line: L estimated from original pattern. Dashed red line: Pointwise mean of half of the simulated patterns. Gray envelopes: MAD. 10\,px correspond to about 1.67' or 24\,pc at an LMC distance of 50\,kpc.
    }
    \label{fig:corr_stars_bubbles}
\end{figure}

Next, bubbles were compared to the stars of \cite{bonanos2009}. In all wavelengths, $L$ exceeds the envelope, and a correlation is found. The envelope is exceeded by a large margin for [\ion{O}{III}], but only slightly for H$\alpha$. The graphs for the LMC are shown in Fig.~\ref{fig:corr_stars_bubbles}. 

\subsection{Supernova remnants and superbubbles}
\label{sec:snr_superbubbles}
The detected bubbles were manually compared to known supernova remnants (SNRs). Of the 59 SNRs and 15 candidates listed by \cite{bozzetto2017}, 10 are detected in \ha, 14 in [\ion{S}{ii}], 7 in [\ion{O}{iii}], and 18 in any wavelength (these numbers include 1, 2, 1, and 3 candidates, respectively). 
The lack of detection is mostly due to the small size (22 cases), which is below the resolution of the image and is often accompanied by lack of any optical counterpart. Supernova remnants are typically found in radio or in X-ray band, in which the emission from the SNR shock is strongest. For those that are in principle large enough, the lack of optical emission will therefore result in nondetection in the optical line emission images. Twenty objects were most likely not detected for this reason. Although the size was large enough or the structure clear, some SNRs were not detected because of neighboring filaments that disturbed the detection (14), many of which were located in the 30 Dor region. \cite{bozzetto2017} used multiwavelength data ranging from radio to X-rays to identify these SNR. It is therefore not surprising that a large number of sources cannot be identified in optical line emission maps alone.

\begin{figure}
    \centering
    \includegraphics[width=.45\textwidth]{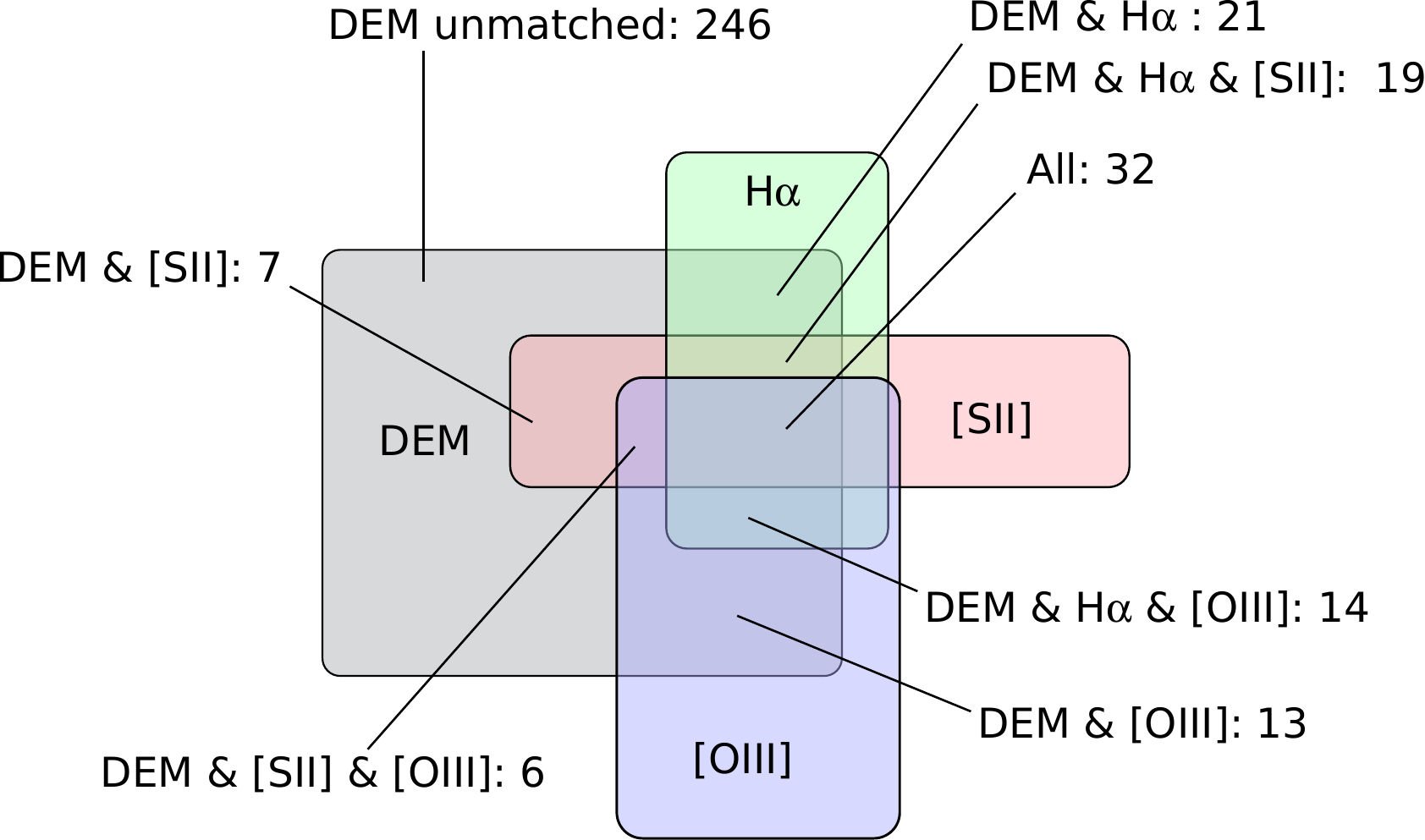}
    \caption{Number of DEM catalog objects that match at least one detected object in a given wavelength(s). Box sizes are not proportional to the total number. The assignments are exclusive, e.g., ``DEM \&\ \ha\ '' does not include ``DEM \&\ \ha\ \&\ [\ion{S}{ii}]''.}
    \label{fig:venndiag_dem}
\end{figure}

\begin{figure}
    \centering
    \includegraphics[trim=0 0 0.5cm 0, clip, width=.49\textwidth]{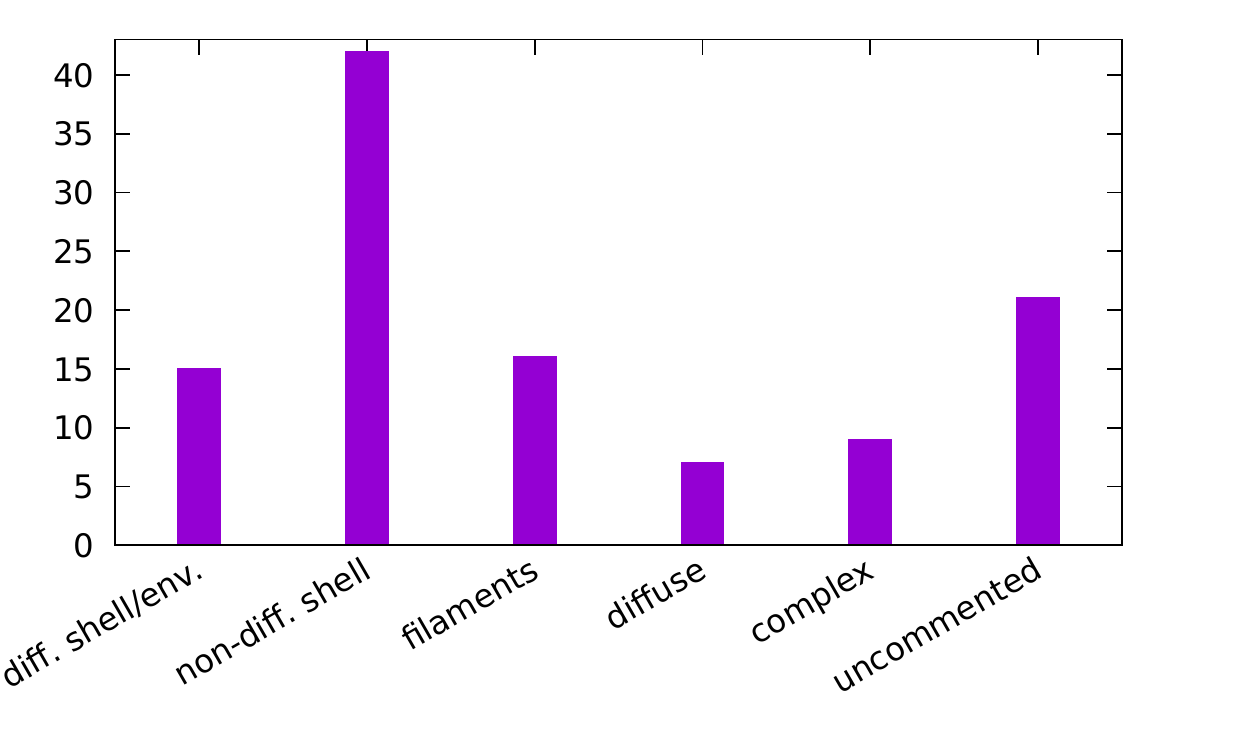}
    \caption{Number of matched DEM objects according to the most important comment by \cite{1976MmRAS..81...89D}: Diffuse shells or envelopes, nondiffuse or filamentary shells or envelopes, non-shell filaments or arcs, diffuse, complex, and uncommented. Objects with several comments are listed as shells if ``shell'' is mentioned and  as filaments if ``filament'' but not ``shell'' is mentioned. The uncommented objects appear to be shell- or envelope-like or are filamentary in our images.}
    \label{fig:hist_dem}
\end{figure}

The bubbles were also compared to the objects in the DEM catalog of emission nebulae detected in \ha+[\ion{N}{ii}] by \cite{1976MmRAS..81...89D}. Objects from this catalog and our bubbles were assumed to refer to the same astrophysical object if their size differed by up to a factor of two and their central separation was smaller than the half of the size of the DEM object. With these criteria, 86 bubbles detected in \ha, 57 detected in [\ion{S}{ii}], and 61 detected in [\ion{O}{iii}] correspond to an object in the above catalog. Thirty-two DEM objects are detected in all wavelengths and 112 in at least one. For more detailed numbers, see Fig.~\ref{fig:venndiag_dem}. The matching objects are usually shells and are often filamentary. Some of our detected objects that are not bubbles correspond to a DEM object. The number of matching objects for the most relevant  comments by \cite{1976MmRAS..81...89D} are shown in Fig.~\ref{fig:hist_dem}. The uncommented objects appear as shells or envelopes or are filamentary in our images.

In a similar manner, we compared the bubbles we detected to a list of 20 LMC superbubbles that
we compiled from \cite{chu_maclow_1990}, \cite{oey_lmc_sb}, and \cite{dunne2001}.  We find
that 13 superbubbles were detected in at least one filter and 5 (DEM L 25,
DEM L 31, DEM L 86, DEM L 192, and DEM L 301) were detected in all filters.  In general, 
these superbubbles are relatively small ($< 6'$ diameter) and have a circular morphology. 
We find that eleven of the thirteen superbubbles were detected in \ha, nine were detected 
in [\ion{S}{ii}], and six were detected in [\ion{O}{iii}].  

Seven of the superbubbles were not detected in any filter.  Five of the seven superbubbles
that were not detected (DEM L 246, DEM L 263, DEM L 269, DEM L 284, and 30 Dor C) are 
located toward the 30 Doradus and the \hii\ regions to the south.  As previously discussed 
in the case of LMC SNRs, some structures are not detected because of confusion from neighboring
filaments and large-scale extended emission.  This is most likely the case for these
superbubbles.  The other two superbubbles that were not detected (DEM L 106 and DEM L 205)
are also projected toward other emission regions that prevent a clean detection. DEM L 106 is located close to a mosaic artifact that disturbs the line density. Images of the individual superbubbles are shown in Appendix~\ref{sec:superbubbles}.

A further attempt was made to determine the number of new detections and false positives. For all detected bubbles, we searched for known objects tagged as \hii\ regions or bubbles in the SIMBAD database \citep{simbad}. The center of the known object had to be within a box half the size of our detected object. 
These are very rough criteria for a quick overview of the number of possibly new objects. In \ha\ ([\ion{S}{ii}], [\ion{O}{iii}]), 60 (61, 31) detected objects have no known counterpart in this selection.

The unmatched objects  clearly show that many are indeed known objects that were not selected by the specific criteria, such as N206, whose center is marked in a different position. Some of our unmatched objects correspond to parts of bubbles, where regions with more curved filaments are detected as a separate object. In general, inaccuracies in size and position complicate the cross-correlation. An accurate determination of the number of new detections versus the number of false positives is not possible in a simple automated way. The necessity of additional data and algorithms to take velocities into account, for example, means that this aim is beyond the scope of this paper.

\subsection{Supergiant shells in the LMC}
\begin{figure*}
    \centering
    \includegraphics[trim=0 1.6cm 0 0,clip,width=\hsize]{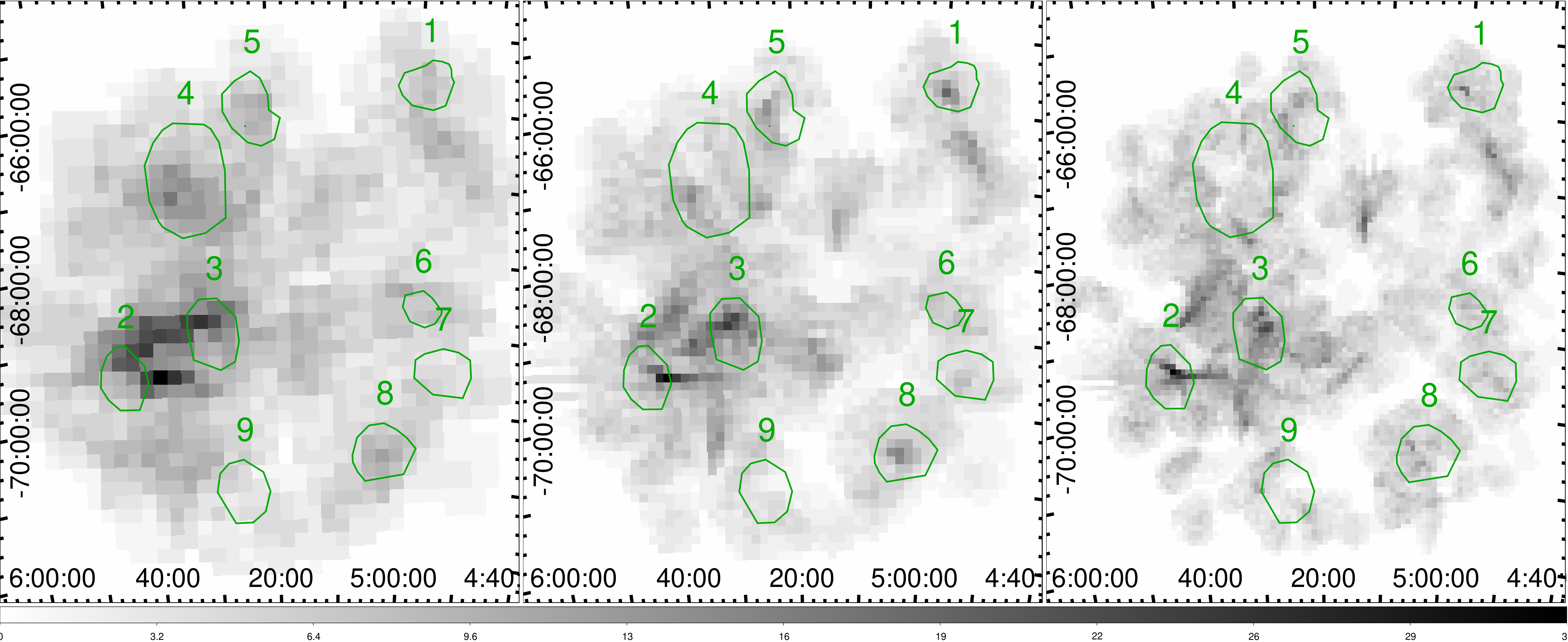}
    \caption{Approximate contours of SGS as in \cite{1980MNRAS.192..365M} in line density images of the LMC in \ha{} with window sizes of 400, 250, and 150\,px. }
    \label{fig:sgs_linedens}
\end{figure*}

\begin{figure*}
    \centering
    \includegraphics[trim=0 1.7cm 0 0,clip,width=\hsize]{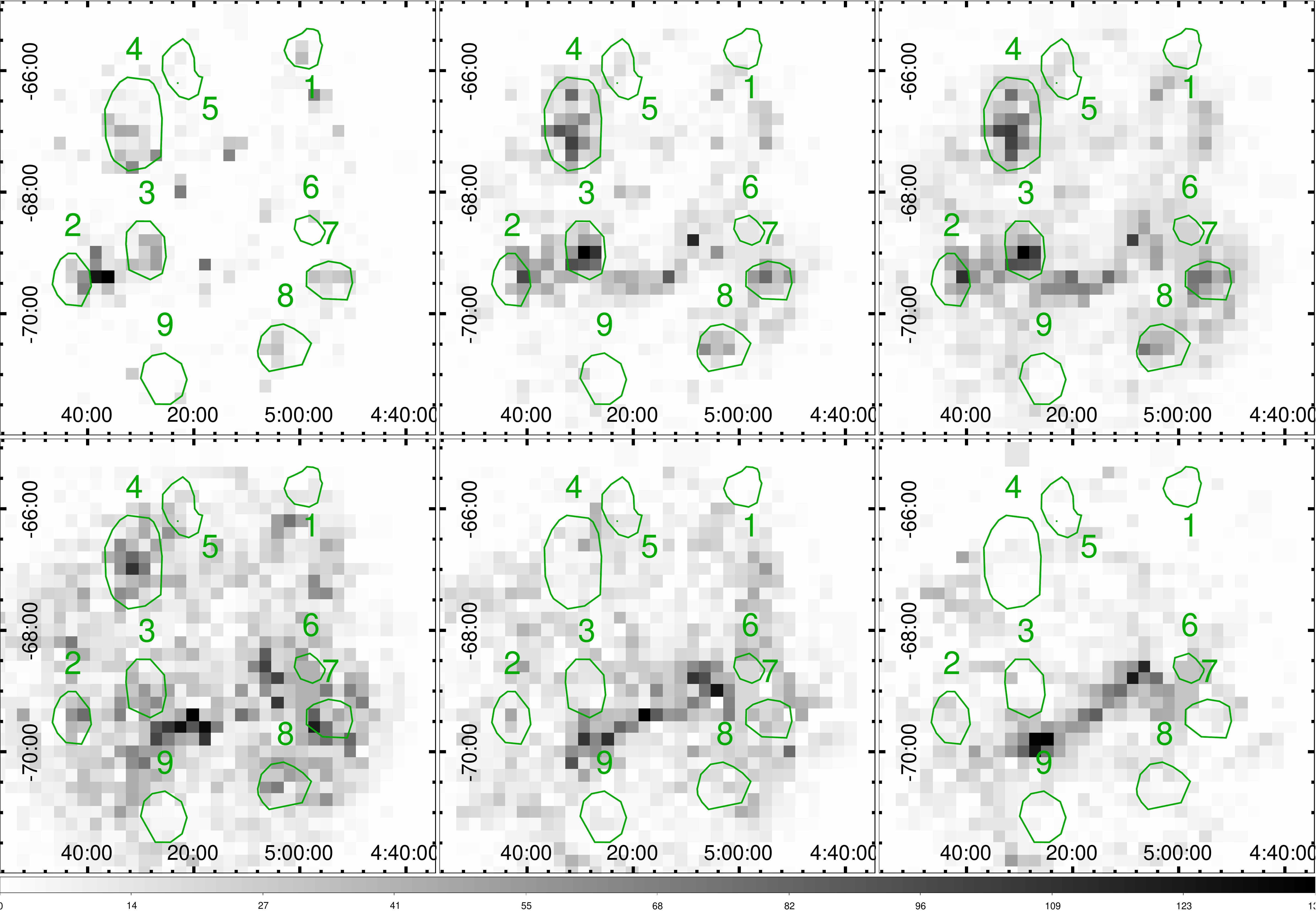}
    \caption{Approximate contours of SGS as in \cite{1980MNRAS.192..365M} on star formation rates in the LMC. Top: 7, 15, 20\,Myr ago. Bottom: 30, 50, 100\,Myr ago, from left to right.}
    \label{fig:sgs_sfr}
\end{figure*}

Early emission line studies of the Magellanic Clouds have revealed nine SGS in the LMC and one in the SMC \citep{1980MNRAS.192..365M}. These SGS are connected structures with a size that is larger by about one order of magnitude or more than that of typical superbubbles, and they have been observed both in optical line emission and in \ion{H}{I} images \citep{1978A&A....68..189G, 1980MNRAS.192..365M, kim1998}.

The origin of the SGSs is not fully understood. They are most likely caused by a superposition of the effects of a large number of massive stars. 
We compared the large-scale line density of the LMC to the distribution of SGSs described by \cite{1980MNRAS.192..365M}. Fig.~\ref{fig:sgs_linedens}
shows the shell contours on the corresponding LMC \ha{} line density at window sizes of 150, 250, and 400\,px. Additionally, they are compared to the star formation rates of the LMC as calculated by \cite{harriszaritsky2009} (see Fig.~\ref{fig:sgs_sfr}). We did not attempt to detect SGSs automatically because there are so few of them. 
Lines perpendicular to a filament indicate that a force is acting on the filament in the direction from a point at which the lines intersect. An excess in line density can thus indicate the position of an energy source, such as the massive star in the center of a stellar bubble or large associations of young stars. 

\subsubsection{Interior enhancement of the line densities}

In smaller SGS (LMC 1, 3, 5, 6, 7, and 8) central enhancements are found in the line densities for window sizes of 150 or 250 px, corresponding to $\sim$500 pc. The comparison to the star formation rate maps shows that there were no clear bursts of star formation in these SGS, with continuous, relatively high star formation since $\sim$30 Myr ago. The line densities of SGS LMC 3 also show a similar dependence on window size and star formation rate, with one small difference: there seemed to be a burst of star formation inside SGS LMC 3 $\sim$15 -- 20 Myr ago.
Inside the largest SGS LMC 4, a maximum in line density is only found for the largest window size. The star formation rate was high between 15 to 30 Myr ago. The young stars formed during this period are most likely responsible for the creation of SGS LMC 4.

Additionally, line density enhancements that are not associated with any of the SGSs listed here are found between SGS 2, 3, and 9 at window sizes of 250 and 150\,px, for instance. This might be a sign for additional SGSs, but this is beyond the scope of this paper, and further investigation is required. 

\subsubsection{Supergiant shell LMC 2}

Supergiant shell LMC 2 shows a different behavior. 
SGS LMC 2 has a strikingly high line density at its western edge between SGS LMC 2 and SGS LMC 3 on larger scales (see also Fig.\,\ref{fig:lmc_lines}, bottom). At the same position, high star formation occurred only $\sim$7 Myr ago. This means that some burst of star formation must have occurred at the position between SGS LMC 3 and SGS LMC 2 several million years ago, which resulted in the formation of SGS LMC 2. The displacement of SGS 2 with respect to the star-forming regions appears to indicate that there must have been a larger-scale velocity field or density gradient in the distribution of matter that caused SGS LMC 2 to expand toward the east.

Studies of the spatial distribution and velocities of atomic hydrogen in the LMC have shown that there are two main components of \hi: the largely extended, disk-filling D component, and the more localized L component, which has a nonzero relative velocity to the D component \citet{1992A&A...263...41L}. Most likely, the two components have collided with each other 10 -- 20 Myr ago. \citet{2017PASJ...69L...5F} analyzed higher-resolution \hi\ data of the eastern part of the LMC, in particular, in the region around the giant \hii\ region 30 Doradus with the young star cluster R136 and south of it. They showed that the L component must have encountered and moved through the D component in this southeastern part of the LMC. The collision of the two large gas clouds, which has started at about the position of 30 Doradus, must have triggered the formation of the extremely massive star cluster R136. The region at the western edge of SGS LMC 2 in which we find high line densities is consistent with this interaction region. We have recently analyzed the properties of the hot interstellar plasma in the entire interaction region using X-ray data. The spectral analysis has shown that the plasma in the region around 30 Doradus and SGS LMC 2 has been heated by the young stars, while the region to the south must have been heated by the cloud collision \citep{knies2021}.

\section{Summary}
We have used MTs to detect and study bubbles and filaments in 
optical emission-line images of the Magellanic Clouds using the MCELS data.
Because astronomical images have continuous pixel values while the 
Minkowski analysis can only be applied to two-valued images with bodies and regions outside a body, 
the observational data had to be sliced into many images using thresholds.
The MT $\psi_2$ was found to be most useful for detecting bubbles.
The detection routine calculates $\psi_2$ over the entire image for a 
specific window size. Higher  $\psi_2$ indicate filaments or parts of a shell 
around a bubble. 
In addition, we used the phase of  $\psi_2$ to draw lines perpendicular to the 
detected filaments and shells. All lines from the shell of a circular bubble should meet in a
small region inside the bubble, thus allowing the identification of
the detected bodies as a shells around one coherent structure. 
In order to find these regions in which lines cross, we calculated line 
densities for each window size used for the calculation of  $\psi_2$.
We detect interstellar bubbles and filaments in the H$\alpha$,
[\ion{S}{II}], and [\ion{O}{III}] images of the LMC and the SMC using the
Minkowski analysis. 

The positions of the detected bubbles were compared in the three bands 
and in the distribution of young massive stars detected in the infrared.
In both cases, the distributions show a significant correlation, confirming
that the detected bubbles are indeed interstellar bubbles. The correlation is
highest between [\ion{O}{III}] and the stars, revealing regions in which
the interstellar gas is photoionized by very young stars. 

The detected bubbles were also compared to known catalogs of emission nebulae, SNRs, and superbubbles. Eighteen of 74 SNRs listed by \cite{bozzetto2017} are detected in all three wavelength bands; those that are not are mostly too small or lack an obvious structure in the emission-line images. Of the 20 superbubbles from \cite{chu_maclow_1990}, \cite{oey_lmc_sb}, and \cite{dunne2001}, 13 are found; the rest are embedded in or are close to other emission regions that disturb the detection. Out of the 358 objects in the DEM catalog of emission nebulae \citep{1976MmRAS..81...89D}, 112 correspond to at least one detected object. Most of these are shell-like or filamentary.

This analysis has shown that MTs are a powerful tool for detecting and characterizing the size and the orientation of interstellar structures over an entire galaxy. Because the analysis  finds elongated structures on different scales, we have detected filaments and shells both in the LMC and the SMC. The LMC is more suited for an analysis of interstellar structures because it has a face-on disk. The structure of the SMC is more irregular and thus more complex. A possible improvement of the bubble-detection routine might be achieved by combining our methods with ML, using the MT approach to preselect candidates and feeding the resulting cutout images to the ML algorithm. Additionally, expanding the search to different wavelengths, especially to the IR, is an obvious starting point for further research.

The line distribution and the line density maps created by drawing lines 
perpendicular to structures with high $\psi_2$ were useful not
only for the detection of interstellar bubbles for which they were initially
calculated, but also for identifying the position of clusters or 
associations of stars, which have formed the bubbles by their radiation
and stellar winds.
By using the line density maps for structures detected
on very large scales ($>$150 px, corresponding to about 300 pc), we
identified regions with high star formation rate in the past, which might
have resulted in the formation of supergiant shells in the LMC. In most cases, these regions are found inside the SGS, as expected.
However, for SGS LMC 2, which is particularly large
and shows large thin filaments, the line density maps indicate that the
stars that created the shell were not located inside the SGS, but 
west of it. This region is located south of the giant \hii\ region 30 Doradus
and has undergone active star formation in the last $\sim$10 Myr. 
SGS LMC 2 has most likely been powered by the young massive stars that were formed during the high star formation episode triggered 
by the collision of large gas clouds in the LMC, as shown by
\citet{2017PASJ...69L...5F}.

In the SMC, similarly to most of the SGSs in the LMC, high line densities
are found inside large bubbles. 
In addition, the line distribution maps have shown
that there are large filaments around the main body of the SMC in the west,
which seem to be aligned vertically, as indicated by the parallel horizontal 
lines, both in the northern and in the southern half of the main body. 
The horizontal lines are especially clearly seen in the [\ion{S}{II}] image 
(Fig.\,\ref{fig:smc_lines_sii}). The origin of these large filaments is unknown.

\begin{acknowledgements}
M.S.\ acknowledges support by the Deutsche Forschungsgemeinschaft
through the Heisenberg professor grants SA 2131/5-1 and 12-1.
MCELS was funded through the support of the Dean B. McLaughlin fund at the University of Michigan and through NSF grant 9540747.
M.A.K.~was supported by the Volkswagenstiftung and by the Princeton University Innovation Fund for New Ideas in the Natural Sciences.
The authors thank Sebastian Kapfer for support with papaya2 and fruitful
discussions, and Fabian Schaller for providing the shapes in Figs.~1 and 2.
\end{acknowledgements}

\bibliography{literature}

\begin{appendix}
\section{Demonstrations of the correlation test}
To demonstrate the viability of the hypothesis test described in Sect.~\ref{sec:stat}, bubbles detected in different wavelengths were compared to each other (where a correlation should be found), and to a pattern generated from the star formation rate 1\,Gyr ago as in \cite{harriszaritsky2009}. In the latter case, no correlation should be found because the massive stars that formed at that time should have exploded long ago and their bubbles vanished.

\begin{figure}[h]
    \centering
    \includegraphics[width=\hsize]{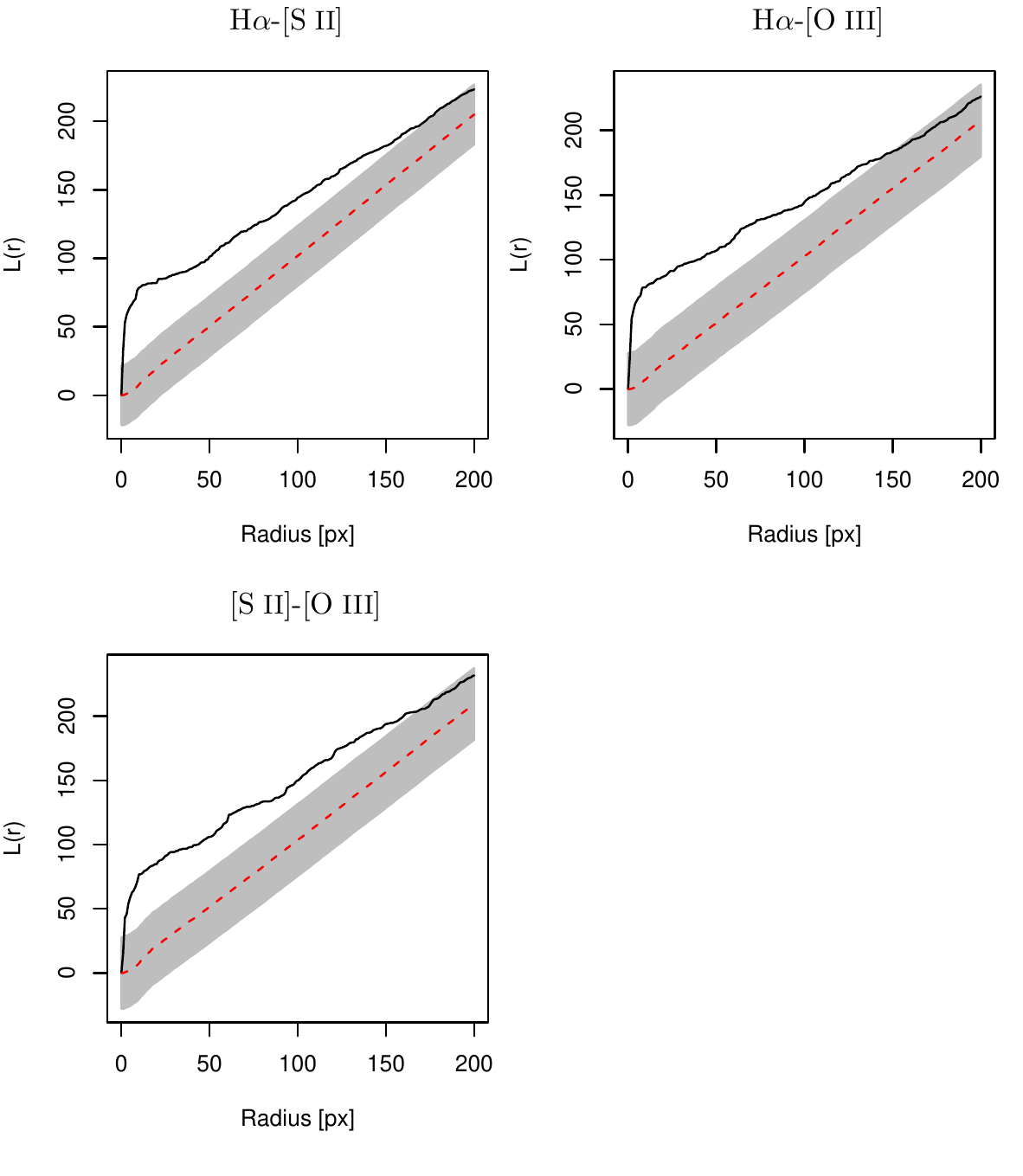}
    \caption{Bivariate Ripley’s L comparing the bubbles detected in the different wavelengths at window sizes 12-70\,px to each other (top left: \ha-\ion{S}{ii}; top right: \ha-\ion{O}{iii}; bottom left: \ion{S}{ii}-\ion{O}{iii}). Solid black line: L estimated from original pattern. Dashed red line: Pointwise mean of half of the simulated patterns. Gray envelopes: MAD. 10\,px corresponds to about 1.67' or 24\,pc at an LMC distance of 50\,kpc.}
    \label{appfig:bubbles_bubbles}
\end{figure}

\begin{figure}[h]
    \centering
    \includegraphics[width=\hsize]{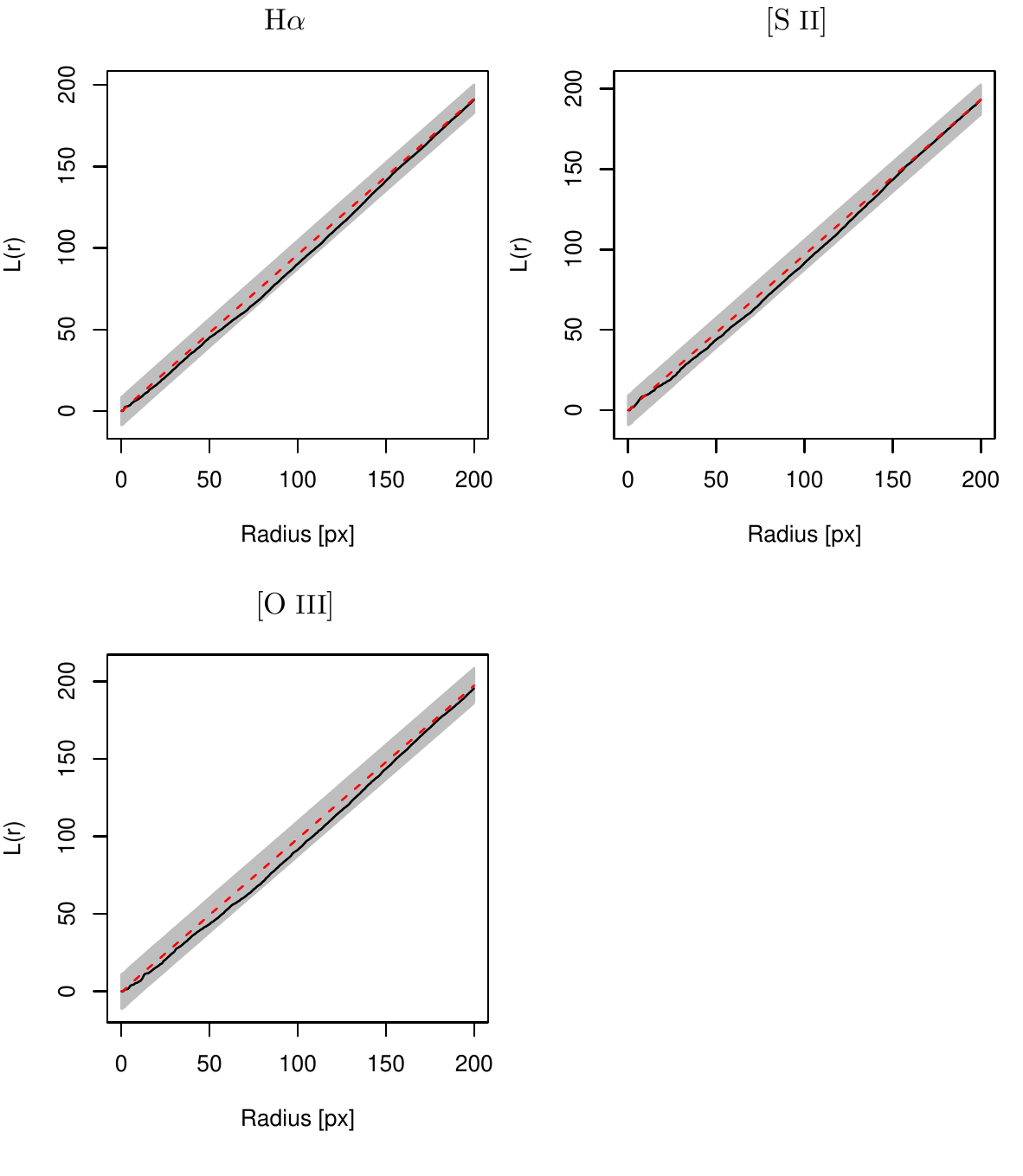}
    \caption{Bivariate Ripley’s L comparing the star formation rate 1\,Gyr ago as calculated by \cite{harriszaritsky2009} and bubbles detected in \ha{} (top left), \ion{S}{ii} (top right), and \ion{O}{iii} (bottom) at window sizes of 12-70\,px. Solid black line: L estimated from original pattern. Dashed red line: Pointwise mean of half of the simulated patterns. Gray envelopes: MAD. 10\,px corresponds to about 1.67' or 24\,pc at an LMC distance of 50\,kpc.}
    \label{appfig:bubbles_sfr}
\end{figure}

The bubble-bubble comparison is shown in Fig.~\ref{appfig:bubbles_bubbles}.  The L-function of the original pattern clearly exceeds the envelopes. The bubbles are correlated. The opposite holds for the bubble-SFR comparison (Fig.~\ref{appfig:bubbles_sfr}), where the stars are uncorrelated to the star formation 1\,Gyr ago.

\clearpage
\section{Superbubble cutout images}
\label{sec:superbubbles}
\begin{figure}[h!]
DEM L 25:\\
\includegraphics[width=\hsize]{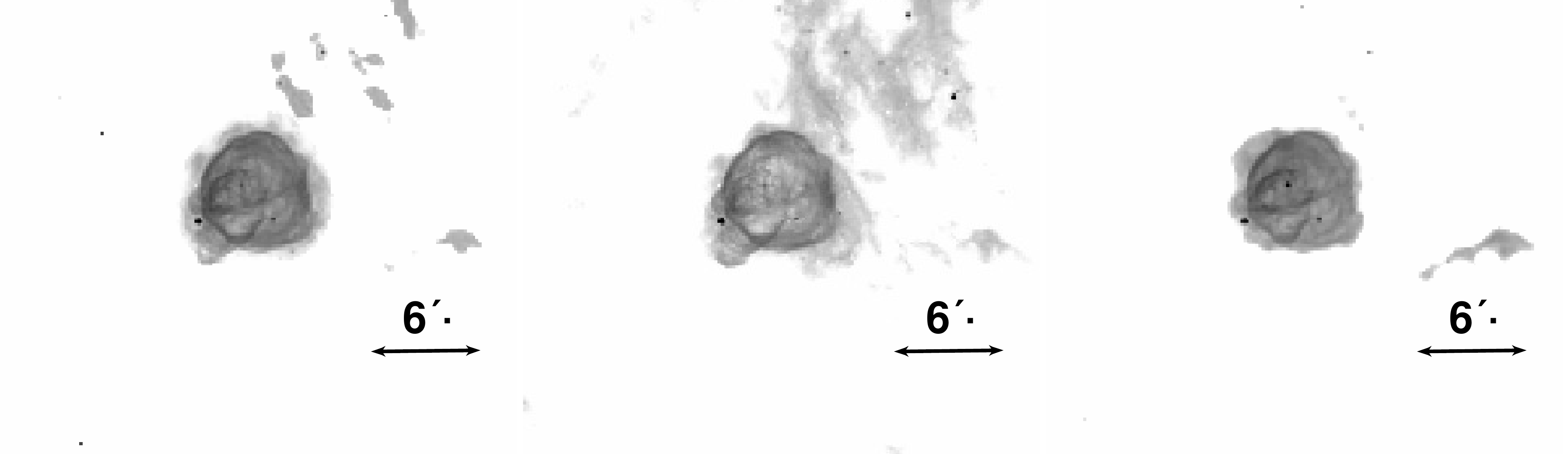}
DEM L 31:\\
\includegraphics[width=\hsize]{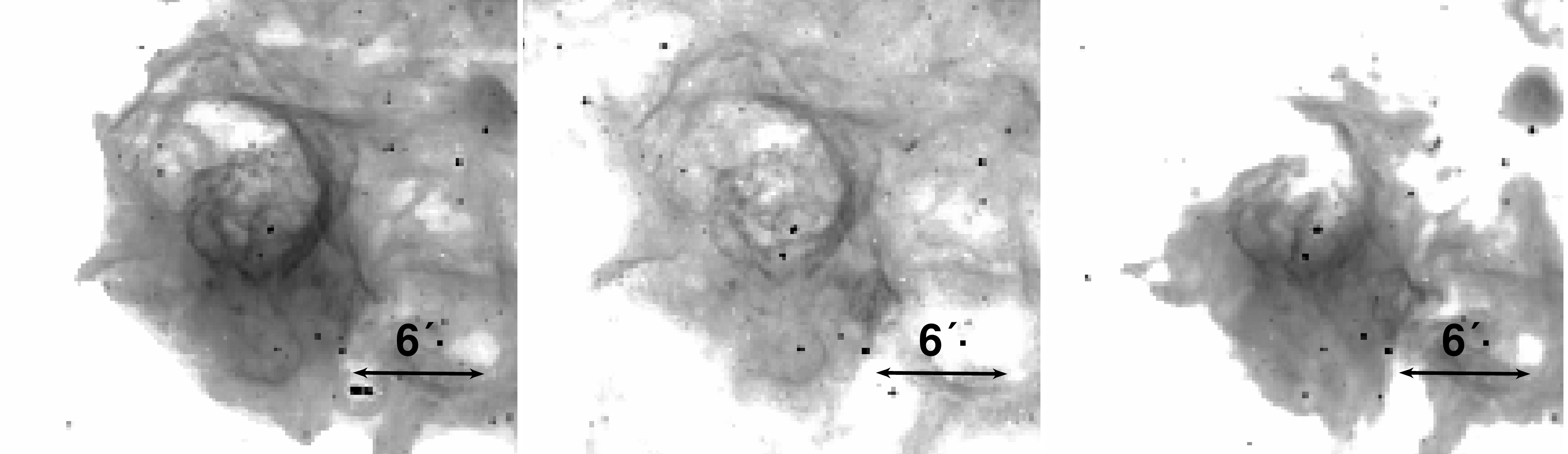}
DEM L 86:\\
\includegraphics[width=\hsize]{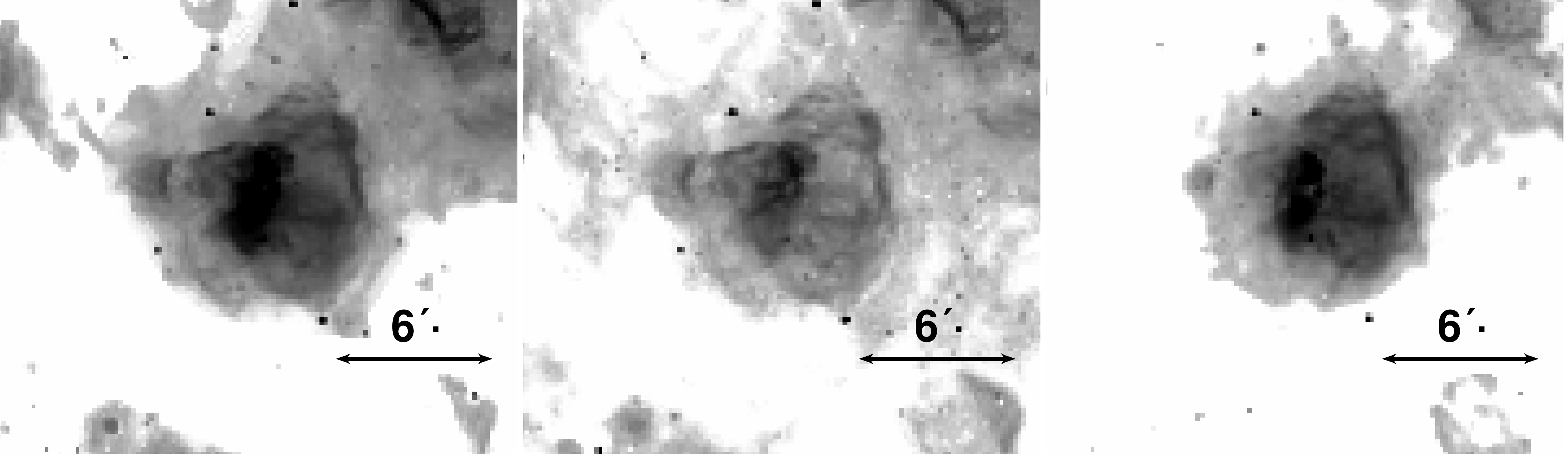}
DEM L 192:\\
\includegraphics[width=\hsize]{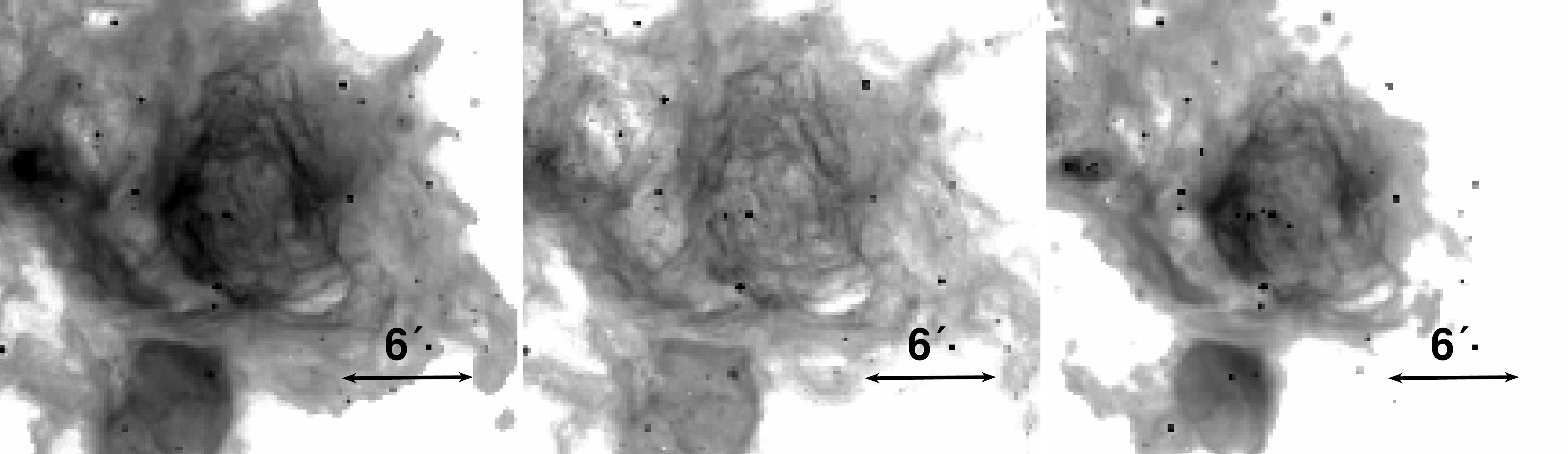}
DEM L 301:\\
\includegraphics[width=\hsize]{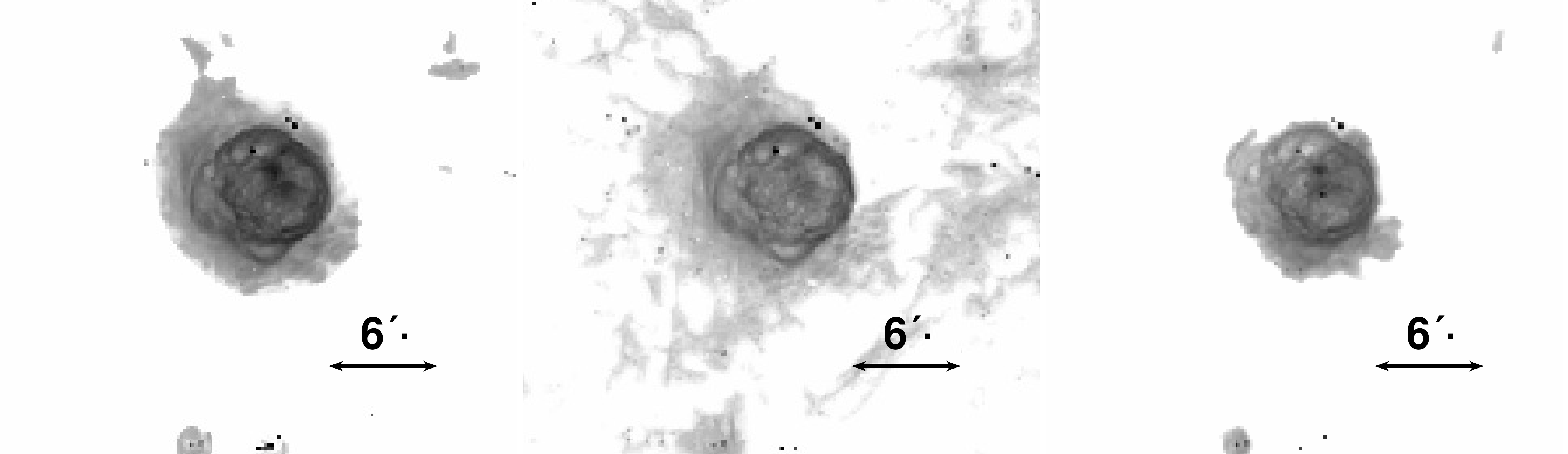}
\caption{Superbubbles detected in all wavelengths depicted in \ha\ (left), [\ion{S}{ii}] (center), and [\ion{O}{iii}] (right). The images are scaled to the min/max thresholds listed in Table~\ref{tab:thresh_lmc}. }
\label{appfig:all}
\end{figure}

\begin{figure}
DEM L 50:\\
\includegraphics[width=\hsize]{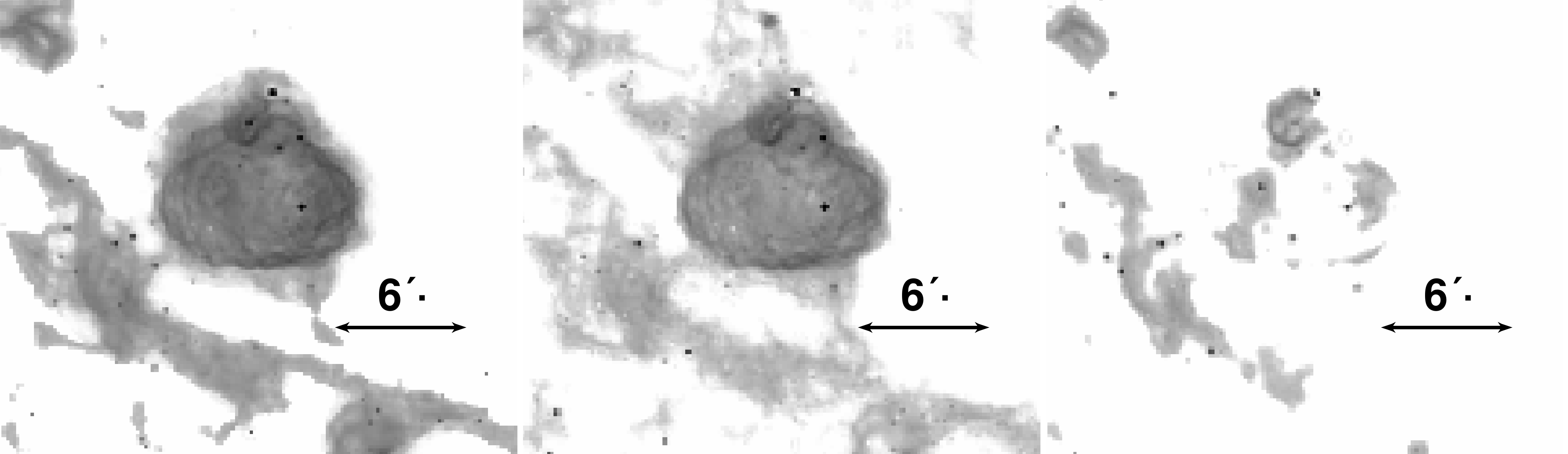}\\
DEM L 221:\\
\includegraphics[width=\hsize]{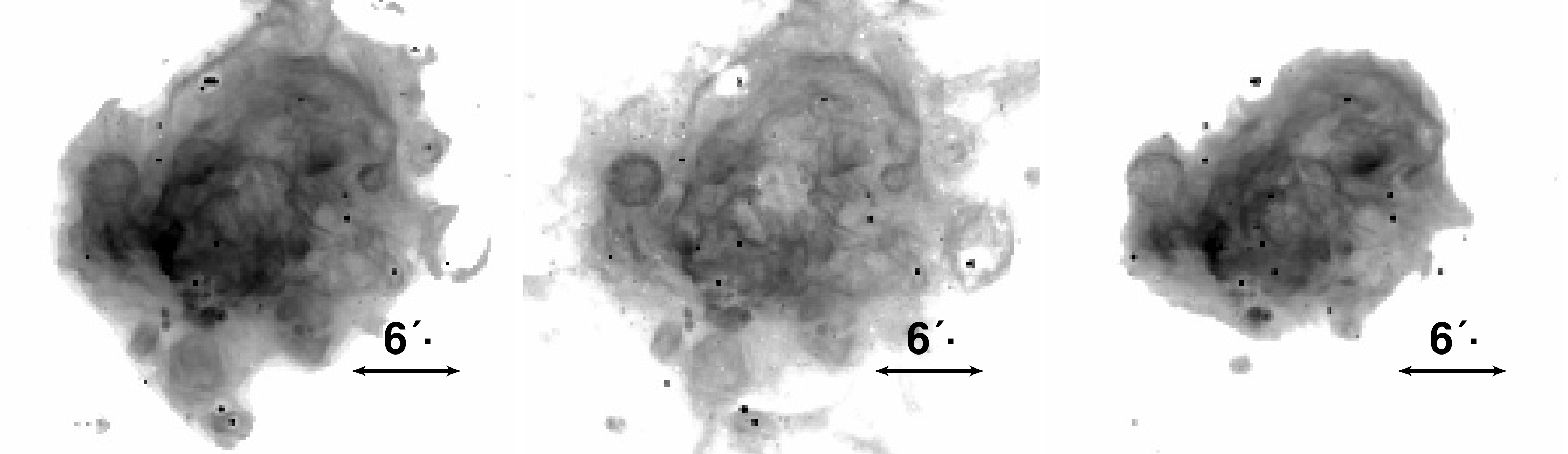}\\
DEM L 229:\\
\includegraphics[width=\hsize]{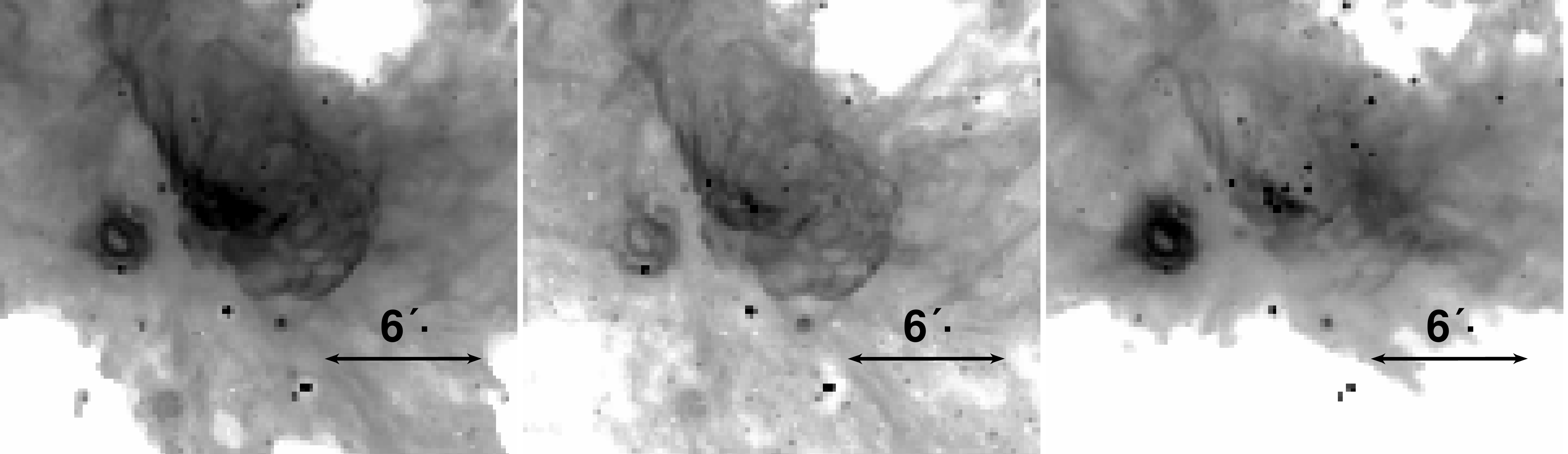}
\caption{Superbubbles detected in \ha\ and [\ion{S}{ii}] depicted in \ha\ (left), [\ion{S}{ii}] (center), and [\ion{O}{iii}] (right). The images are scaled to the min/max thresholds listed in Table~\ref{tab:thresh_lmc}. }
\label{appfig:hasii}
\end{figure}

\begin{figure}
DEM L 84:\\
\includegraphics[width=\hsize]{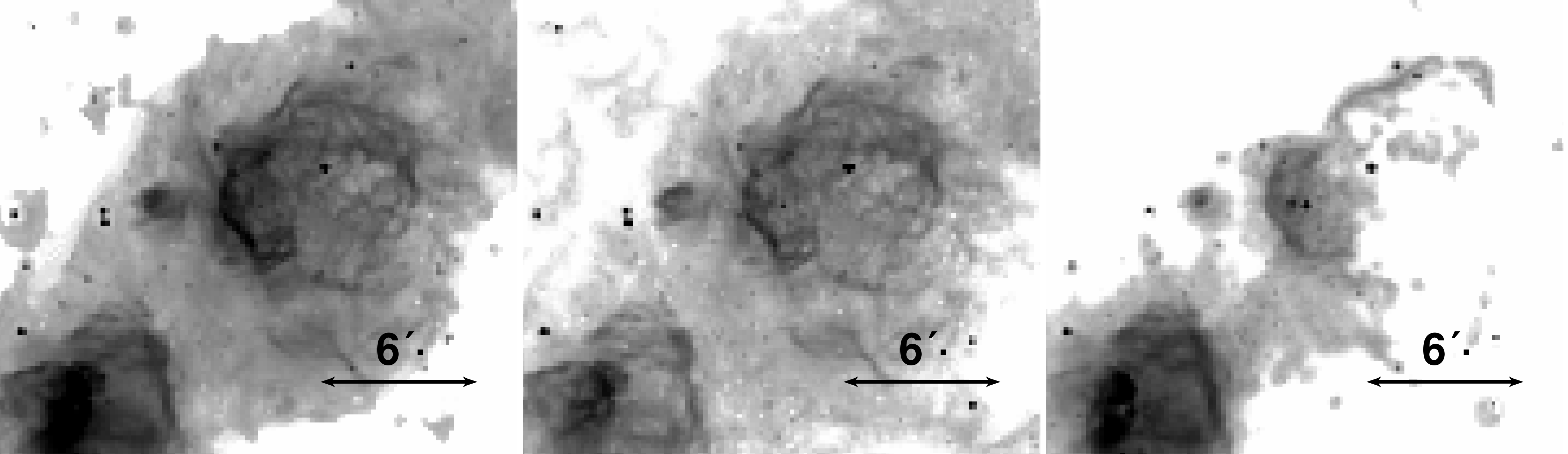}\\
DEM L 152:\\
\includegraphics[width=\hsize]{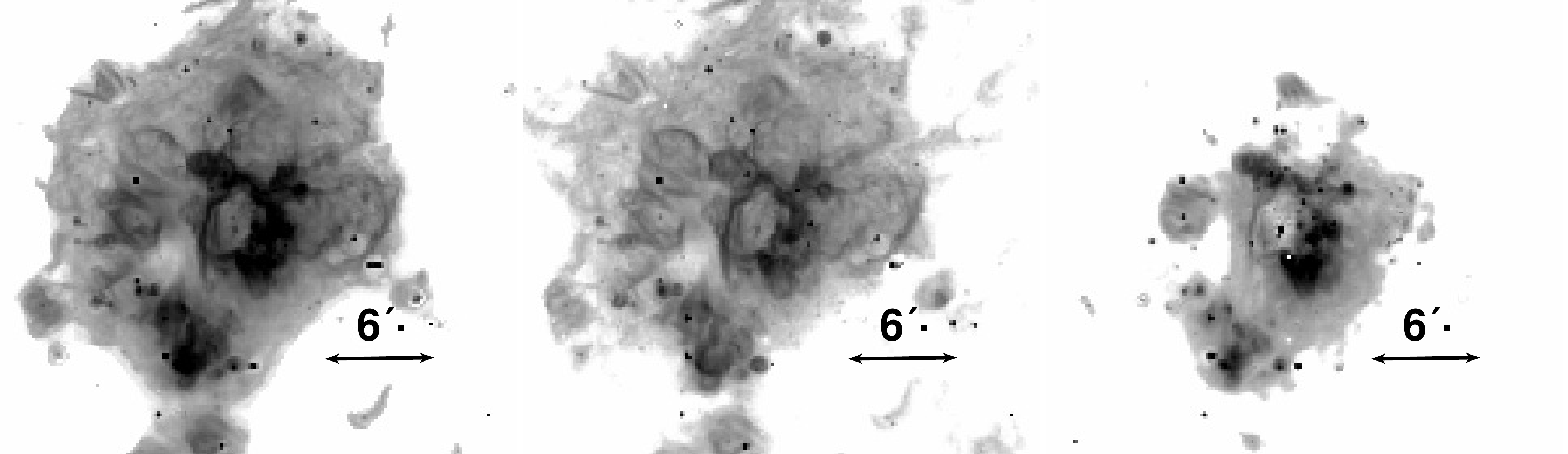}\\
DEM L 226:\\
\includegraphics[width=\hsize]{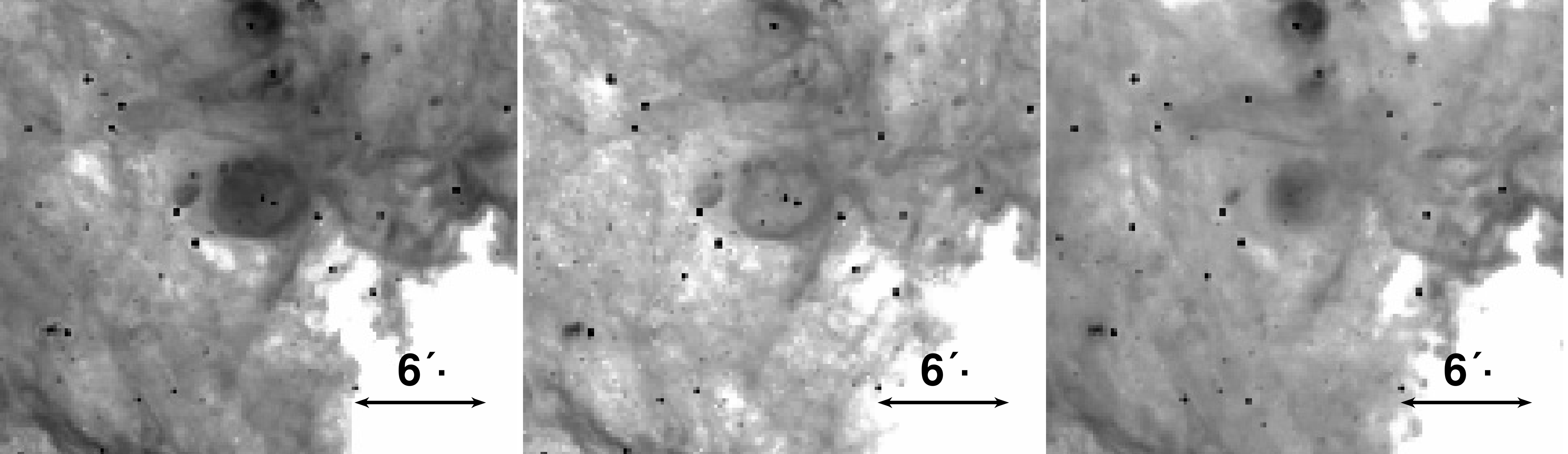}
\caption{Superbubbles detected in \ha\ depicted in \ha\ (left), [\ion{S}{ii}] (center), and [\ion{O}{iii}] (right). The images are scaled to the min/max thresholds listed in Table~\ref{tab:thresh_lmc}. }
\label{appfig:ha}
\end{figure}

\begin{figure}
\includegraphics[width=\hsize]{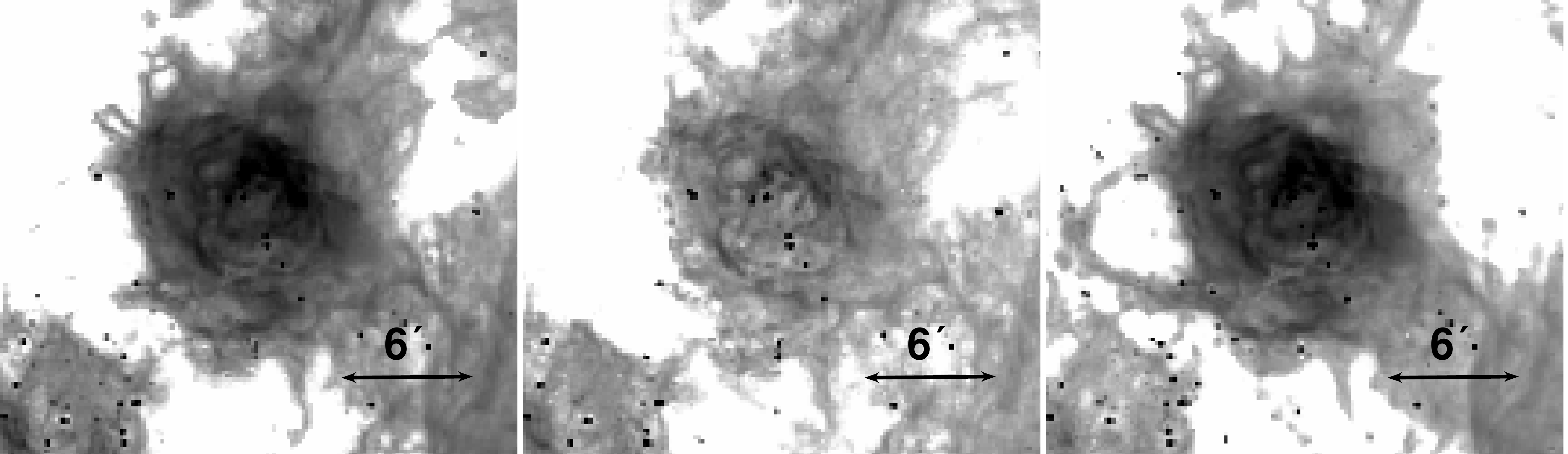}
\caption{Superbubble DEM L 199 detected in [\ion{S}{ii}] depicted in \ha\ (left), [\ion{S}{ii}] (center), and [\ion{O}{iii}] (right). The images are scaled to the min/max thresholds listed in Table~\ref{tab:thresh_lmc}. }
\label{appfig:sii}
\end{figure}

\begin{figure}
\includegraphics[width=\hsize]{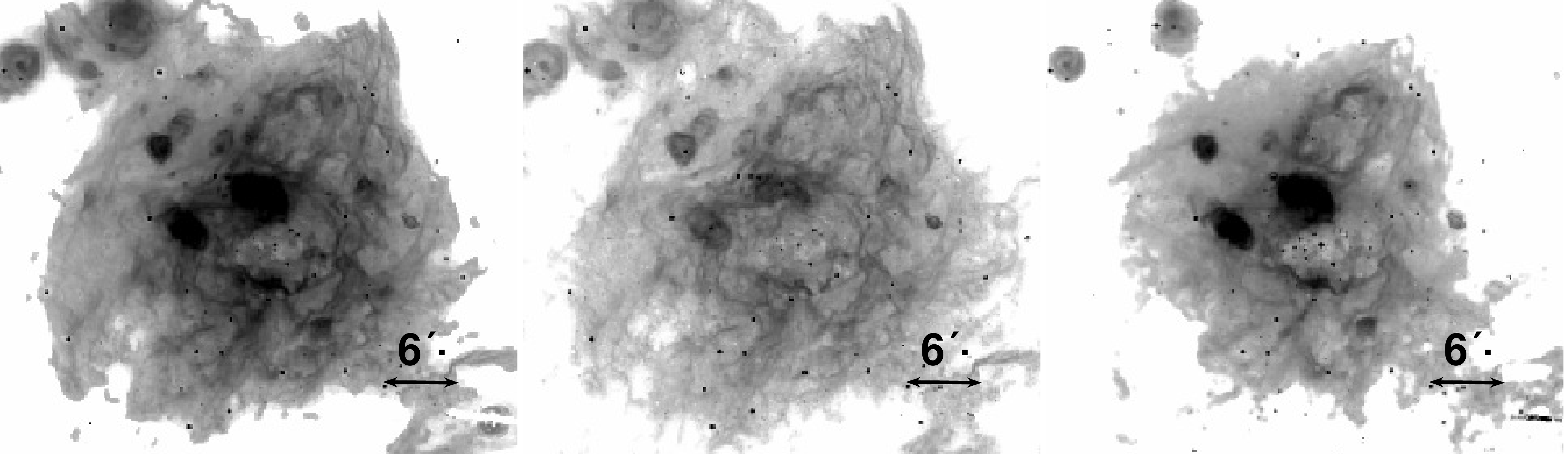}
\caption{Superbubble DEM L 34 detected in [\ion{O}{iii}] depicted in \ha\ (left), [\ion{S}{ii}] (center), and [\ion{O}{iii}] (right). The images are scaled to the min/max thresholds listed in Table~\ref{tab:thresh_lmc}. }
\label{appfig:oiii}
\end{figure}

In this section, images of the 20 superbubbles discussed in Sect.~\ref{sec:snr_superbubbles} are shown.
A clear
round structure is visible in all images for the objects detected in all wavelengths (Fig.~\ref{appfig:all}). If a bubble is detected in some but not all wavelengths (\ha\ and [\ion{S}{ii}]: Fig.~\ref{appfig:hasii}, \ha\ only: Fig.~\ref{appfig:ha}, [\ion{S}{ii}] only: Fig.~\ref{appfig:sii}, [\ion{O}{iii}] only: Fig.\ref{appfig:oiii}), this structure is not present or less clear in the other wavelengths, although the exact reason is not always obvious.

\begin{figure}
DEM L 106:\\
\includegraphics[width=\hsize]{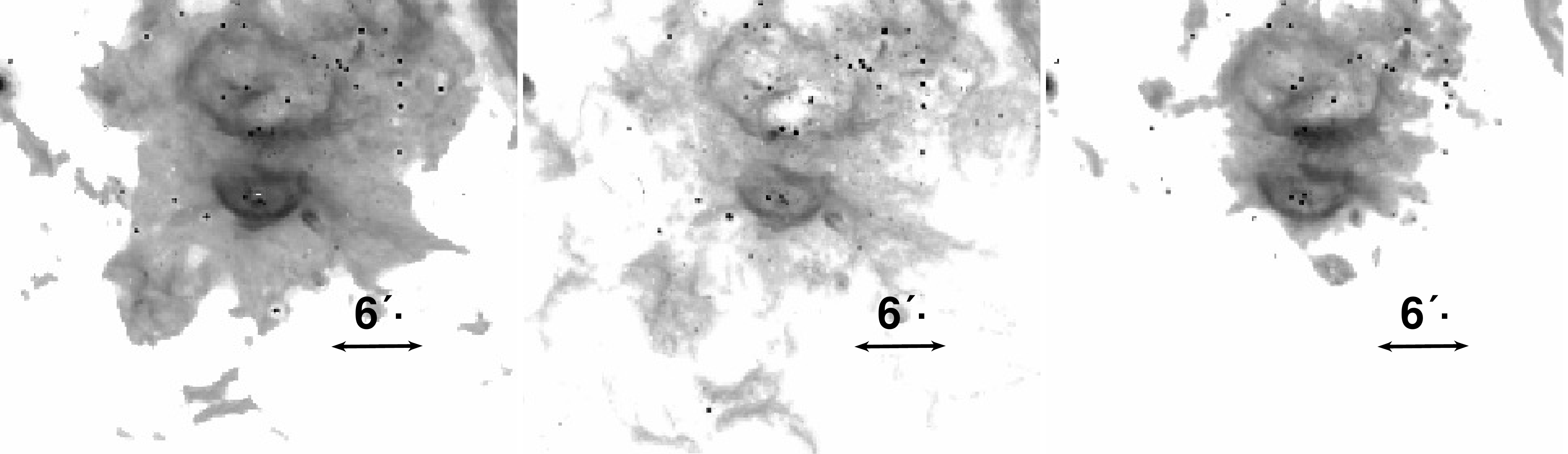}\\
DEM L 205:\\
\includegraphics[width=\hsize]{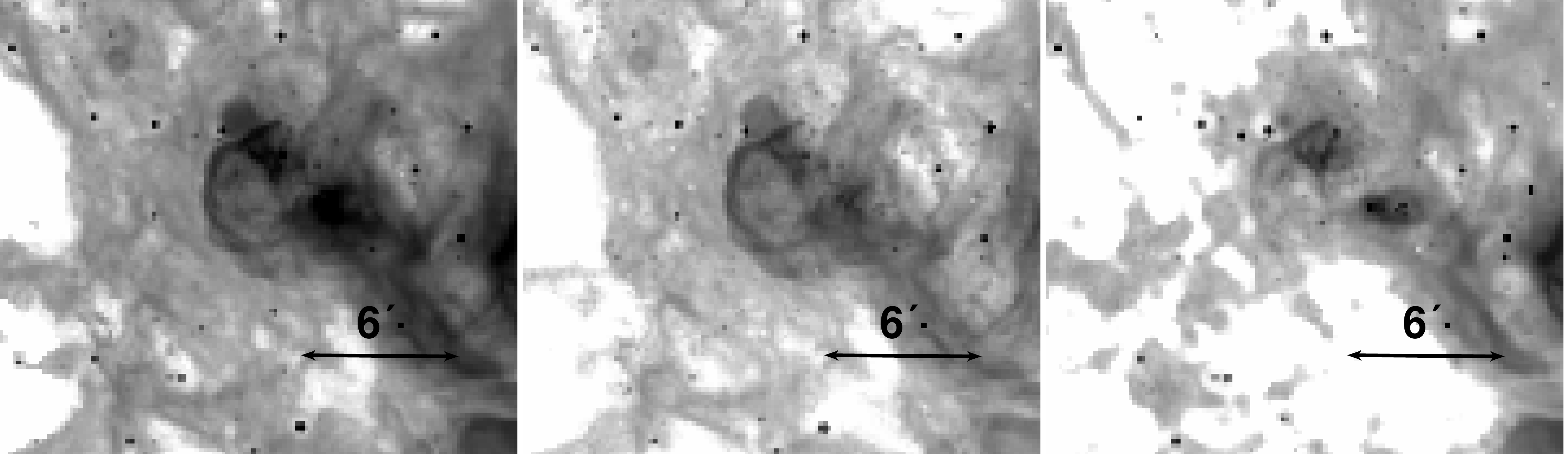}\\
DEM L 246:\\
\includegraphics[width=\hsize]{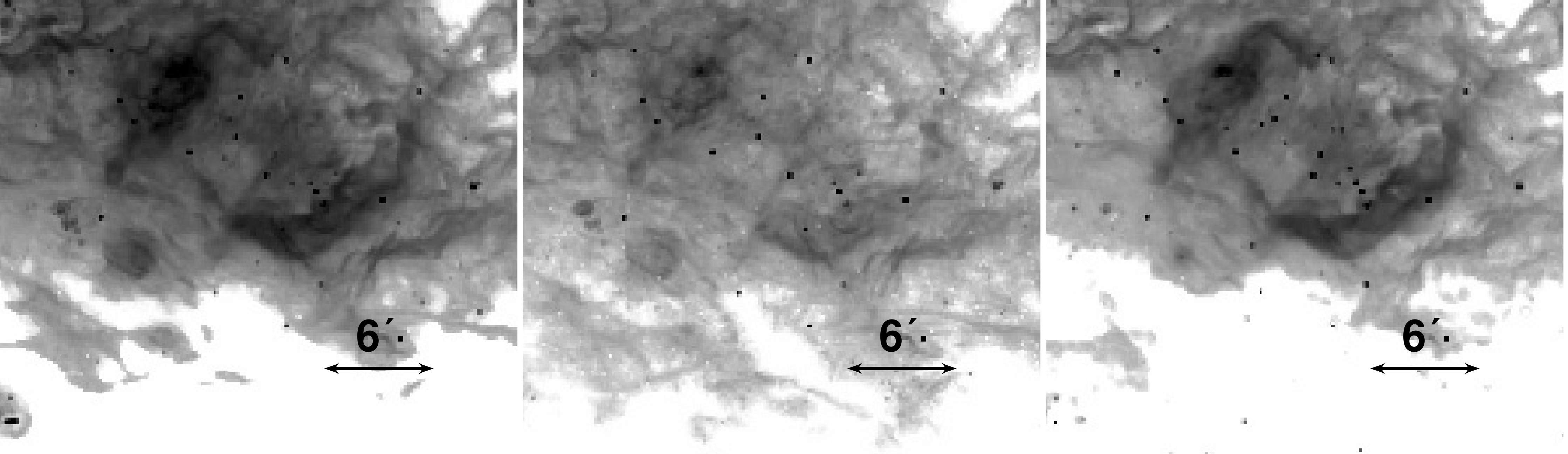}\\
DEM L 263:\\
\includegraphics[width=\hsize]{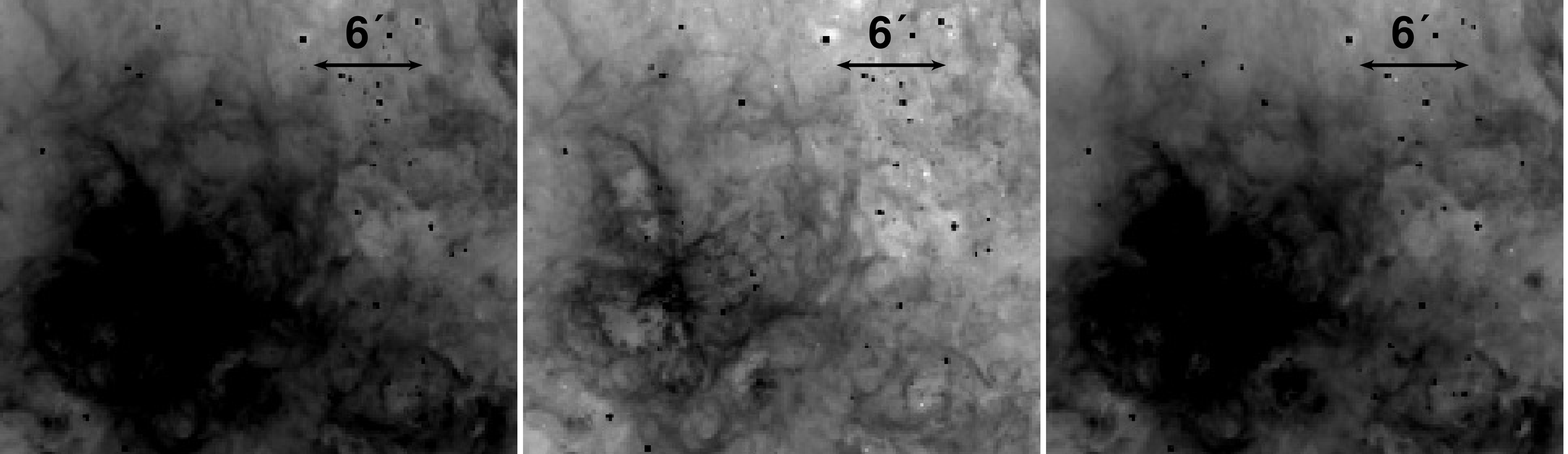}\\
{DEM L 269:\\
\includegraphics[width=\hsize]{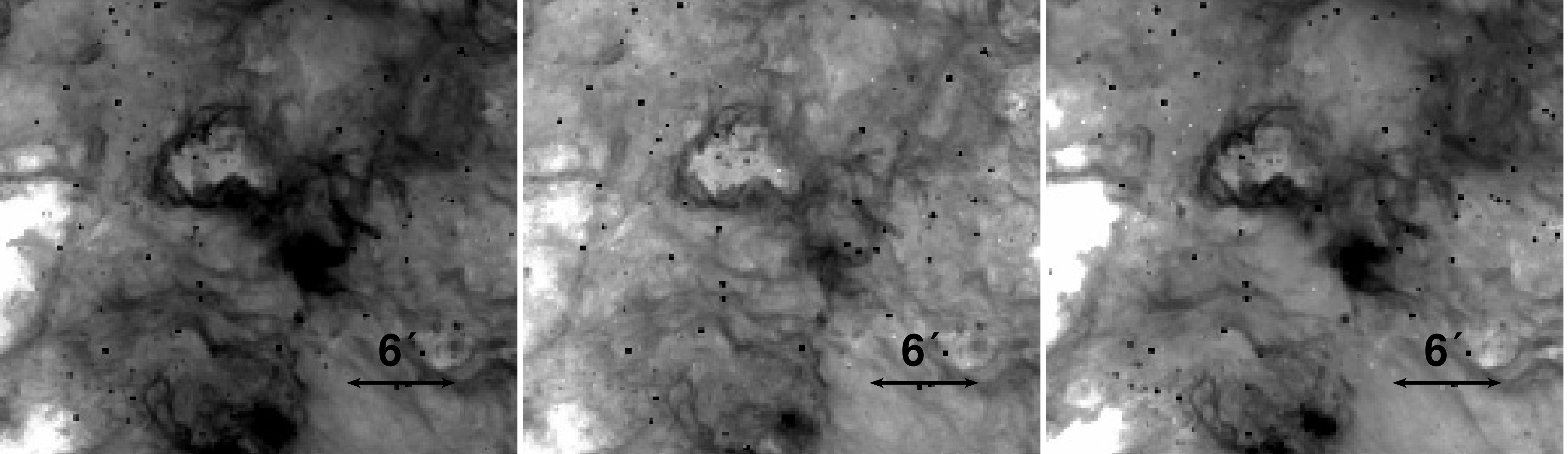}}\\
DEM L 284:\\
\includegraphics[width=\hsize]{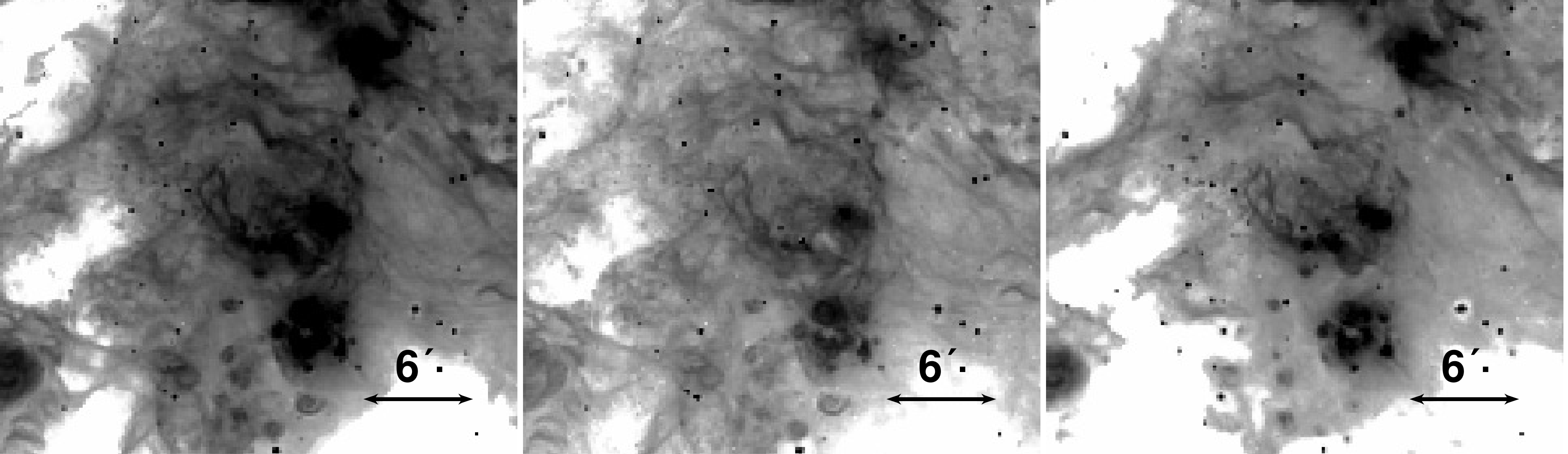}\\
30 Dor C:\\
\includegraphics[width=\hsize]{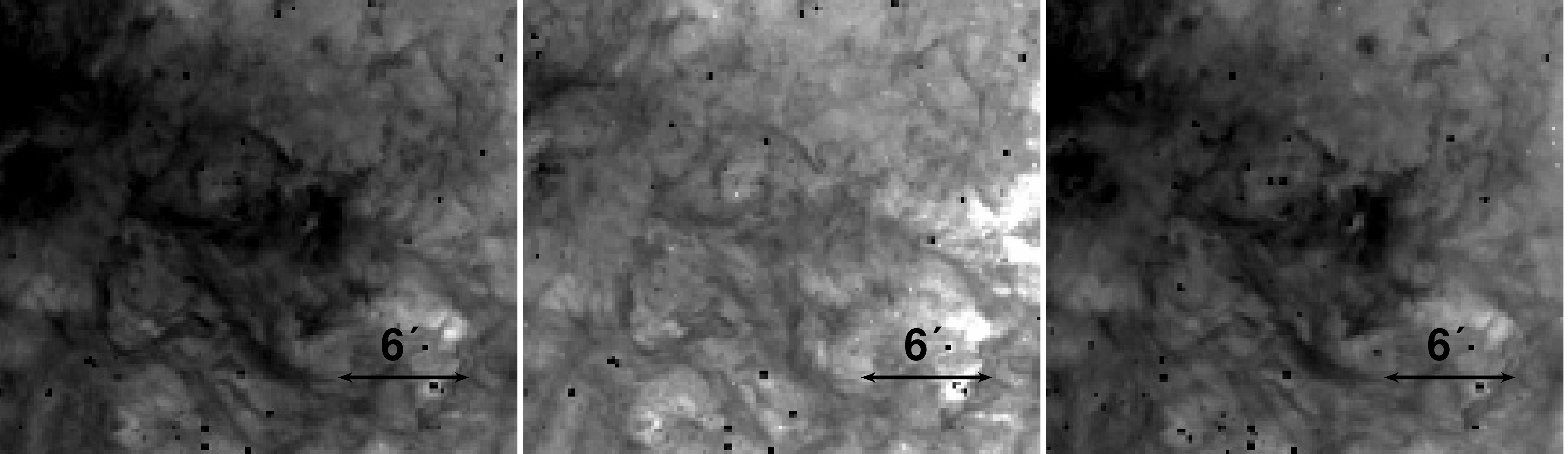}
\caption{Undetected superbubbles in \ha\ (left), [\ion{S}{ii}] (center), and [\ion{O}{iii}] (right). The images are scaled to the min/max thresholds listed in Table~\ref{tab:thresh_lmc}. }
\label{appfig:undet}
\end{figure}

The completely undetected objects (Fig.~\ref{appfig:undet}) are usually embedded in more complex structures and/or are not visible as round structures in our data. DEM L 106 is located near mosaic artifacts that disturb the detection.



\end{appendix}

\end{document}